\documentclass[preprint]{aastex}
\usepackage{apjfonts}
\usepackage{amsmath}
\usepackage{graphicx}
\usepackage{natbib}
\usepackage{longtable}
\usepackage{xcolor}
\shorttitle{Twin Neutron Stars}
\shortauthors{Hwang et. al.}

\def\pasa{\rm{PASA}}
\def\nar{\rm{New Astr. Rev.}}

\begin{document}

\title{Stability and Coalescence of Massive Twin Binaries}

\author{J. Hwang,$^{1,3}$ J. C. Lombardi Jr.,$^2$ F. A. Rasio,$^{1,3}$ \& V. Kalogera$^{1,3}$}
\affil{$^1$Department of Physics and Astronomy, Northwestern University, Evanston, IL 60208, USA}
\affil{$^2$Department of Physics, Allegheny College, Meadville, PA 16335, USA}
\affil{$^3$Center for Interdisciplinary Exploration and Research in Astrophysics (CIERA), Northwestern University, Evanston, IL 60208, USA}

\begin{abstract}
Massive stars are usually found in binaries, and binaries with periods less than 10 days may have a preference for near equal component masses (``twins''). In this paper we investigate the evolution of massive twin binaries all the way to contact and the possibility that these systems can be progenitors of double neutron star binaries.
The small orbital separations of observed double neutron star binaries suggest that the progenitor systems underwent a common envelope phase at least once during their evolution.
\citet{1998ApJ...506..780B} proposed that massive binary twins will undergo a common envelope evolution while both components are ascending the red giant branch (RGB) or asymptotic giant branch (AGB) simultaneously, also known as double-core evolution.
Using models generated from the stellar evolution code {\it EZ} ({\it Evolve Zero Age Main Sequence}), we determine the range of mass ratios resulting in a contact binary with both components simultaneously ascending the RGB or AGB as a function of the difference in birth times, $\Delta\tau$.
We find that, even for a generous $\Delta\tau=5$ Myr, the minimum mass ratio $q_\mathrm{min}=0.933$ for an $8\ M_\odot$ primary and increases for larger mass primaries.
We use a smoothed particle hydrodynamics ({\it SPH}) code, {\it StarSmasher}, to study specifically the evolution of $q=1$ common envelope systems as a function of initial component mass, age, and orbital separation.
We also consider a $q=0.997$ system to test the effect of relaxing the constraint of strictly identical components.
We find the dynamical stability limit, the largest orbital separation where the binary becomes dynamically unstable, as a function of the component mass and age. 
Finally, we calculate the efficiency of ejecting matter during the inspiral phase to extrapolate the properties of the remnant binary from our numerical results, assuming the common envelope is completely ejected.
We find that for the nominal core masses, there is a minimum orbital separation for a given component mass such that the helium cores survive common envelope evolution in a tightly bound binary and are viable progenitors for double neutron stars.
\end{abstract}

\keywords{binaries: close, binaries: general, hydrodynamics, instabilities, methods: numerical, stars: general} 

\section{Introduction}
\label{Sec:Intro}
We follow up on a previous study by \citet{2011ApJ...737...49L}, hereafter referred to as {\it Paper 1}, that investigated the hydrodynamic evolution of twin stars (mass ratio $q=1$) in a contact binary using polytropic models for the stellar envelopes.
We begin by refining the parameter space, specifically the limits on the mass ratio and component age.
We also update our stellar models, generated from a stellar evolution code in place of polytropic models $(\Gamma=5/3)$, and equation of state (see \S\ref{SSec:StarCrash}), as the mass transfer and common envelope evolution of the binary system are highly sensitive to the density and entropy profiles of the stellar envelopes and radiation pressure becomes significant in the envelopes of massive giants ($M_*\gtrsim14\ M_\odot$).
For our primary runs we restrict our parameter space to massive twin binaries ($8\ M_\odot < M_\mathrm{*}<20\ M_\odot$, $q=1$) where both stars are simultaneously ascending the RGB and the system will evolve into a helium star binary within a common envelope (described in \S\ref{Sec:Parameters}).
As a preliminary study, we also include a set of calculations with $q=0.997$ to test the importance of strictly identical components (described in \S\ref{SSec:q1_Models}).
We organize our binaries by component age and initial mass, vary the degree of contact (discussed in \S \ref{Sec:Numerical}), then dynamically evolve the system and follow the mass flow between the two stars, the common envelope, and the ejecta.
To determine if an unstable system survives as a tightly bound binary or undergoes a merger, we use energetics to extrapolate the final orbital separation after assuming that the system fully ejects the remaining gas in the common envelope.
The energy required to eject the remaining gas depends on the efficiency of ejection and the core mass (see \S\ref{SSec:Remnant_Properties}), and is extracted from the orbital energy of the helium cores, resulting in a decrease of the orbital separation.

Forming a double neutron star system may be difficult under the standard model. The systems we observe have small orbital separations, suggesting that the progenitor binary underwent an inspiral during a common envelope phase, reducing the initial orbital separation to the currently observed value. The result of an inspiral, either a merger or a tightly bound binary, depends very sensitively on the age and core mass of the stars and the initial orbital separation. 
\citet{2010NewAR..54...65T} suggest that the survival of the binary through the common envelope phase is more likely if at least one star in the binary is along the RGB or AGB, otherwise the binary components will likely merge, resulting in a system with one compact object. For an in-depth discussion of common envelope evolution in binaries we refer the reader to \citet{2000ARA&A..38..113T} and \citet{2010ApJ...720.1752P}.

In a progenitor system with a mass ratio far from unity, the primary will undergo a supernova before the secondary, and after the secondary ascends the RGB or AGB, the remnant neutron star begins to orbit within the gas envelope of the secondary, which may lead to a period of hypercritical accretion \citep{1998ApJ...506..780B}.
\citet{1998ApJ...506..780B} suggest that if the time between the two supernovae is large enough, the first resultant neutron star accumulates enough mass to collapse into a black hole, excluding these systems from being progenitors of double neutron star binaries.
\citet{1995ApJ...440..270B} proposes that progenitor systems with a mass ratio close enough to unity will enter a common envelope phase where both stars have evolved off the main sequence and eject the common envelope before the formation of the first neutron star.
\citet{2006MNRAS.368.1742D} predicts a double neutron star birth rate through this channel, also known as double-core evolution, at $~0.16-24\ \mathrm{Myr}^{-1}$.
\citet{1995ApJ...440..270B} assumes an equation of state with a maximum neutron star mass of roughly  $1.5\ M_\odot$, shown to be an underestimate by the observation of a $2.0\ M_\odot$ neutron star \citep{2010Natur.467.1081D}.
However, the constraint on the maximum timescale in the presence of hypercritical accretion may still place strong restrictions on the mass ratio of the progenitor system.

Observational studies of massive star populations have shown mixed results on the frequency of twins. \citet{1979AJ.....84..401L} were the first to suggest that close binaries may have a narrow peak at $q\simeq1$. \citet{2006ApJ...639L..67P} analyze the binary parameters of 50 eclipsing binaries (21 detached, 28 semi-detached, and 1 contact) in the Small Magellanic Cloud and suggest that the data show a twin population ($q>0.95$) of $20-25\%$. \citet{2006ApJ...639L..67P} observe that these twin binary systems exhibit a preference towards a lower eccentricity and are only preferentially found in low period populations ($P<1000$ days). 
\citet{2012Sci...337..444S} find that roughly $71\%$ of all massive stars ($m>15\ M_\odot$) underwent mass transfer and report a power law for number density with respect to the mass ratio with an exponent of $-0.1\pm0.6$.
The observational evidence that does support the preference for $q\sim1$ in binaries suggests that close binaries have a much higher occurrence of twins than long-period binary systems, indicating that the mechanism causing a preference for mass ratios near unity likely occurs in binary systems created during formation rather than capture.
\citet{2000MNRAS.314...33B} shows that accretion from a circumbinary disk tends to drive the mass ratio towards unity for massive, close binaries.
In addition, \citet{2007ApJ...661.1034K} argue that pre-main sequence mass transfer in a close binary consisting of two massive protostars tends to both circularize the orbits and push the mass ratio towards $q=1$, leading to a twin population among massive stars ($M_*\ge5-10\ M_\odot$) in a tight binary ($a\le25$ AU).
For further discussion of the ``twin'' population in binaries we refer the reader to \citet{2006A&A...457..629L} and \citet{2000A&A...360..997T}.

In massive-star binaries the common envelope phase begins when the mass transfer timescale decreases below that of the stability timescale, defined as the time required for the angular momentum redistribution between the spin and the orbit to bring the components of the system into a state of synchronous rotation.
When this occurs, both stars fill their Roche lobes and the system becomes dynamically unstable. Binaries with $q=1$ can still be in equilibrium when gas reaches the $L_\mathrm{1}$ Lagrangian point due to the symmetry of these systems; a contact binary configuration is formed instead.
These systems can become unstable, however, once mass outflows through the $L_\mathrm{2}$ and $L_\mathrm{3}$ Lagrangian points (see {\it Paper 1}), during which, the the helium cores undergo an inspiral as the energy and angular momentum of the orbiting cores are lost to the surrounding gas.
The increase in energy causes the outermost gas to become unbound from the system, forming an equatorial outflow around the helium cores.
The decrease in orbital separation of the helium cores causes spiral shocks which generate an additional outwards force, further expelling gas from the center of the system \citep{2010NewAR..54...65T}.
In a surviving binary, the orbital separation begins to stabilize due to a lack of gas around the cores and the final separation may decrease by two orders of magnitudes from the original separation \citep{2000ARA&A..38..113T}.

The rest of the paper is organized as follows: In \S\ref{Sec:Parameters} we discuss our choice of initial conditions, specifically the mass ratios and ages of the components.
In \S\ref{Sec:Numerical} we go over the numerical methods used in the code as well as the changes from {\it Paper 1}.
In \S\ref{Sec:Results} we discuss the results of the numerical runs and in \S\ref{Sec:Analysis} we use the results to extrapolate the orbital parameters of the remnant systems.
We summarize our conclusions in \S\ref{Sec:Conclusions}.
\pagebreak

\section{Discussion of Parameter Space and Methods}
\label{Sec:Parameters}
\subsection{Stellar Models}
\label{SSec:Stellar_Models}
We are primarily interested in binary configurations that will evolve into two tightly bound helium cores orbiting within a common envelope and constrained our study to binary systems where both stars have evolved a definable core, an inner region of depleted Hydrogen ($X<0.01$), and a gaseous envelope.
We use a stellar evolution code, {\it EZ} \citep{2004PASP..116..699P}, to produce a library of stellar models with masses between $7.50$ and $21.50\ M_\odot$ in mass increments of $0.01\ M_\odot$. We assume that the two stars in a binary are coeval to a degree, $\Delta\tau$, and refine our parameter space by eliminating mass combinations where the two stars do not have an overlapping RGB after allowing for the most generous $\Delta\tau$.

\begin{figure*}[htp]
\begin{center}
\begin{tabular}{cc}
\includegraphics[width=12.0cm]{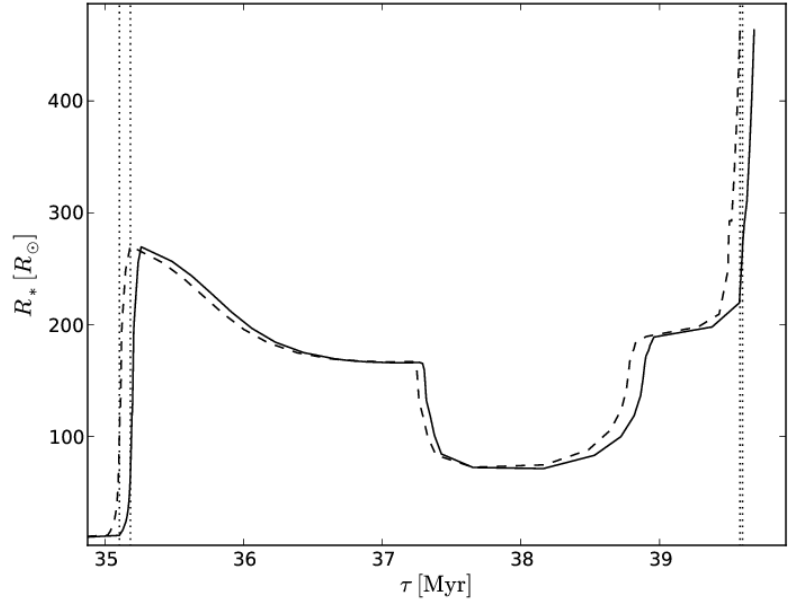}\\
\end{tabular}
\end{center}
\caption{Radius as a function of age, $\tau$, for two isolated stars, one of mass $8.00\ M_\odot$ (solid curve) and the other of mass $8.01\ M_\odot$ (dashed curve), showing two bounded areas as described in \S\ref{SSec:Stellar_Models}.
The first bounded region, $\tau_\mathrm{RGB}<\tau\le\tau_\mathrm{max}$, begins when the secondary begins to ascend the RGB and ends when the primary star's radius reaches the local maximum, $R_\mathrm{max}$, on the RGB.
The second bounded region, $\tau_\mathrm{AGB}<\tau<\tau_\mathrm{end}$, begins when the primary star's radius first exceeds $R_\mathrm{max}$ and ends when the primary undergoes a supernova.
We find that $\tau_\mathrm{max}-\tau_\mathrm{RGB}>\tau_\mathrm{end}-\tau_\mathrm{AGB}$.}
\label{Fig:RGB_8.0_R}
\end{figure*}

We constrain the age of the binary by requiring that either component overflows its Roche Lobe.
We designate four critical ages: $\tau_\mathrm{RGB}$, when the secondary begins ascending the RGB, $\tau_\mathrm{max}$, corresponding to the point along the RGB when the primary reaches a local maximum radius ($R_\mathrm{max}$), $\tau_\mathrm{AGB}$, when the primary has a radius larger than $R_\mathrm{max}$ while ascending the AGB, and $\tau_\mathrm{end}$, when the primary reaches the end of the AGB.
We generate two bounded regions shown in Figure~\ref{Fig:RGB_8.0_R}: $\tau_\mathrm{RGB}<\tau\le\tau_\mathrm{max}$ and $\tau_\mathrm{AGB}<\tau<\tau_\mathrm{end}$.
A binary stable at $\tau_\mathrm{max}$ will likely remain stable in the range $\tau_\mathrm{max}<\tau<\tau_\mathrm{AGB}$, as the primary is closest to Roche Lobe overflow at at $\tau_\mathrm{max}$.
We note also that expanding our study to include systems where both components are ascending the AGB may increase the maximum orbital separation resulting in hydrodynamic instability, but requires a very narrowly tuned set of initial conditions to even form a contact binary.
Specifically, the initial orbital separation must be large enough such that the binary does not undergo hydrodynamic instability during the RGB and small enough such that the binary enters a contact configuration, leading to a tight constraint on the allowed mass ratio and age difference of the components (see Fig.~\ref{Fig:RGB_8.0_R} and Table~\ref{TBL:min_q}), as well as the maximum degree of contact.
We disregard potential contact configurations where the primary is on the AGB and the secondary is on the RGB, as this configuration requires the same narrowly tuned set of initial conditions to even form a contact binary as the double-AGB case, and there is no guarantee that two stars with very different mass profiles will form a double-core system.

\begin{figure*}[htp]
 \begin{center}
  \includegraphics[width=8.0cm]{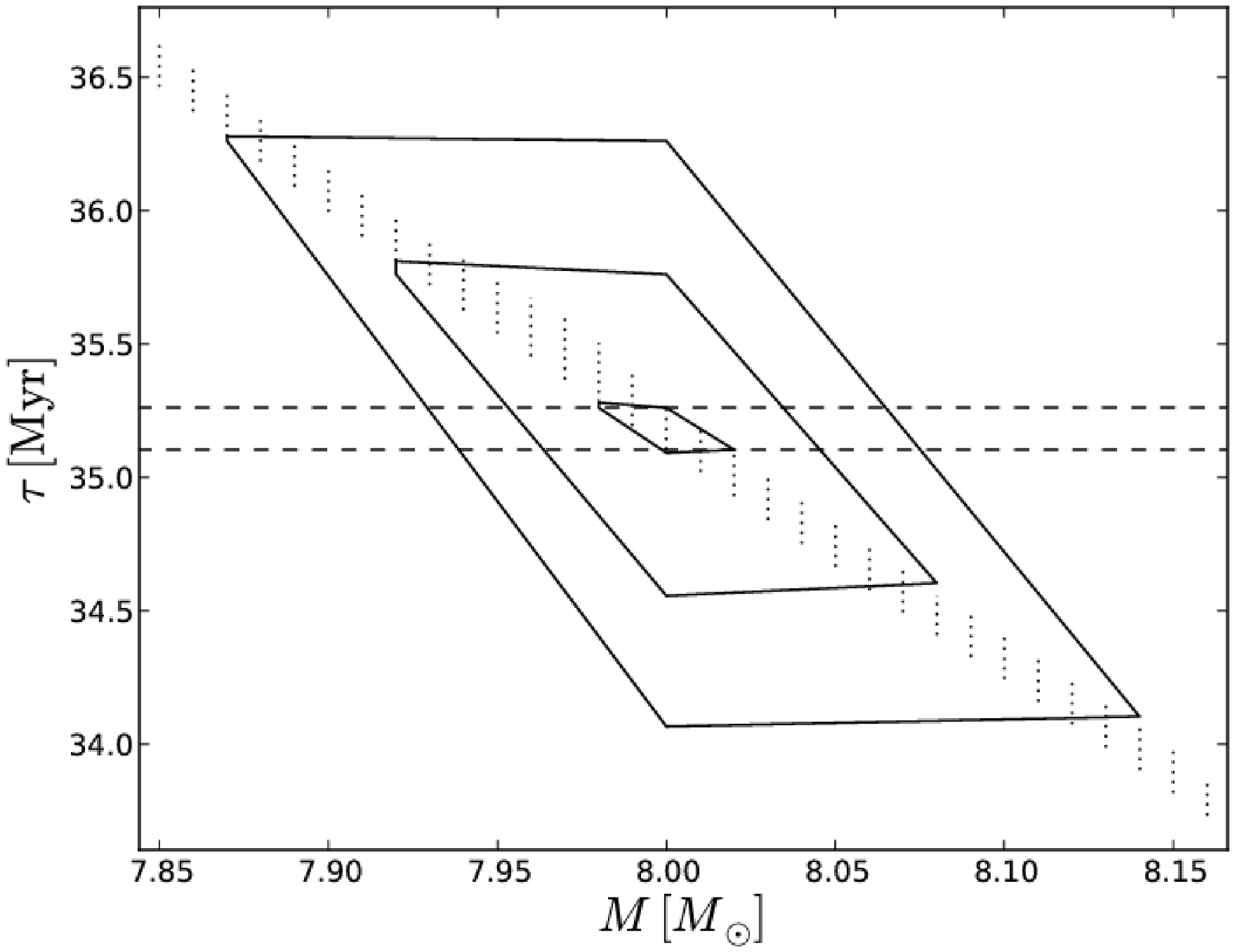}
  \includegraphics[width=8.0cm]{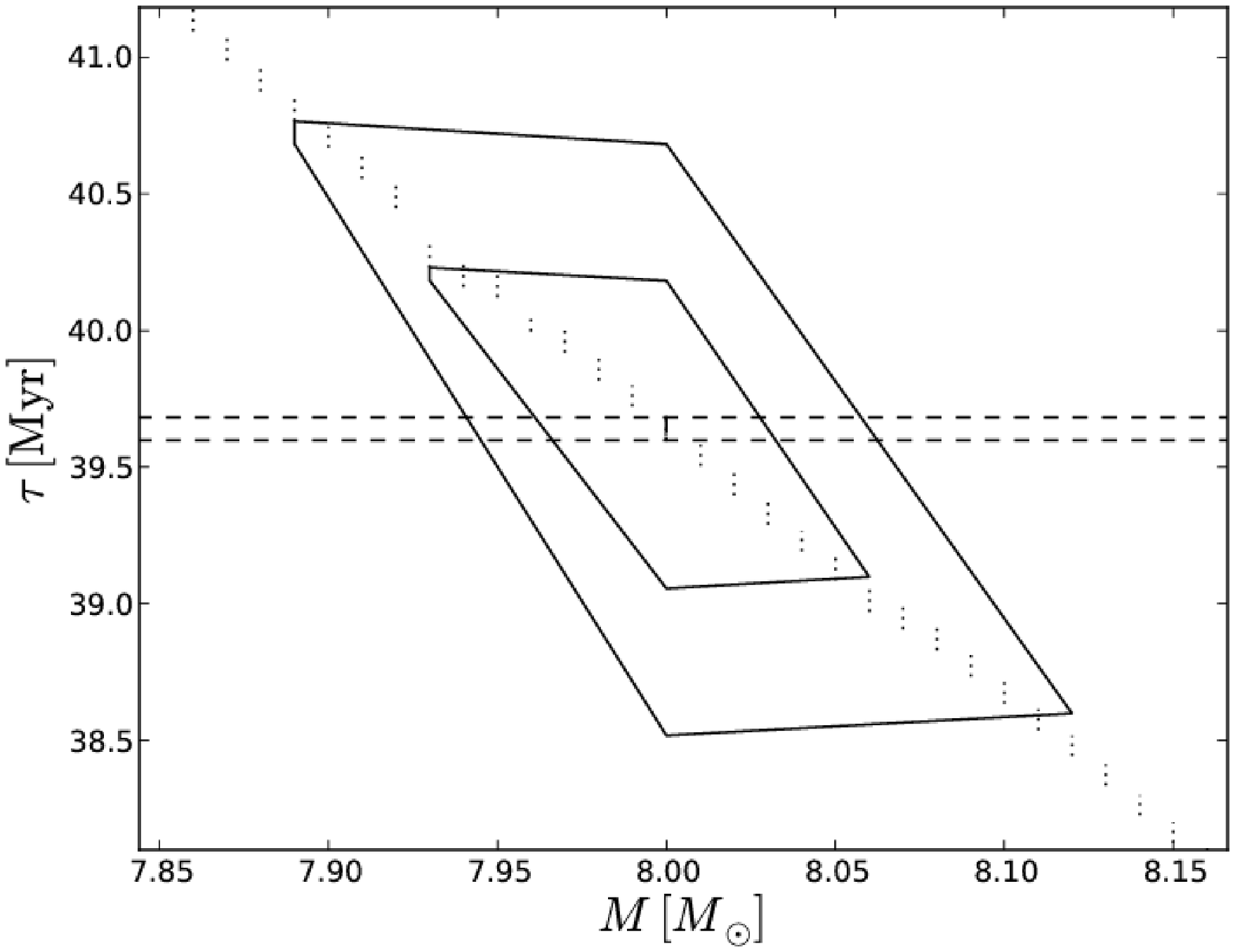}
 \end{center}
\caption{Ages for a binary where $\tau_\mathrm{RGB}<\tau_\mathrm{max}$ (left) and $\tau_\mathrm{AGB}<\tau_\mathrm{end}$ (right) as a function of the companion mass for a system with an $8.00\ M_\odot$ component, assuming both stars are coeval to a value $\Delta\tau$, referenced in Table~\ref{TBL:min_q}.
The horizontal dashed lines show $\tau_\mathrm{RGB}$ and $\tau_\mathrm{max}$ (left) and $\tau_\mathrm{AGB}$ and $\tau_\mathrm{end}$ (right) of the $8.00\ M_\odot$ component and the vertical dotted lines show $\tau_\mathrm{RGB}$ and $\tau_\mathrm{max}$ (left) and $\tau_\mathrm{AGB}$ and $\tau_\mathrm{end}$ (right) of the companion, defined in \S \ref{SSec:Stellar_Models}.
The area enclosed by the solid lines designate configurations where $\tau_\mathrm{RGB}<\tau\le\tau_\mathrm{max}$ (left) and $\tau_\mathrm{AGB}<\tau<\tau_\mathrm{end}$ (right). The largest region corresponds to $\Delta\tau=1.0\ \mathrm{Myr}$, the second region to $\Delta\tau=0.5\ \mathrm{Myr}$, and the smallest region to a strictly coeval binary.
We find that even allowing for $\Delta\tau=5.0\ \mathrm{Myr}$, $q_\mathrm{min}=0.933$}
\label{Fig:min_q}
\end{figure*}

In order to investigate the limits assuming non-coeval evolution, we find the minimum mass ratio such that both stars have simultaneously evolved onto the RGB or AGB after allowing for differences in birth times up to $\Delta\tau$.
The minimum values for $q$ are calculated assuming that the primary is born at a time $\Delta\tau$ after the secondary, allowing for the greatest variance in mass ratios, while still fulfilling the criteria that $\tau_\mathrm{RGB}<\tau_\mathrm{max}$ or $\tau_\mathrm{AGB}<\tau_\mathrm{end}$. We use $0 < \Delta\tau < 5$ Myr, considering the observation by \citet{2008Natur.453.1079S}, showing strong evidence of a $300$ kyr difference in birth times between two newly formed stars with $q\simeq1$, and observations of high-mass stars in clusters \citep{2000AJ....119.2214M} suggesting that high-mass stars evolve coevally up to a difference of $\Delta\tau<1$ Myr.
Figure~\ref{Fig:min_q} shows the valid companion mass and system age for an $8.0\ M_\odot$ star at $35.26$ Myr as a function of $\Delta\tau$ where $\tau_\mathrm{RGB}<\tau_\mathrm{max}$ and $\tau_\mathrm{AGB}<\tau_\mathrm{end}$. We find that in the strictly coeval case, $q_\mathrm{min}=0.996$, and with $\Delta\tau=5\ \mathrm{Myr}$, $q_\mathrm{min}=0.936$.
Table~\ref{TBL:min_q} summarizes the minimum allowed mass ratios for an $8.0\ M_\odot$ primary component as a function of $\Delta\tau$.
For more massive primary components the minimum mass ratio is even closer to unity as the radial expansion timescales are more compact, resulting in an even smaller overlapping region.
Allowing for $\Delta\tau=5\ \mathrm{Myr}$, we found $q_\mathrm{min}=0.965$ for a $14.0\ M_\odot$ primary component and $q_\mathrm{min}=0.975$ for a $20.0\ M_\odot$ primary component.
Our results differ from the analysis in \citet{2006MNRAS.368.1742D}, which finds that larger component masses relax the constraint on $q_\mathrm{min}$, primarily from omitting cases where the primary is ascending the AGB while the secondary is on the RGB and cases where $\tau_\mathrm{max}<\tau<\tau_\mathrm{AGB}$.

\begin{deluxetable}{cccc}
\tablewidth{7cm}
\tabletypesize{\small}
\tablecolumns{4}
\tablecaption{Minimum mass ratios with an $8.0\ M_\odot$ primary $(q=m_\mathrm{2}/m_\mathrm{1}<1)$\label{TBL:min_q}}
\tablehead{
    \colhead{Condition} & \colhead{$\Delta\tau\ [\mathrm{Myr}]$} & \colhead{$m_\mathrm{min}\ [M_\odot]$} & \colhead{$q_\mathrm{min}$}
   }
\startdata
$\tau_\mathrm{RGB}<\tau_\mathrm{max}$ & $0$   & $7.97$ & $0.996$\\
$\tau_\mathrm{RGB}<\tau_\mathrm{max}$ & $0.1$ & $7.96$ & $0.995$\\
$\tau_\mathrm{RGB}<\tau_\mathrm{max}$ & $0.5$ & $7.91$ & $0.989$\\
$\tau_\mathrm{RGB}<\tau_\mathrm{max}$ & $1.0$ & $7.86$ & $0.983$\\
$\tau_\mathrm{RGB}<\tau_\mathrm{max}$ & $5.0$ & $7.46$ & $0.936$\\
\\
$\tau_\mathrm{AGB}<\tau_\mathrm{end}$ & $0$   & $7.99$ & $0.999$\\
$\tau_\mathrm{AGB}<\tau_\mathrm{end}$ & $0.1$ & $7.97$ & $0.996$\\
$\tau_\mathrm{AGB}<\tau_\mathrm{end}$ & $0.5$ & $7.92$ & $0.990$\\
$\tau_\mathrm{AGB}<\tau_\mathrm{end}$ & $1.0$ & $7.88$ & $0.985$\\
$\tau_\mathrm{AGB}<\tau_\mathrm{end}$ & $5.0$ & $7.20$ & $0.943$\\
\enddata
\tablecomments{Here we summarize the results constraining the minimum mass ratio assuming that $\tau_\mathrm{RGB}<\tau_\mathrm{max}$ or $\tau_\mathrm{AGB}<\tau_\mathrm{end}$ for an $8.0\ M_\odot$ primary. $\Delta\tau$ is the maximum difference in birth times between the two components, $m_\mathrm{min}$ is the minimum mass of the secondary such that $\tau_\mathrm{RGB}<\tau_\mathrm{max}$ or $\tau_\mathrm{AGB}<\tau_\mathrm{end}$, and $q_\mathrm{min}$ is the corresponding mass ratio. We note that the range of mass ratios resulting in $\tau_\mathrm{RGB}<\tau_\mathrm{max}$ is larger than that for $\tau_\mathrm{AGB}<\tau_\mathrm{end}$.}
\end{deluxetable}

\subsection{Binary Twin Models}
\label{SSec:q1_Models}
In what follows we focus on coeval, $q=1$ binary systems with component birth masses $m_\mathrm{i}=8.0\ M_\odot$, $m_\mathrm{i}=14.0\ M_\odot$, and $m_\mathrm{i}=20.0\ M_\odot$ in a circular orbit ($e=0$).
We first discuss the timescales associated with this problem in order to organize the different evolution mechanisms of our binary system.
Hydrodynamic mass transfer occurs roughly on the orbital period, $P_\mathrm{orb} \simeq 2.8\, a^{3/2}\,M_\mathrm{tot}^{-1/2}\,\,\mathrm{hr}$, where $a$ is the orbital separation in units of $R_\odot$ and $M_\mathrm{tot}$ is the mass of the binary in units of $M_\odot$. The dynamical timescale is $\tau_\mathrm{dyn} \simeq0.44\,R_\mathrm{*}^{3/2}\,M_\mathrm{*}^{-1/2}\,\,\mathrm{hr}$, where $R_\mathrm{*}$ is the radius of the star in units of $R_\odot$ and $M_\mathrm{*}$ is the mass of the star in units of $M_\odot$.
We estimate the nuclear timescale for stellar evolution as $\tau_\mathrm{nuc} \simeq 1\times10^{10}\,M_\mathrm{*}\,L_\mathrm{*}^{-1}\,\,\mathrm{yr}$, where $L_\mathrm{*}$ is the luminosity of the star in units of $L_\odot$.
In our binary system $\tau_\mathrm{nuc} \gg \tau_\mathrm{dyn}$ and $\tau_\mathrm{dyn}\sim P_\mathrm{orb}$ near contact so we are able to neglect stellar evolution and evolve the system dynamically.
We recognize that the structure of contact binaries is still an unsolved problem \citep{2003ASPC..293...76W}, and although the systems studied here will have been in a contact configuration for many thermal timescales, we make the necessary approximation that their thermal structure has not significantly deviated from the combined thermal structure of two individual stars.

We choose several ages, $\tau$, for each component mass, shown in Table~\ref{TBL:Core_Masses}, and generate binaries with different initial orbital separations, $a_\mathrm{i}$, for each age.
Because we are interested in finding the dynamical stability limit, $a_\mathrm{crit}(\tau)$, the maximum orbital separation where the hydrodynamic interaction between the stars causes the binary to eject the common envelope and enter an inspiral phase, we can constrain $a_\mathrm{i}$ by requiring that the system be in contact but not already at the Roche limit.
We sample different degrees of contact, $\eta(a_\mathrm{i})$ (see \S\ref{SSec:Binary_Scan}), within this range and dynamically integrate these systems.
We are primarily interested in the final orbital separation, the characteristics of the mass flow, and the stability of the system.
From these results we find the dynamical stability limit for each component age and initial mass.
Likewise, for each orbital separation that would become unstable there exists a critical evolutionary age which the binary experiences an inspiral, found using the inverse relation $\tau_\mathrm{crit}(a_\mathrm{i})$.

As a preliminary study to test the impact of strictly identical components on the stability of the system we also generate a binary using two stars with the same initial mass at different evolutionary ages ($q=0.997$).
This model corresponds to two stars forming from the same cloud at slightly different times.

\section{Numerical Methods}
\label{Sec:Numerical}

\subsection{Hydrodynamic Calculations}
\label{SSec:StarCrash}
{\it SPH} is a Lagrangian particle method that approximates a continuous fluid as discrete nodes, each carrying various parameters such as mass, position, velocity, pressure, and temperature.
In an {\it SPH} simulation the resolution scales with the particle density. Since we are interested in the mass flow between two stars we use an equal number density, non-equal mass particle distribution when generating the single-star models to maintain a higher resolution near the lower mass-density surface where the mass transfer occurs ({\it Paper 1}).
We calculate the pressure for each particle, $p_i$, taking into account both the radiation pressure and ideal gas, \begin{equation}p_i = \frac{\rho_ikT_i}{\mu_i}+\frac{1}{3}aT_i^4,\end{equation} where $\rho_i$, $T_i$, and $\mu_i$ are the density, temperature, and mean molecular mass of the particle $i$, $k$ is the Boltzmann constant, and $a$ is the radiation constant.
The temperature of the particle is determined by solving \begin{equation}u_i=\frac{3}{2}\frac{kT_i}{\mu_i}+\frac{aT_i^4}{\rho_i}.\end{equation}

For the hydrodynamic calculations we use an {\it SPH} code, {\it StarSmasher} (originally {\it StarCrash}) developed originally by \citet{1991PhDT........11R}, updated and maintained as described in \citet{1999JCoPh.152..687L} and \citet{2000PhRvD..62f4012F}. The code now implements variational equations of motion and libraries to calculate the gravitational forces between particles using direct summation on NVIDIA graphics cards as described in \citet{2010MNRAS.402..105G}.
Using a direct summation instead of a tree-based algorithm for gravity increases the accuracy of the gravity calculations at the cost of speed \citep{2010ProCS...1.1119G}.
The code uses a cubic spline \citep{1985A&A...149..135M} for the smoothing kernel and an artificial viscosity prescription coupled with a Balsara Switch \citep{1995JCoPh.121..357B} to prevent unphysical interparticle penetration (see \S\ref{A:Artificial Viscosity}).

While scanning the binary equilibrium sequence (see \S\ref{SSec:Binary_Scan}) we apply an artificial relaxation force to the equations of motion to add a drag term to the calculated accelerations.
The relaxation force is calculated at each time step as $\dot{\bf v} = -{\bf v}/t_\mathrm{relax}$, where $\dot{\bf v}$ is the time derivative of the velocity vector, ${\bf v}$, and $t_\mathrm{relax}$ is a parameter that determines the magnitude of the relaxation force.
We set $t_\mathrm{relax}$ to a value at least as large as the oscillation period of the corresponding single-star models, which are on the order of the global dynamical timescale.
While relaxing our single-star models of the $14.0\ M_\odot$ and $20.0\ M_\odot$ stars, we set $t_\mathrm{relax}$ to an arbitrarily large value and allow the artificial viscosity alone to dampen spurious fluid oscillations; in order to preserve the entropy profile in these stars, we do not reintroduce the small amount of kinetic energy removed as internal energy.
For a more in-depth overview of {\it SPH} and {\it StarSmasher}, see \citet{2006ApJ...640..441L}.

\subsection{Single-Star Model}
\label{SSec:Single_Star_Model}
We use {\it EZ} to generate a library of stellar models with masses varying between $7.50\ M_\odot$ and $21.50\ M_\odot$ in steps of $0.01\ M_\odot$.
For our single-star {\it SPH} models we use three initial masses, $8.00\ M_\odot$, $14.00\ M_\odot$, and $20.00\ M_\odot$, to generate initial conditions from the stellar evolution profiles at several ages in the range $\tau_\mathrm{RGB}<\tau\le\tau_\mathrm{max}$ (described in \S\ref{SSec:Stellar_Models}).
At each timestep {\it EZ} calculates the stellar parameters along the radial grid as a function of the enclosed mass. The parameters we extract for use in our models are pressure, density, enclosed mass, and elemental abundances (Hydrogen through Neon).
We convert the stellar profiles into {\it SPH} models with $N\simeq10^5$ particles in an unequal mass, constant number-density, hexagonal close-packed lattice.
An additional core particle is placed at the center of each star, with a mass determined by the central density and resolution of the model.
The softening length of the core particle sets the resolution limit of our simulations: gravity is softened according to the mass distribution defined by the {\it SPH} kernel at distances less than twice the softening length.
We rerun a subset of calculations varying the number of particles to determine the effect this resolution limit has on our results.
Specifically, we test configurations with $N\simeq10^4$, $N\simeq5\times10^4$, and $N\simeq7.5\times10^4$ (see \S\ref{SSec:Resolution_Test}).

\begin{figure*}[htp]
\begin{center}
\begin{tabular}{cc}
\includegraphics[width=8.0cm]{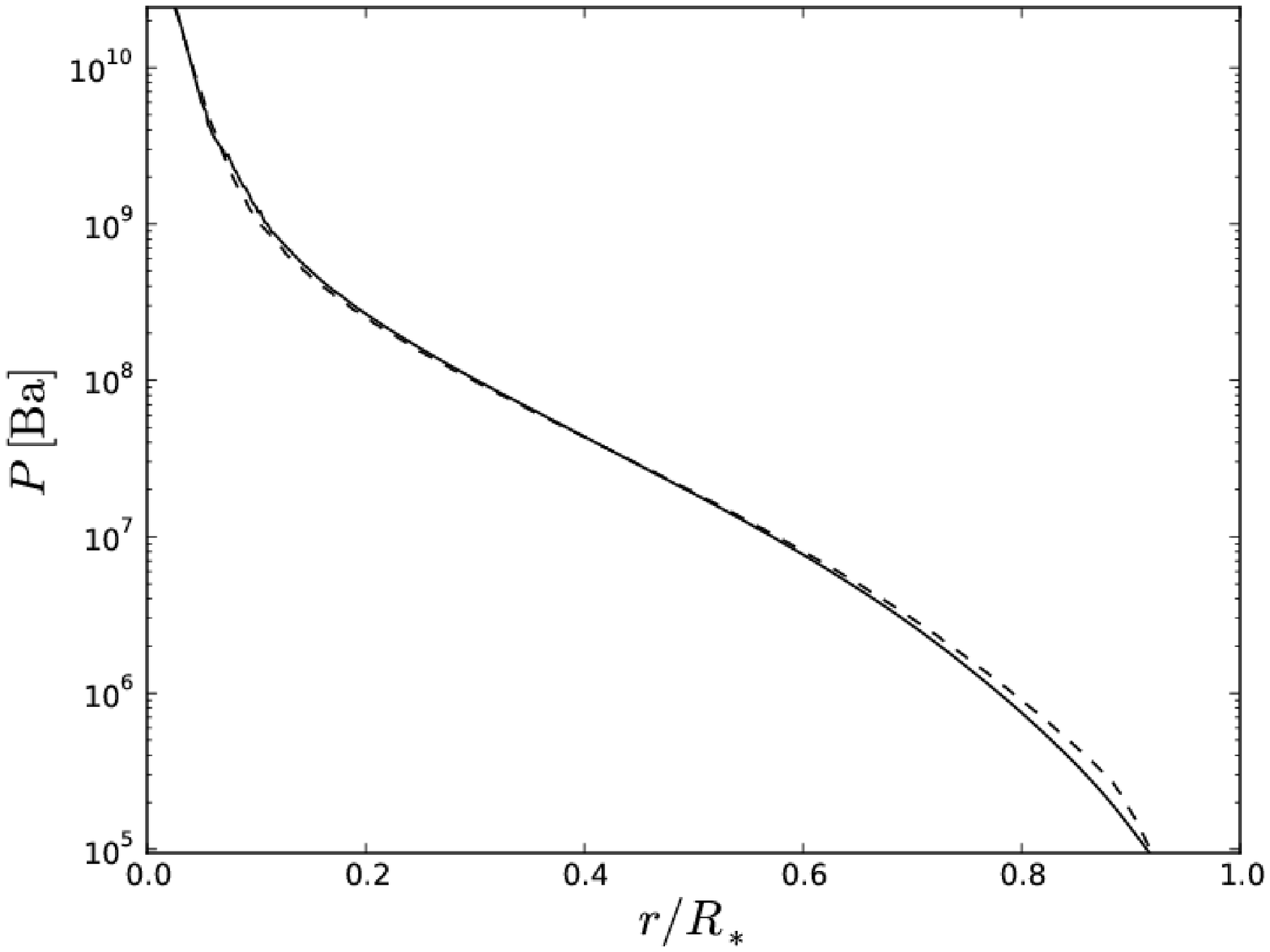}&
\includegraphics[width=8.0cm]{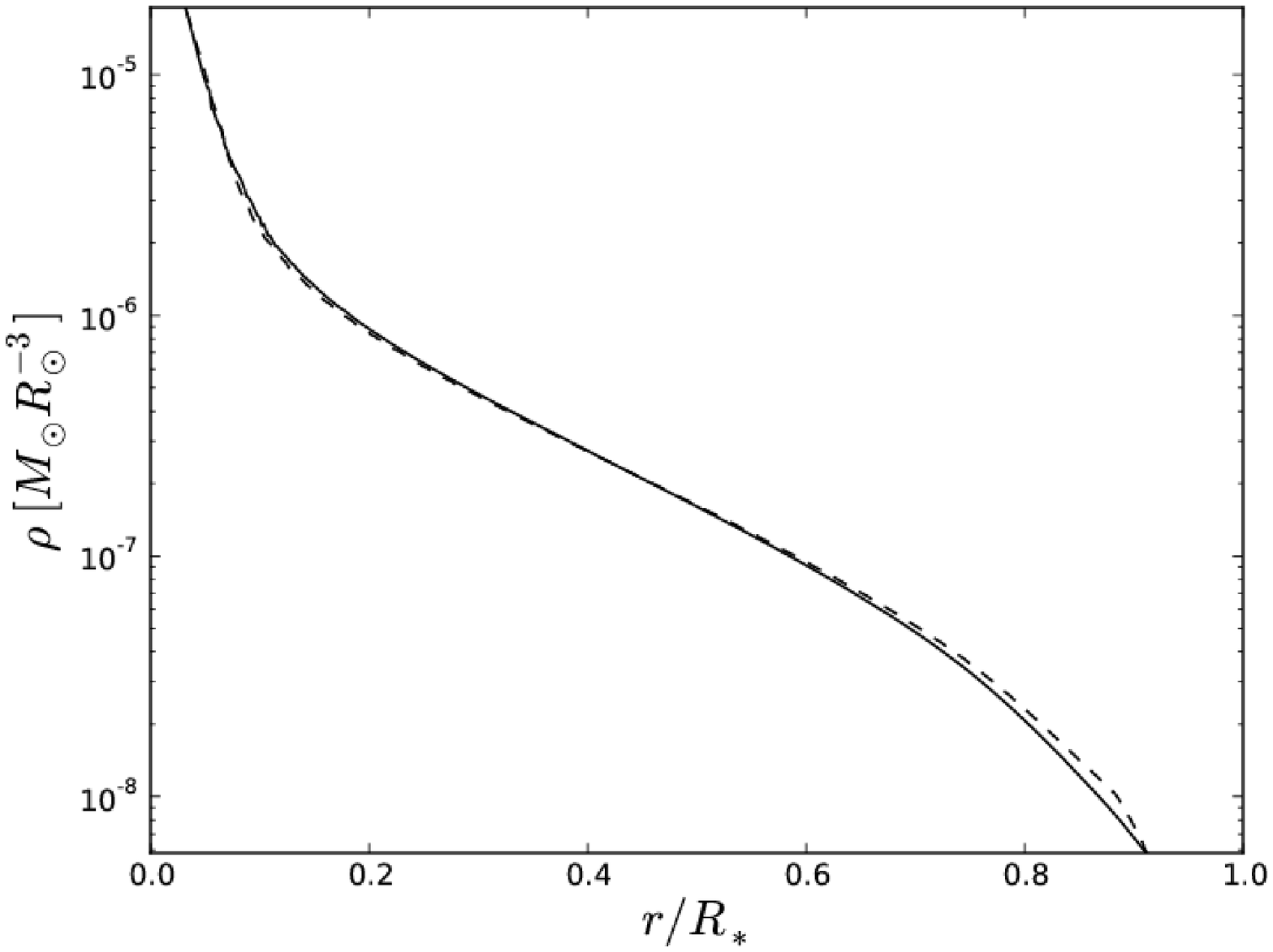}\\
\\
\includegraphics[width=8.0cm]{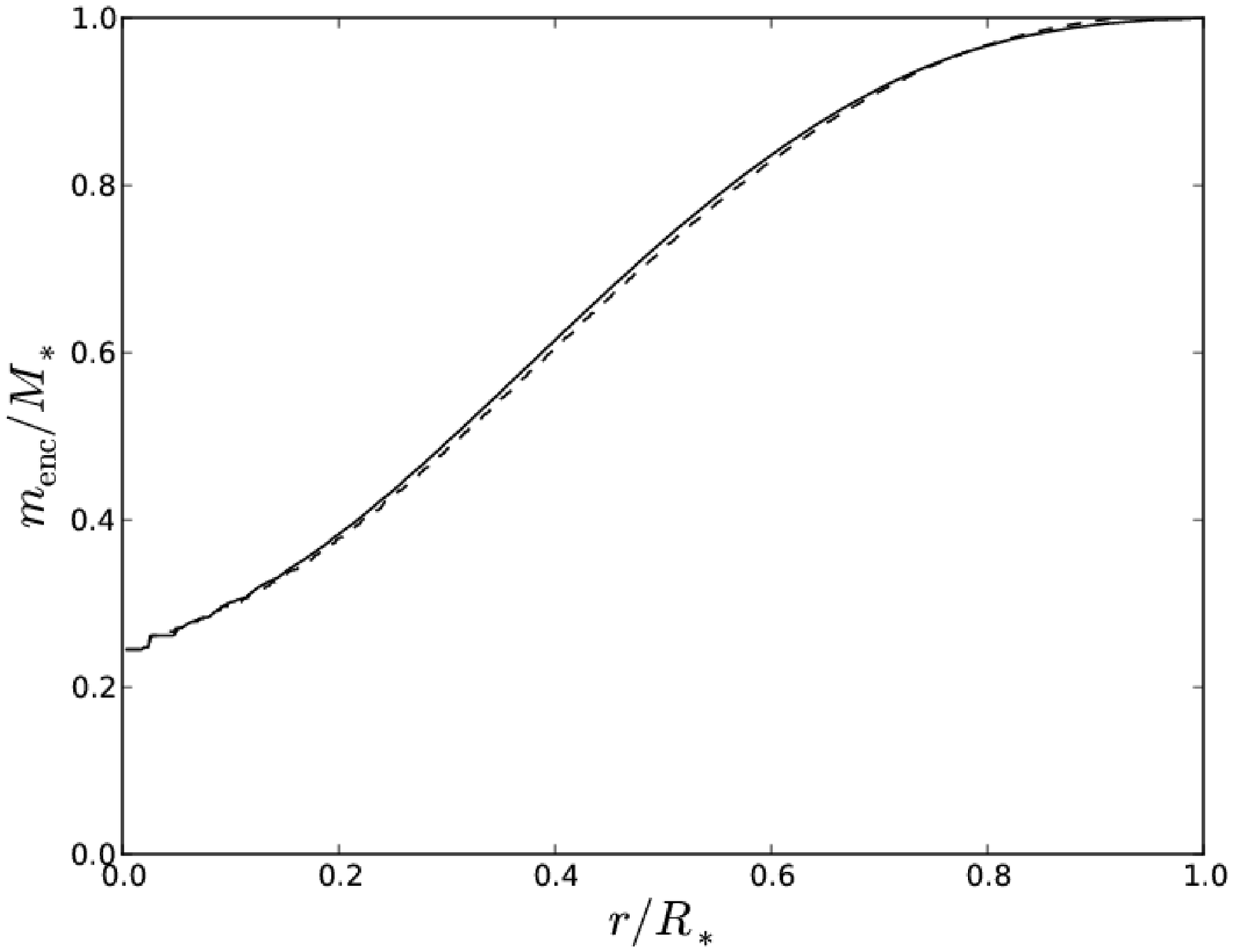}&
\includegraphics[width=8.0cm]{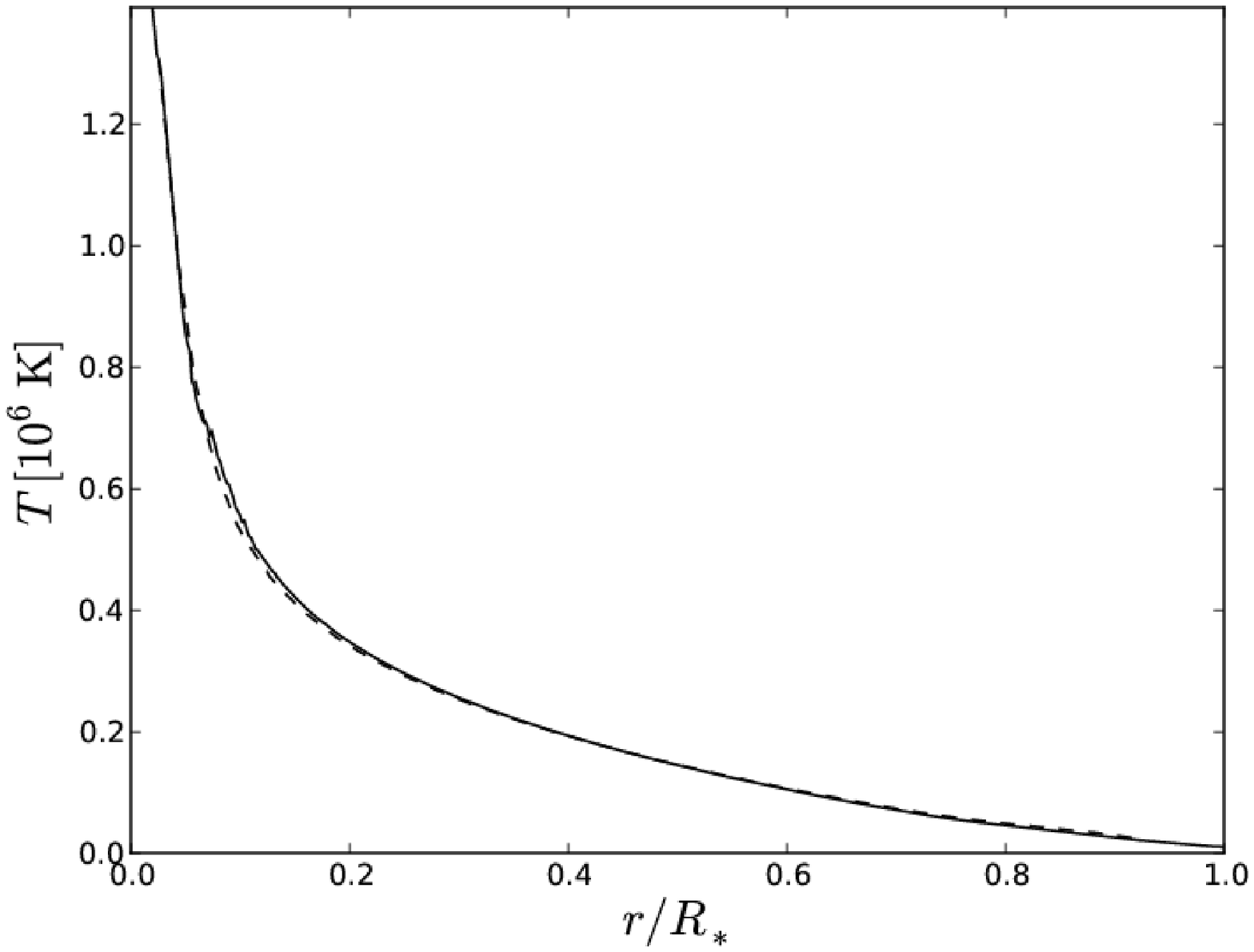}\\
\end{tabular}
\end{center}
\caption{Pressure (top left), density (top right), enclosed mass (bottom left), and temperature (bottom right) as a function of radius for a $8.0\ M_\odot$ star at $35.26$ Myr. The dashed lines represent the initial profile generated by {\it EZ} and the solid lines show the profiles after the model has been relaxed in {\it StarSmasher}. We find that the stellar profiles after the relaxation run agree reasonably well with the initial stellar profiles generated by {\it EZ}.}
\label{Fig:8.0_3526_Profile}
\end{figure*}

For the $14.0\ M_\odot$ and $20.0\ M_\odot$ models we set the optimal number of neighbors to roughly $20$ and for the $8.0\ M_\odot$ models we set the optimal number of neighbors to roughly $40$.
Using a lower number of neighbors for the higher mass models is necessary to adequately resolve the high density and pressure gradients near the core.
In all models the softening length is larger than the physical core radii by at least an order of magnitude.
In Table~\ref{TBL:Core_Masses} and Table~\ref{TBL:Core_Radii} we compare the numerical core mass and radius to the core masses and radii determined using various prescriptions \citep{2001A&A...369..170T} described in \S\ref{SSSec:Core_Masses}.
We relax the systems by dynamically evolving the single-star models, allowing the artificial viscosity to dampen the oscillations.
We monitor the pressure, temperature, density, and mass profiles as a function of radius and find that the relaxed parameter profiles agree reasonably well with the initial parameter profiles.
Comparisons of the parameter profiles from the stellar evolution code and the models after the relaxation are shown in Figure~\ref{Fig:8.0_3526_Profile} for an $8.0\ M_\odot$ star at $35.26$ Myr.

\subsubsection{Core Mass Determination}
\label{SSSec:Core_Masses}
\begin{figure*}[htp]
\begin{center}
\begin{tabular}{cc}
\includegraphics[width=8.0cm]{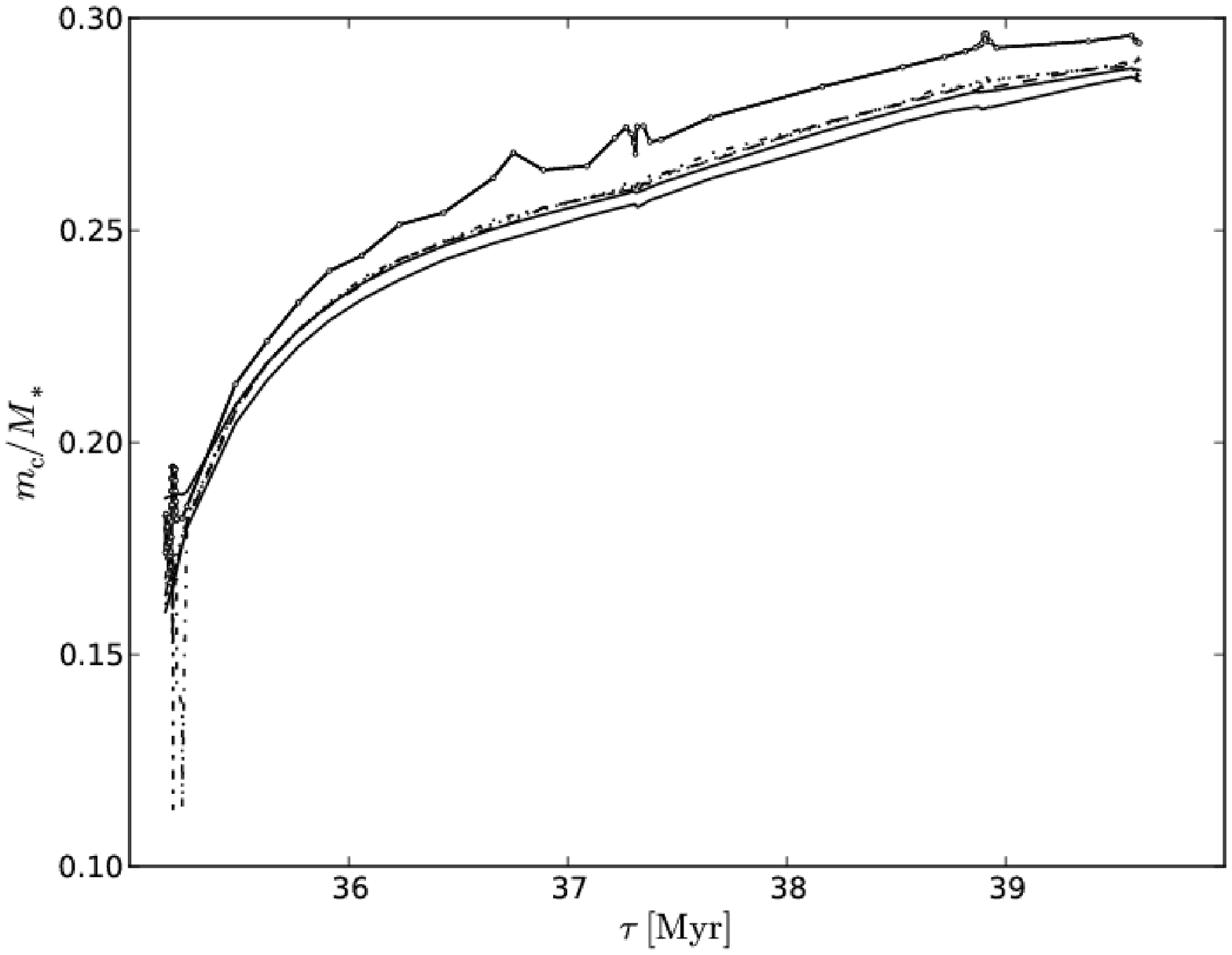}
\includegraphics[width=8.0cm]{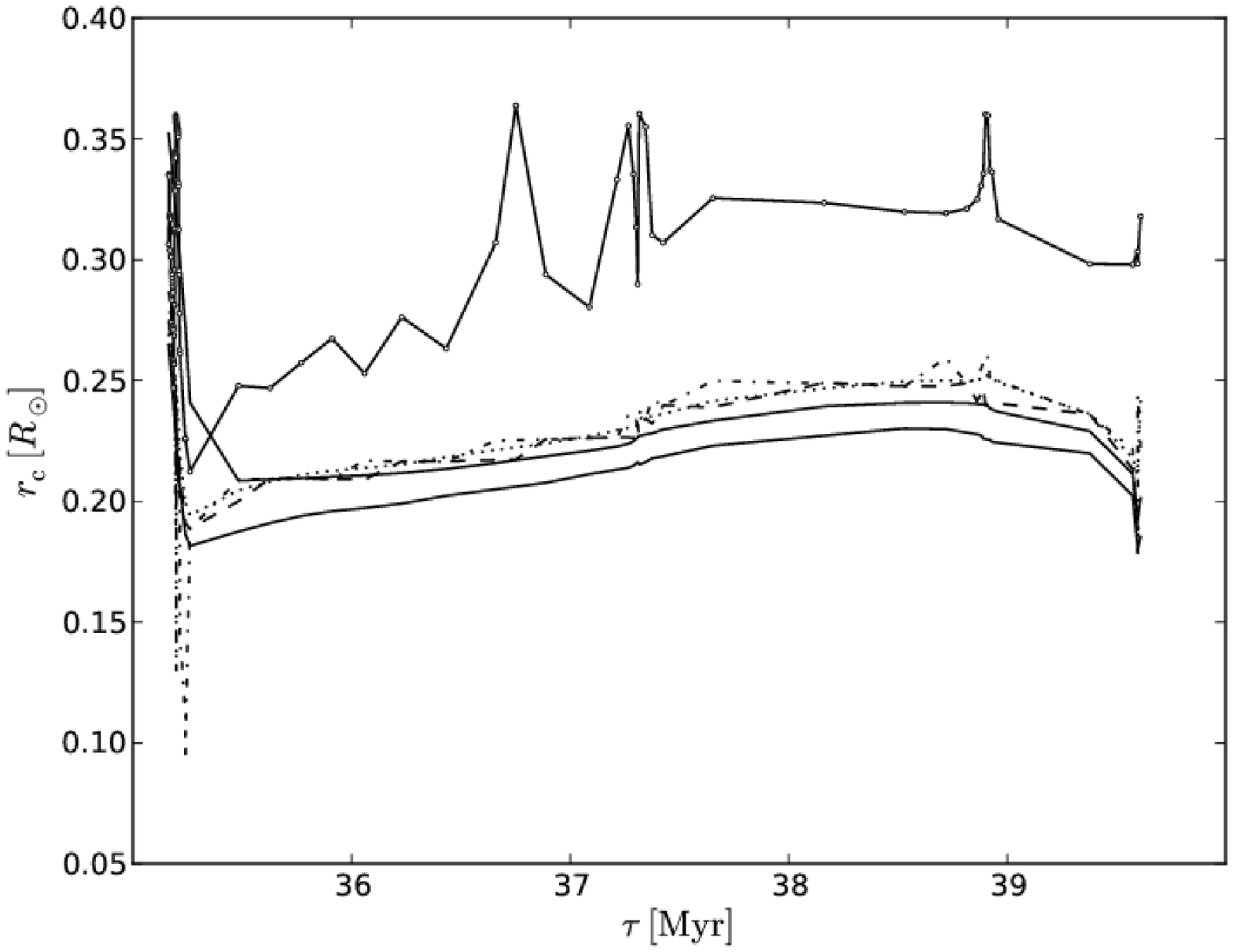}
\end{tabular}
\end{center}
\caption{Core masses, $m_\mathrm{c}$, and radii, $r_\mathrm{c}$, as a function of age for an $8.0\ M_\odot$ star during the RGB and AGB. The solid lines show the core masses calculated using methods (1) and (2), the dashed line method (3), the dotted line method (4), the dash-dotted line method (5), and the solid line with open circles method (6), as described in \S\ref{SSSec:Core_Masses}. The core radii and masses calculated from method (6) are an overestimate as the core is defined as the transition between the convective and radiative layer in the stellar envelope and in massive stars this transition does not fully reach the hydrogen shell \citep{2001A&A...369..170T}.}
\label{Fig:Core_Mass_t}
\end{figure*}

Figure~\ref{Fig:Core_Mass_t} shows the calculated core masses and radii from a variety of methods discussed in \citet{2001A&A...369..170T}.
In Tables~\ref{TBL:Core_Masses} and \ref{TBL:Core_Radii}, we compare these values to the resolution-bound core masses and softening length of the core-particle used in our {\it SPH} simulations.
Using a larger core radius leads to less mass in the common envelope ejected prior to forming the remnant binary, resulting in a larger final orbital separation. 
Conversely, a smaller core radius reduces the minimum stable orbital separation.
We investigate the sensitivity of the results on the core radius in \S\ref{SSec:Remnant_Properties} and find that there is a constraint on both the maximum and minimum core mass leading to a surviving binary.

Method (1) defines the core as the region having a hydrogen abundance less than $1\%$, while method (2) defines the core as having a hydrogen abundance less than $10\%$.
Method (3) defines the bifurcation where the energy generation peaks, method (4) defines the bifurcation where where $d^2\log\rho/dm^2=0$, method (5) defines the bifurcation where $\sinh^{-1}(\Delta W)$ transfers from a steeply increasing function to a shallow function, where $\Delta W = E_\mathrm{int}+E_\mathrm{grav}$ \citep{1994MNRAS.270..121H}, and method (6) defines the bifurcation where the entropy transfers from a steeply increasing function to a shallow function.
We find that, at our standard resolution of $N=10^5$ particles, the masses of the core particles in our simulations are within twice of the maximum calculated core masses while the softening lengths of the core particles are more than an order of magnitude larger than the calculated core radii due to a combination of the steep density gradient of the core and the resolution limit set by the number of particles.

\begin{deluxetable}{cccccccccccc}
\tablewidth{15.5cm}
\tabletypesize{\footnotesize}
\tablecolumns{11}
\tablecaption{Mass properties of single-star models\label{TBL:Core_Masses}}
\tablehead{
    \colhead{$Run$} & \colhead{$m_\mathrm{i}\ [M_\odot]$} & \colhead{$\tau$ [Myr]} & \colhead{$m_\mathrm{f}\ [M_\odot]$} & \colhead{$m_\mathrm{c,s}/m_\mathrm{f}$} & \colhead{$m_\mathrm{c,1}/m_\mathrm{f}$} & \colhead{$m_\mathrm{c,2}/m_\mathrm{f}$} & \colhead{$m_\mathrm{c,3}/m_\mathrm{f}$} & \colhead{$m_\mathrm{c,4}/m_\mathrm{f}$} & \colhead{$m_\mathrm{c,5}/m_\mathrm{f}$} & \colhead{$m_\mathrm{c,6}/m_\mathrm{f}$}}
\startdata
$1-5$   & $8.0$  & $35.22$ & $7.98$  & $0.264$ & $0.170$ & $0.188$ & $0.171$ & $0.173$ & $0.174$ & $0.182$ & \\
$6-10$  & $8.0$  & $35.24$ & $7.98$  & $0.246$ & $0.175$ & $0.188$ & $0.176$ & $0.178$ & $0.162$ & $0.182$ & \\
$11-13$ & $8.0$  & $35.26$ & $7.97$  & $0.245$ & $0.180$ & $0.188$ & $0.181$ & $0.182$ & $0.183$ & $0.185$ & \\
$14-18$ & $14.0$ & $13.30$ & $13.96$ & $0.351$ & $0.224$ & $0.251$ & $0.224$ & $0.224$ & $0.224$ & $0.213$ & \\
$19-24$ & $14.0$ & $13.32$ & $13.95$ & $0.335$ & $0.234$ & $0.252$ & $0.236$ & $0.236$ & $0.238$ & $0.243$ & \\
$25-28$ & $20.0$ & $8.487$ & $19.94$ & $0.411$ & $0.271$ & $0.298$ & $0.273$ & $0.271$ & $0.273$ & $0.283$ & \\
$28-31$ & $20.0$ & $8.514$ & $19.89$ & $0.422$ & $0.292$ & $0.302$ & $0.296$ & $0.296$ & $0.298$ & $0.309$ &
\enddata

\tablecomments{$m_\mathrm{i}$ is the mass of the star at birth, $\tau$ is the star's age, $m_\mathrm{f}$ is the mass of the star at $\tau$, $m_\mathrm{c,s}$ is the mass of the core particle calculated by {\it StarSmasher} while generating the {\it SPH} model at our standard resolution of $N=10^5$ particles, and $m_{\mathrm{c,}i}$ are the calculated core masses, found using the methods described in \S\ref{SSec:Single_Star_Model}, where $i$ designates the method used. In this paper we refer to the core masses calculated by method (1) as the nominal core masses.}
\end{deluxetable}

\begin{deluxetable}{cccccccccccc}
\tablewidth{17cm}
\tabletypesize{\footnotesize}
\tablecolumns{11}
\tablecaption{Radius properties of single-star models\label{TBL:Core_Radii}}
\tablehead{
    \colhead{$Run$} & \colhead{$m_\mathrm{i}\ [M_\odot]$} & \colhead{$\tau$ [Myr]} & \colhead{$R_\mathrm{*}\ [R_\odot]$} & \colhead{$r_\mathrm{c,s}\ [R_\odot]$} & \colhead{$r_\mathrm{c,1}\ [R_\odot]$} & \colhead{$r_\mathrm{c,2}\ [R_\odot]$} & \colhead{$r_\mathrm{c,3}\ [R_\odot]$} & \colhead{$r_\mathrm{c,4}\ [R_\odot]$} & \colhead{$r_\mathrm{c,5}\ [R_\odot]$} & \colhead{$r_\mathrm{c,6}\ [R_\odot]$}}
\startdata
$1-6$   & $8.0$  & $35.22$ & $211.3$ & $7.036$ & $0.206$ & $0.300$ & $0.211$ & $0.219$ & $0.221$ & $0.261$ & \\
$6-10$  & $8.0$  & $35.24$ & $255.4$ & $8.503$ & $0.187$ & $0.273$ & $0.191$ & $0.200$ & $0.145$ & $0.226$ & \\
$11-13$ & $8.0$  & $35.26$ & $269.5$ & $8.973$ & $0.182$ & $0.241$ & $0.189$ & $0.194$ & $0.197$ & $0.212$ & \\
$14-18$ & $14.0$ & $13.30$ & $533.9$ & $9.406$ & $0.344$ & $0.558$ & $0.340$ & $0.344$ & $0.340$ & $0.295$ & \\
$19-24$ & $14.0$ & $13.32$ & $667.9$ & $11.766$& $0.311$ & $0.462$ & $0.321$ & $0.324$ & $0.335$ & $0.376$ & \\
$25-28$ & $20.0$ & $8.487$ & $1026.5$& $18.083$& $0.429$ & $0.691$ & $0.439$ & $0.432$ & $0.439$ & $0.514$ & \\
$29-31$ & $20.0$ & $8.514$ & $1051.2$& $18.519$& $0.424$ & $0.502$ & $0.447$ & $0.450$ & $0.465$ & $0.577$ &
\enddata
\tablecomments{$m_\mathrm{i}$ is the mass of the star at birth, $\tau$ is the star's age, $R_\mathrm{*}$ is the radius of the star at $\tau$, $r_\mathrm{c,s}$ is the softening length of the core particle calculated by {\it StarSmasher} while generating the {\it SPH} model at our standard resolution of $N=10^5$ particles, and $r_\mathrm{c,i}$ are the calculated core radii, found using the methods described in \S\ref{SSec:Single_Star_Model}, where $i$ designates the method used. In this paper we primarily refer to the core radii calculated by method (1).}
\end{deluxetable}

\subsubsection{Comparison to Polytropes}
\label{SSec:Polytrope_Comparison}
\begin{figure*}[htp]
\begin{center}
\begin{tabular}{cc}
\includegraphics[width=8.0cm]{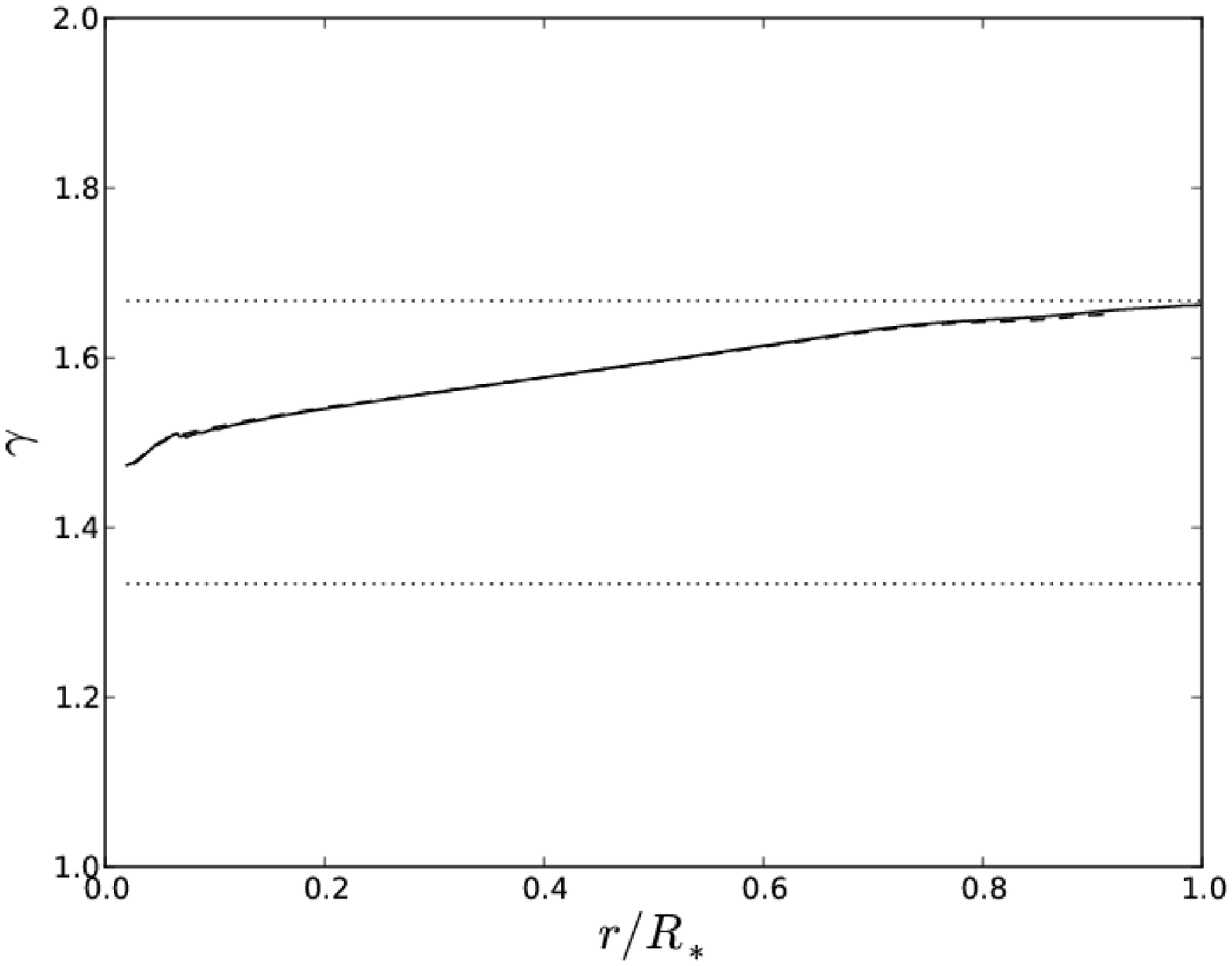}&
\includegraphics[width=8.0cm]{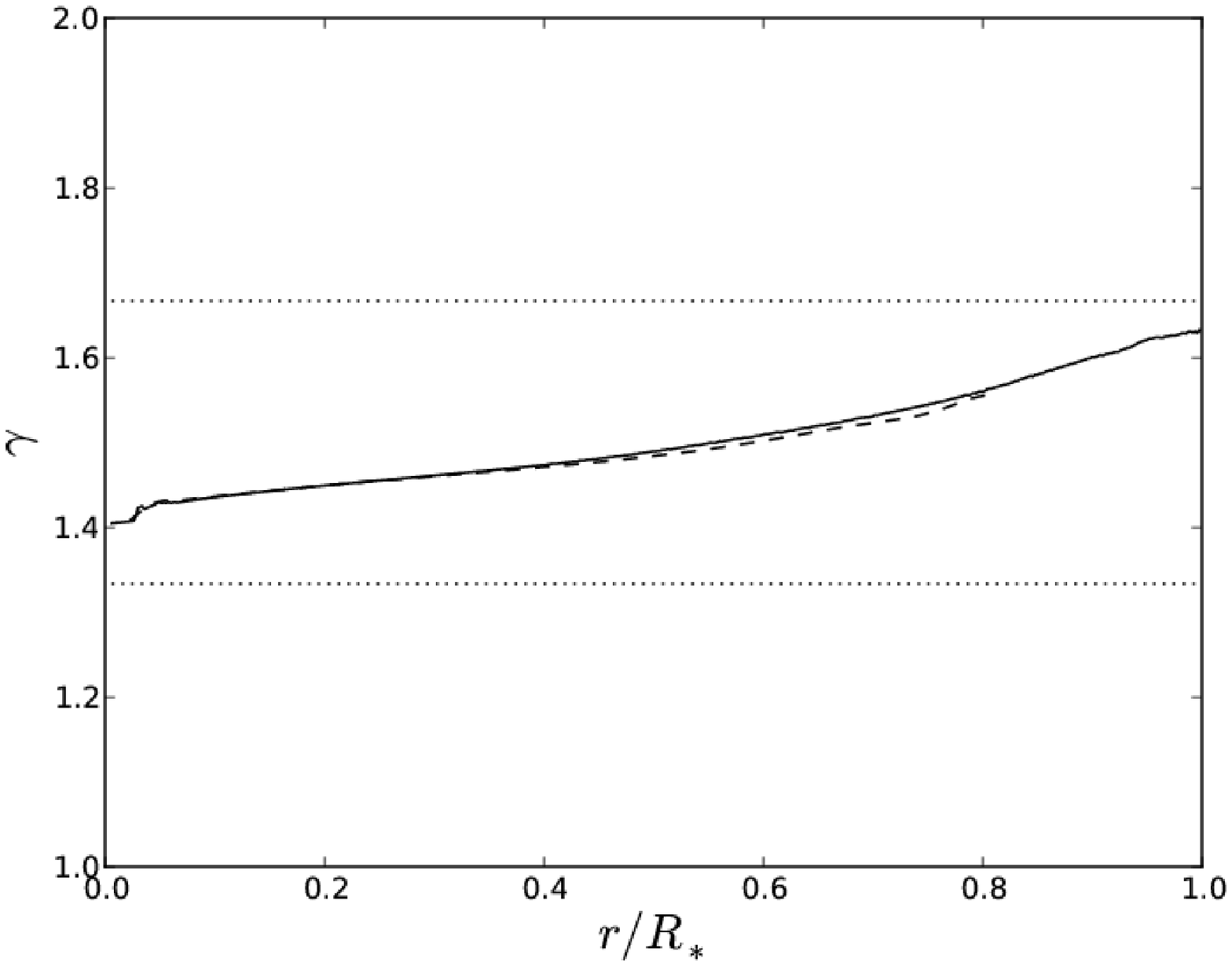}\\
\end{tabular}
\end{center}
\caption{Adiabatic index, calculated using the ratio of gas pressure to the total pressure, as a function of radius for an $8.0\ M_\odot$ star at $35.26$ Myr (left) and a $20.0\ M_\odot$ star at $8.514$ Myr (right). The dashed lines represent the initial profile generated by {\it EZ} and the solid lines show the profiles after the model has been relaxed in {\it StarSmasher}. The two dotted lines show commonly chosen values used for stellar models, $\Gamma=4/3$ and $\Gamma=5/3$. {\it Paper 1} uses polytropes and condensed polytropes with $\Gamma=5/3$.}
\label{Fig:Gamma}
\end{figure*}

{\it Paper 1} used polytropic models with adiabatic index $\Gamma=5/3$ ($n=1.5$) to construct the single-star models \citep{1987ApJ...318..794H}.
The stellar models generated using {\it EZ} have an adiabatic index that changes with the radius and age of the star.
Because common envelope evolution is very sensitive to the density and entropy profiles of the stars, small deviations in $m(r)$ may lead to a significantly different remnant binary.
Comparisons of $\Gamma(r)$ from the stellar evolution code and our stellar models after the system has been relaxed (see \S\ref{SSec:Single_Star_Model}) are shown in Figure~\ref{Fig:Gamma} alongside commonly used values for polytropic models.

\pagebreak

\subsection{Binary Scan}
\label{SSec:Binary_Scan}
We generate a circular, co-rotating (i.e. spin-synchronized) binary entering the common envelope phase by placing two single-star models on the {\it x}-axis.
We find the separation where the star's radius is roughly equal to the effective Roche lobe radius and increase this by a factor of a few for the initial orbital separation to ensure the stars are able to settle into an equilibrium without risk of Roche lobe overflow \citep{1983ApJ...268..368E}.
The stars are then slowly scanned from this initial orbital separation using the relaxation force described in \S \ref{SSec:StarCrash} until an {\it SPH} particle is within one smoothing length of an outer Lagrangian point, $L_\mathrm{2}$ or $L_\mathrm{3}$, signaling the end of dynamically stable mass transfer.
By allowing the scan to proceed over many dynamical timescales, the model remains very close to equilibrium throughout the scan from the initial to the final orbital separation.
This process generates a series of models at different separations, $a_\mathrm{i}$, that are used as initial conditions for our hydrodynamic calculations of contact binary systems.

\begin{figure*}[htp]
\begin{center}
\begin{tabular}{cc}
\includegraphics[width=16cm]{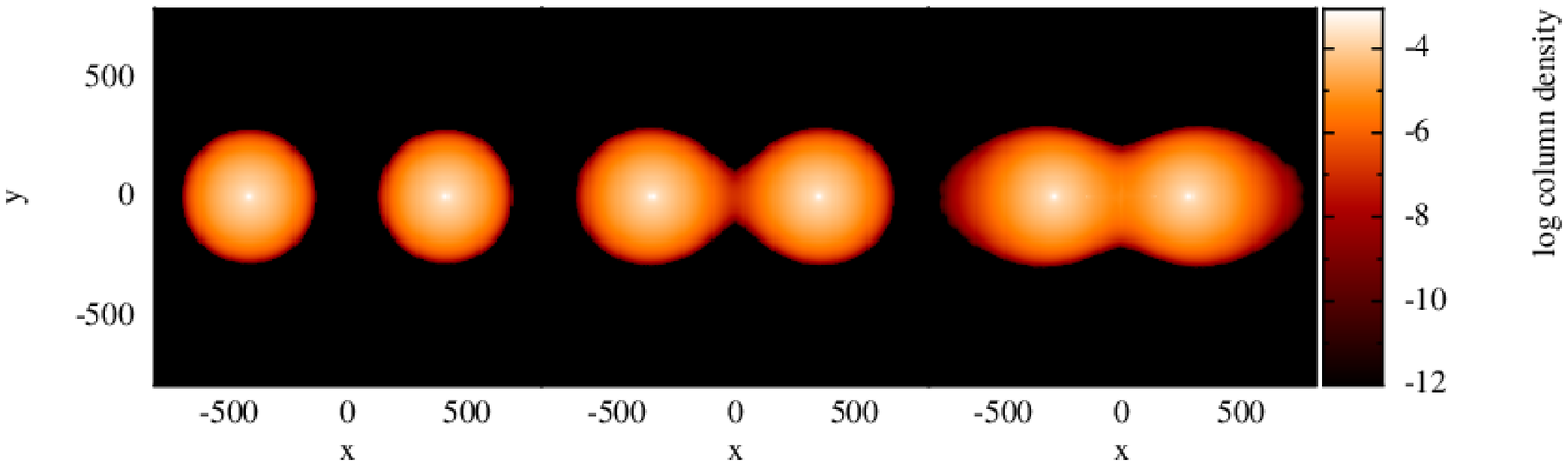}\\
\\
\includegraphics[width=16cm]{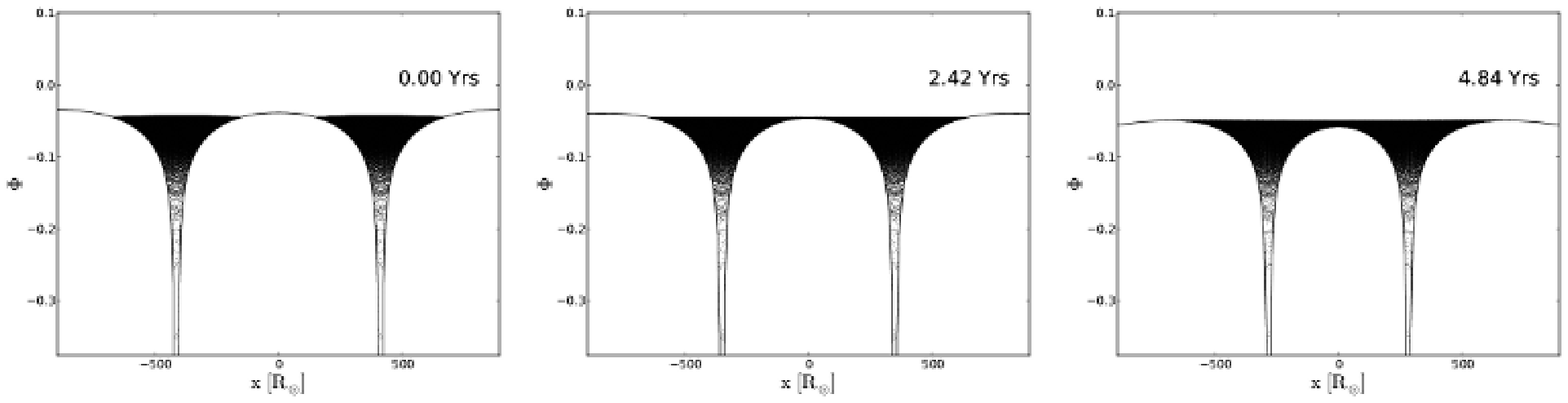}\\
\end{tabular}
\end{center}
\caption{Snapshots of the binary scan. Each set of three plots show the binary at separations of $a_\mathrm{i} = 825.0\ R_\odot$ ($\eta = -1.148$), $a_\mathrm{i} = 697.9\ R_\odot$ ($\eta = 0.120$), and $a_\mathrm{i} = 564.9\ R_\odot$ ($\eta = 1.002$). (Top) Series of logarithm of column density plots in the orbital plane. The axes are scaled to $R_\odot$, and the column density has units of $M_\odot R_\odot^{-2}$. (Bottom) Series of effective potential plots in the {\it x}-axis (defined as the axis connecting the two cores) for a $8.0\ M_\odot$ component binary at $35.26$ Myr.
The time labelled in the upper right shows that the scan proceeds slowly (that is, over many dynamical timescales), allowing the system to remain near equilibrium throughout the calculation.}
\label{Fig:8.0M_Scan}
\end{figure*}

\begin{figure*}[htp]
\begin{center}
\begin{tabular}{cc}
\includegraphics[width=8.0cm]{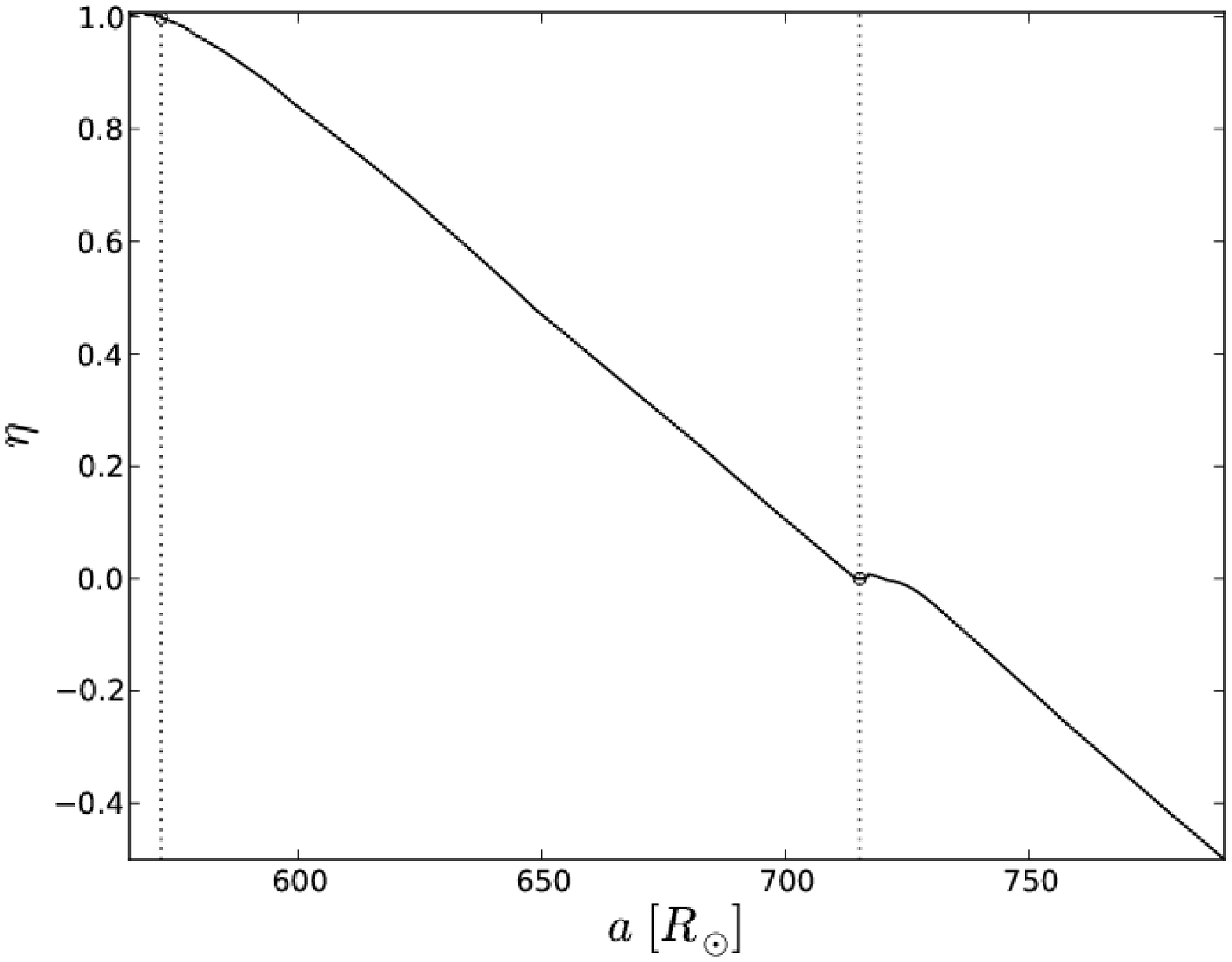}
\includegraphics[width=8.0cm]{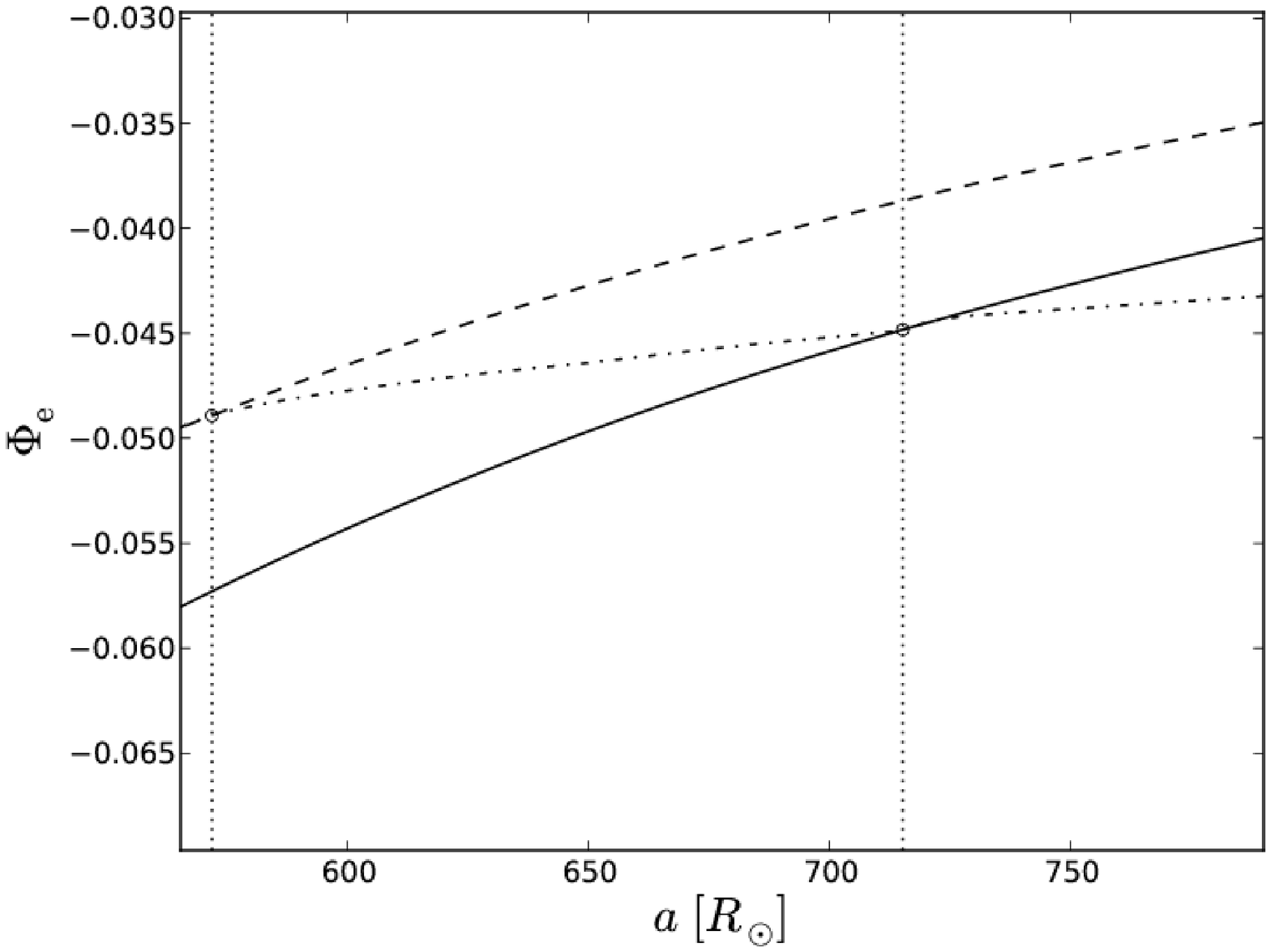}
\end{tabular}
\end{center}
\caption{(Left) Degree of contact, $\eta$, as a function of the center of mass separation, $a$, for a $q=1$ binary with $8.0\ M_\odot$ components at $35.26$ Myr.
(Right) Various effective potentials, $\Phi_\mathrm{e}$, as a function of $r_\mathrm{sep}$ for a $q=1$ binary with $8.0\ M_\odot$ components, where the solid line shows $\Phi_\mathrm{i}$, the maximum effective potential along the {\it x}-axis (defined as the line between the two cores), the dashed line shows $\Phi_\mathrm{o}$, the maximum effective potential between the two cores, and the dotted line shows $\Phi_\mathrm{s}$, the maximum effective potential of a particle within one smoothing length of the {\it x}-axis.
The two vertical lines show the separation at first contact ($\eta=0$) and at the Roche limit ($\eta=1$).}
\label{Fig:8M_Eta_r}
\end{figure*}

The initial and final separations of each scan are tabulated in Table~\ref{TBL:Scan_Parameters}.
We take note of several events that occur during the scan by calculating the degree of contact, $\eta(a_\mathrm{i})$, defined as \citep{1995ApJ...438..887R} \begin{equation}\label{EQ:DoC}\eta=\frac{\Phi_\mathrm{s}-\Phi_\mathrm{i}}{\Phi_\mathrm{o}-\Phi_\mathrm{i}},\end{equation} where $\Phi_\mathrm{s}$ is the maximum effective potential of a particle within one smoothing length of the {\it x}-axis, $\Phi_\mathrm{o}$ is the maximum effective potential along the {\it x}-axis, and $\Phi_\mathrm{i}$ is the maximum effective potential between the two cores (at $x=0$ in a $q=1$ binary).
The indices $\mathrm{s}$, $\mathrm{o}$, and $\mathrm{i}$ stand for surface, inner Lagrangian point, and outer Lagrangian point, respectively.
The effective potential of a point on the {\it x}-axis is calculated using the spline-softened form of the gravitational potential \citep{1989ApJS...70..419H} \begin{equation}\Phi_\mathrm{e}(x)=-\frac{1}{2}\Omega^2x^2-\sum_i^{\mathrm{N}}Gm_if(r_i,h_i),\end{equation} where $\Omega$ is the angular velocity, $\mathrm{N}$ is the number of particles, $m_i$ is the mass of the particle, $r_i$ is the distance of the particle from the point, $h_i$ is the smoothing length of the particle, and \[f(r,h)= \begin{cases} 
      -\frac{2}{h}\left(\frac{1}{3}u^2-\frac{3}{20}u^4+\frac{1}{20}u^5\right)+\frac{7}{5h} & 0 \leq u \leq 1 \\
      -\frac{1}{15r}-\frac{1}{h}\left(\frac{4}{3}u^2-u^3+\frac{3}{10}u^4-\frac{1}{30}u^4-\frac{1}{30}u^5\right)+\frac{8}{5h} & 1\leq u\leq 2 \\
      \frac{1}{r} & u\geq 2
   \end{cases}
,\] where $u=r/h$.
The effective potential of a particle is calculated by \begin{equation}\Phi_\mathrm{e}=\Phi_\mathrm{grav}-\frac{1}{2}\Omega^2(x^2+y^2),\end{equation} where $\Phi_\mathrm{grav}$ is the gravitational potential of the particle.
The degree of contact $\eta=0$ at first contact, while $\eta=1$ at the Roche limit.
Figure~\ref{Fig:8.0M_Scan} shows column density snapshots for one of our scans, as well as the associated effective potential energies of the particles as a function of $x$, the coordinate along the axis connecting the centers of the stars.
Figure~\ref{Fig:8M_Eta_r} shows the degree of contact and effective potential of the system as a function of the orbital separation during the same binary scan.

The code uses snapshots from the single-star models to generate the components used in the binary system. The code offsets one star in the positive {\it x} direction and one star in the negative {\it x} direction such that the orbital separation is equal to the initial separation specified for the scan and the center of mass for the system lies at the origin.
In the $q=1$ case we use the same snapshot for both stellar models which generates a binary system symmetric about the $y-z$ plane. The code decreases the orbital separation from the initial value, $a(0)$, down to the final separation, $a(t_\mathrm{scan})$, where the orbital separation at time $t$ is described as \begin{equation}\label{EQ:ScanRate}a(t)=a(0)\left(\frac{a(t_\mathrm{scan})}{a(0)}\right)^{t/t_\mathrm{scan}}.\end{equation}
For the $8.0\ M_\odot$ component binary we set $t_\mathrm{scan}=2.5\ \mathrm{yr}$, for the $14.0\ M_\odot$ component binary we set $t_\mathrm{scan}=22.7\ \mathrm{yr}$, and for the $20.0\ M_\odot$ component binary we set $t_\mathrm{scan}=45.5\ \mathrm{yr}$

Specifically for $q=1$ systems, the system is still in a stable equilibrium through first contact at $\eta=0$, when a particle passes through the inner Lagrangian point, $L_\mathrm{1}$.
The Roche limit occurs at $\eta = 1$ when a particle passes through an outer Lagrangian point, $L_\mathrm{2}$ or $L_\mathrm{3}$, and we consider the system to be dynamically unstable and end the binary scan.

We examine the total energy as a function of time to test for secular instability, which occurs when the total energy reaches a local minimum.
We do not observe secular instability in any of the binary scans, consistent with {\it Paper 1}, which found that the maximum core mass to experience secular instability is $0.15\ M_\mathrm{*}$, lower than any of the core masses used in our binary scans.

\begin{deluxetable}{ccccccccccc}
\centering
\tablewidth{17.5cm}
\tabletypesize{\small}
\tablecolumns{10}
\tablecaption{Properties of binary scans\label{TBL:Scan_Parameters}}
\tablehead{
    \colhead{$m_\mathrm{i}\ [M_\odot]$} & \colhead{$\tau\ [\mathrm{Myr}]$} & \colhead{$a_{t=0}\ [R_\odot]$} & \colhead{$P_{t=0}\ [\mathrm{yr}]$} & \colhead{$a_{t=t_\mathrm{scan}}\ [R_\odot]$} & \colhead{$P_{t=t_\mathrm{scan}}\ [\mathrm{yr}]$} & \colhead{$a_{\eta=0}\ [R_\odot]$} & \colhead{$P_{\eta=0}\ [\mathrm{yr}]$} & \colhead{$a_{\eta=1}\ [R_\odot]$} & \colhead{$P_{\eta=1}\ [\mathrm{yr}]$}}
\startdata
$8.0$  & $35.22$  & $600$  & $1.17$  & $449$  & $0.75$ & $574$  & $1.09$ & $452$  & $0.76$ &  \\
$8.0$  & $35.24$  & $800$  & $1.80$  & $535$  & $0.97$ & $683$  & $1.41$ & $542$  & $0.99$ &  \\
$8.0$  & $35.26$  & $825$  & $1.88$  & $565$  & $1.05$ & $716$  & $1.52$ & $572$  & $1.07$ &  \\
$14.0$ & $13.30$  & $2200$ & $6.19$  & $1168$ & $2.38$ & $1461$ & $3.35$ & $1174$ & $2.40$ &  \\
$14.0$ & $13.32$  & $2400$ & $6.20$  & $1450$ & $2.76$ & $1825$ & $3.99$ & $1466$ & $2.80$ &  \\
$20.0$ & $8.487$  & $6000$ & $23.35$ & $2357$ & $5.71$ & $2891$ & $7.79$ & $2373$ & $5.77$ &  \\
$20.0$ & $8.514$  & $6000$ & $23.38$ & $1804$ & $4.02$ & $3014$ & $8.31$ & $2414$ & $5.93$ &  \\
\enddata
\tablecomments{$m_\mathrm{i}$ is the initial mass of each component at birth, $\tau$ is the age of the components, $a_{t=0}$ and $P_{t=0}$ are the initial separation and orbital period of the binary during the scan, $a_{t_\mathrm{scan}}$ and $P_{t_\mathrm{scan}}$ are the final separation and orbital period of the binary during the scan, $a_{\eta=0}$ and $P_{\eta=0}$ are the separation and period at first contact, and $a_{\eta=1}$ and $P_{\eta=1}$ are the separation and period at the Roche limit. The orbital period is calculated as $P=2\pi/\Omega_\mathrm{orb}$, where $\Omega_\mathrm{orb}$ is the angular velocity.}
\end{deluxetable}
\pagebreak

\subsection{Dynamical Calculations}
\label{SSec:Dynamical_Run}
From each binary scan we choose a series of separations to use as the initial conditions for our hydrodynamic calculations. We are primarily interested in finding the dynamical stability limit, $a_\mathrm{crit}$, for each mass and age combination, where systems with $a_\mathrm{i}<a_\mathrm{crit}$ undergo an inspiral and systems with $a_\mathrm{i}\ge a_\mathrm{crit}$ are stable.
The models are evolved in the corotating frame, taking into account both centripetal and Coriolis forces: doing so keeps particle coordinate velocities to a minimum and therefore minimizes any spurious effects from artificial viscosity.

We determine that a system is stable if we do not observe an inspiral during the dynamical calculation and if the core separation versus time shows only stable oscillations around a consistent limiting value.
These oscillations in stable systems are due to the impossibility of obtaining perfect initial conditions (including not only the structure but also the proper orbital frequency) and can be broken down into two components.
The shorter period fluctuations occur because the stars are in slightly non-circular orbits, while the longer period fluctuations come from oscillations in the fluid configuration.
We designate the stable run with the deepest degree of contact as the dynamical stability limit, and characterize these runs further in \S\ref{SSec:StabilityLimit}.

In an unstable binary we observe several distinct stages of common envelope evolution \citep{2001ASPC..229..239P}, starting with the {\it loss of co-rotation}, triggered by mass loss at the $L_\mathrm{2}$ and $L_\mathrm{3}$ Lagrangian points, leading to energy and angular momentum transfer from the orbiting cores to the surrounding gas.
This is followed by the {\it plunge-in}, caused by the cores depositing frictional energy into the envelope, unbinding mass in the envelope, and resulting in a rapid decrease in core separation on the dynamical timescale.
Finally, after enough orbital energy from the cores has been deposited to the envelope, the system begins the {\it slow spiral-in}, where enough of the envelope has been ejected that the rate of decrease in the core separation becomes stable on the dynamical timescale.
We end the simulations after the binary enters the {\it slow spiral-in} and extrapolate the properties of the remnant binary, assuming that the entire envelope is eventually ejected (see \S\ref{SSec:Remnant_Properties}).
For a more in-depth review of common envelope evolution, we refer the readers to \citet{2013A&ARv..21...59I}.

\begin{deluxetable}{cccccccc}
\tabletypesize{\scriptsize}
\tablewidth{13cm}
\tablecolumns{8}
\tablecaption{Parameters of $q=1$ dynamical calculations\label{TBL:Parameters}}
\tablehead{
    \colhead{$Run$} & \colhead{$a_\mathrm{i}\ [R_\odot]$} & \colhead{$P_\mathrm{i}\ [\mathrm{yr}]$} & \colhead{$\eta$} & \colhead{$Run$} & \colhead{$a_\mathrm{i}\ [R_\odot]$} & \colhead{$P_\mathrm{i}\ [\mathrm{yr}]$} & \colhead{$\eta$}}
\startdata
$m_\mathrm{i}= 8.0\ M_\odot$&$\tau=35.22\mathrm{\ Myr}$&&&$m_\mathrm{i}=14.0\ M_\odot$&$\tau=13.30\mathrm{\ Myr}$&&\\
$1$  & $466.5$ & $0.79$ & $0.873$                         & $16$ & $1238$  & $2.60$ & $0.761$\\
$2$  & $462.0$ & $0.78$ & $0.915$                         & $17$ & $1207$  & $2.50$ & $0.885$\\
$3$  & $457.9$ & $0.77$ & $0.958$                         & $18$ & $1198$  & $2.47$ & $0.940$\\
$4$  & $454.6$ & $0.76$ & $0.985$                         & $19$ & $1188$  & $2.44$ & $0.973$\\
$5$  & $451.2$ & $0.75$ & $1.005$                         & $20$ & $1178$  & $2.41$ & $0.996$\\
     &         &        &                                 &      &        &        &        \\
$m_\mathrm{i}= 8.0\ M_\odot$&$\tau=35.24\mathrm{\ Myr}$&&&$m_\mathrm{i}=14.0\ M_\odot$&$\tau=13.32\mathrm{\ Myr}$&&\\
$6$  & $566.8$ & $1.06$ & $0.837$                         & $21$ & $1535$ & $3.59$ & $0.794$\\
$7$  & $563.8$ & $1.05$ & $0.873$                         & $22$ & $1512$ & $3.51$ & $0.839$\\
$8$  & $558.0$ & $1.04$ & $0.914$                         & $23$ & $1505$ & $3.48$ & $0.869$\\
$9$  & $552.2$ & $1.02$ & $0.952$                         & $24$ & $1500$ & $3.47$ & $0.884$\\
$10$ & $546.4$ & $1.00$ & $0.984$                         & $25$ & $1489$ & $3.43$ & $0.918$\\
$11$ & $540.7$ & $0.99$ & $1.002$                         & $26$ & $1477$ & $3.39$ & $0.978$\\
     &         &        &                                 &      &        &        &        \\
$m_\mathrm{i}= 8.0\ M_\odot$&$\tau=35.26\mathrm{\ Myr}$&&&$m_\mathrm{i}=20.0\ M_\odot$&$\tau=8.487\mathrm{\ Myr}$&&\\
$12$ & $606.1$ & $1.17$ & $0.799$                         & $27$ & $2497$ & $6.26$ & $0.786$\\
$13$ & $598.2$ & $1.15$ & $0.853$                         & $28$ & $2486$ & $6.17$ & $0.817$\\
$14$ & $592.9$ & $1.13$ & $0.886$                         & $29$ & $2469$ & $6.13$ & $0.850$\\
$15$ & $580.0$ & $1.10$ & $0.963$                         & $30$ & $2442$ & $6.03$ & $0.892$\\
     &         &        &                                 & $31$ & $2415$ & $5.93$ & $0.942$\\
     &         &        &                                 &      &        &        &        \\
     &         &        &                                 &$m_\mathrm{i}=20.0\ M_\odot$&$\tau=8.514\mathrm{\ Myr}$&&\\
     &         &        &                                 & $32$ & $2534$ & $6.38$ & $0.805$\\
     &         &        &                                 & $33$ & $2522$ & $6.34$ & $0.834$\\
     &         &        &                                 & $34$ & $2491$ & $6.22$ & $0.890$\\
     &         &        &                                 & $35$ & $2461$ & $6.11$ & $0.937$\\
\enddata
\tablecomments{$m_\mathrm{i}$ is the initial mass of each component at birth, $\tau$ is the age of the components, $a_\mathrm{i}$ is the initial orbital separation, $P_\mathrm{i}$ is the initial orbital period, and $\eta$ is the degree of contact.}
\end{deluxetable}
\pagebreak

\section{Results}
\label{Sec:Results}
\begin{figure*}[htp]
\begin{center}
\begin{tabular}{cc}
\includegraphics[width=8.0cm]{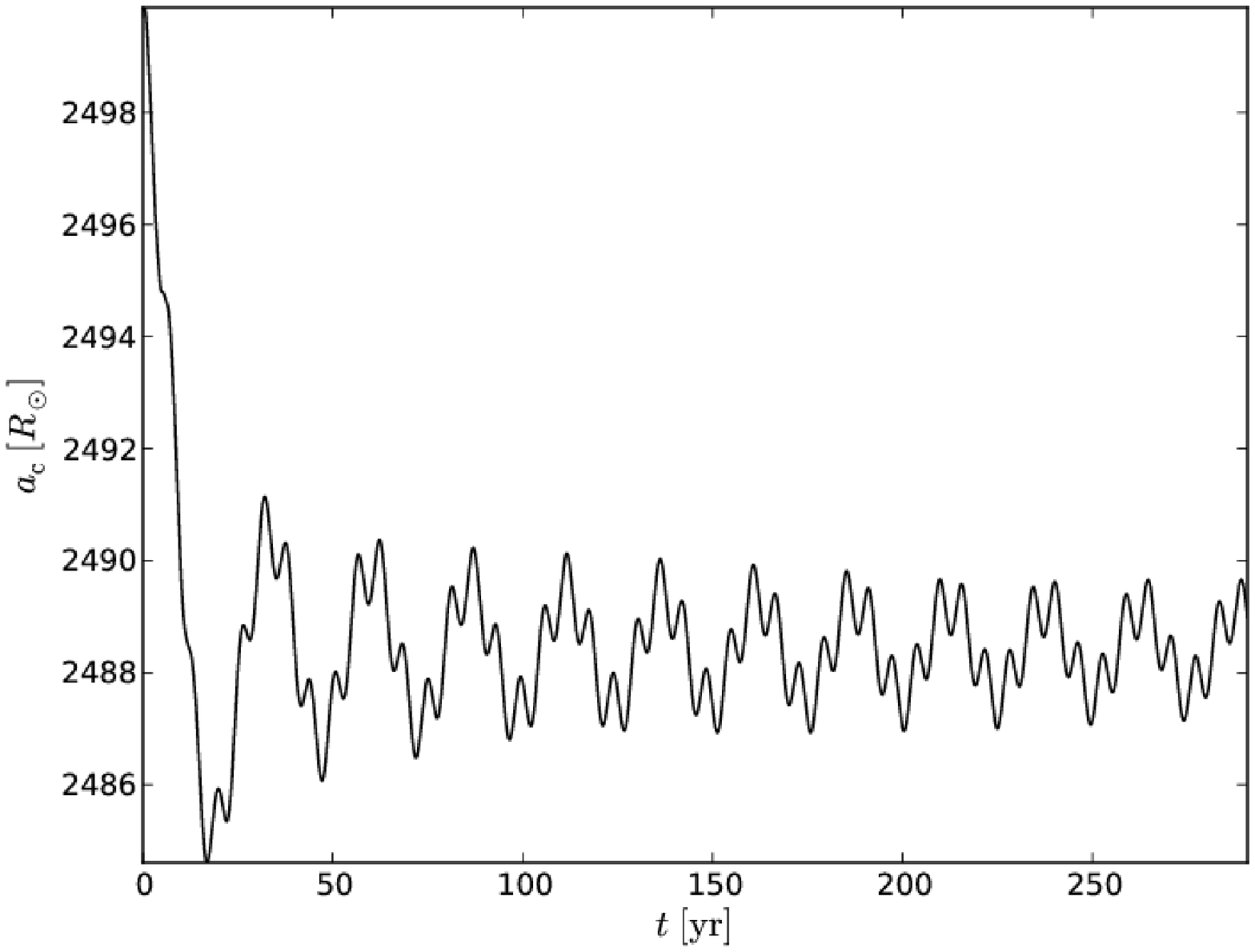}
\includegraphics[width=8.0cm]{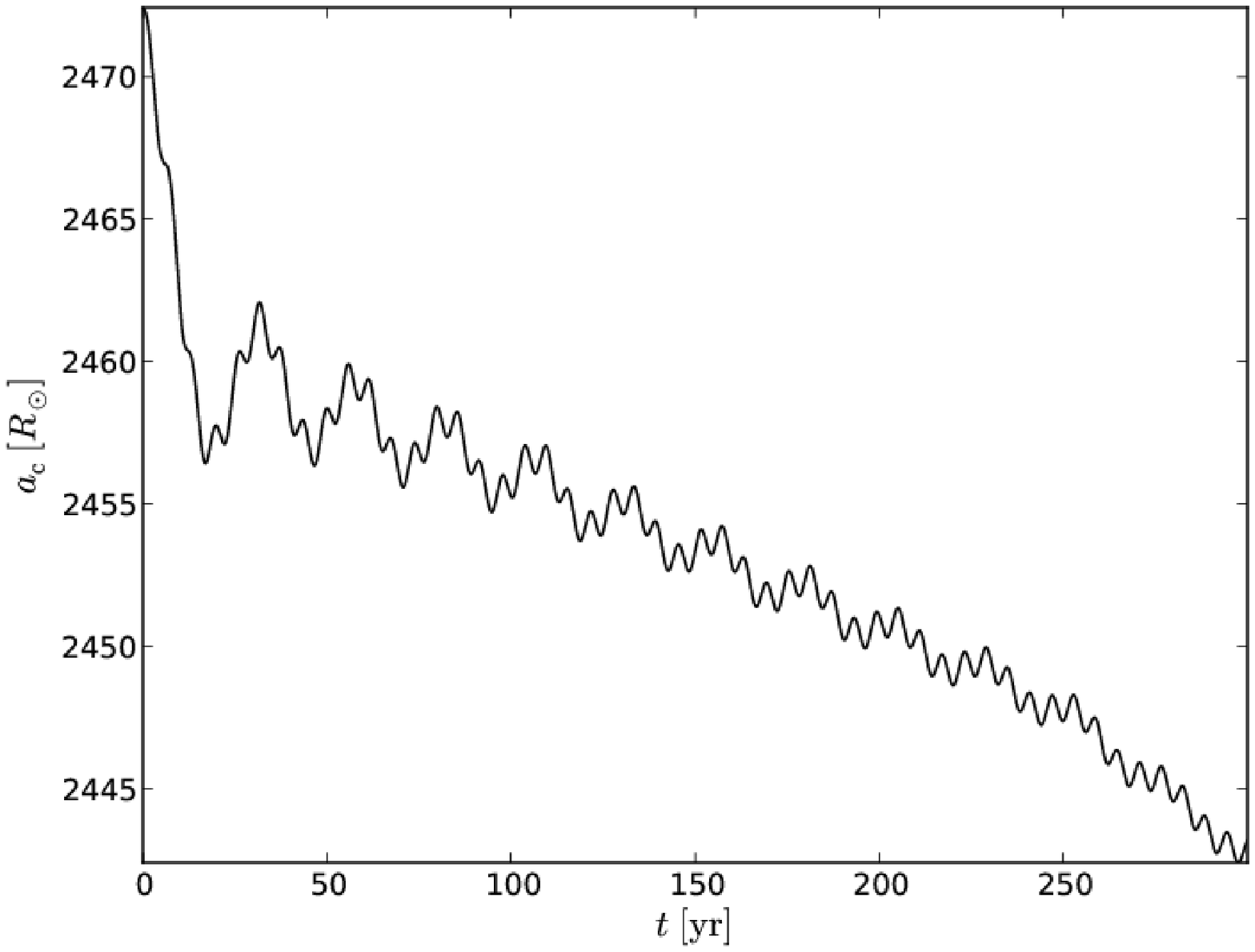}
\end{tabular}
\end{center}
\caption{Core separation as a function of time for a $20.0\ M_\odot$ star at 8.487 Myr at $\eta=0.786$ (left) and $\eta=0.850$ (right). The stable configuration (left) exhibits an initial drop before oscillating around a stable value. The unstable configuration (right) exhibits a similar initial drop, but the core separation does not fully recover and the sinusoidal oscillation amplitude quickly decays. The unstable configuration eventually underwent an inspiral later in the calculation. We use this behavior to determine the stability of a binary, as the actual {\it plunge-in} could occur past the integration time of the calculation.}
\label{Fig:Rcore_stable}
\end{figure*}

\begin{figure*}[htp]
\begin{center}
\begin{tabular}{cc}
\includegraphics[width=8.0cm]{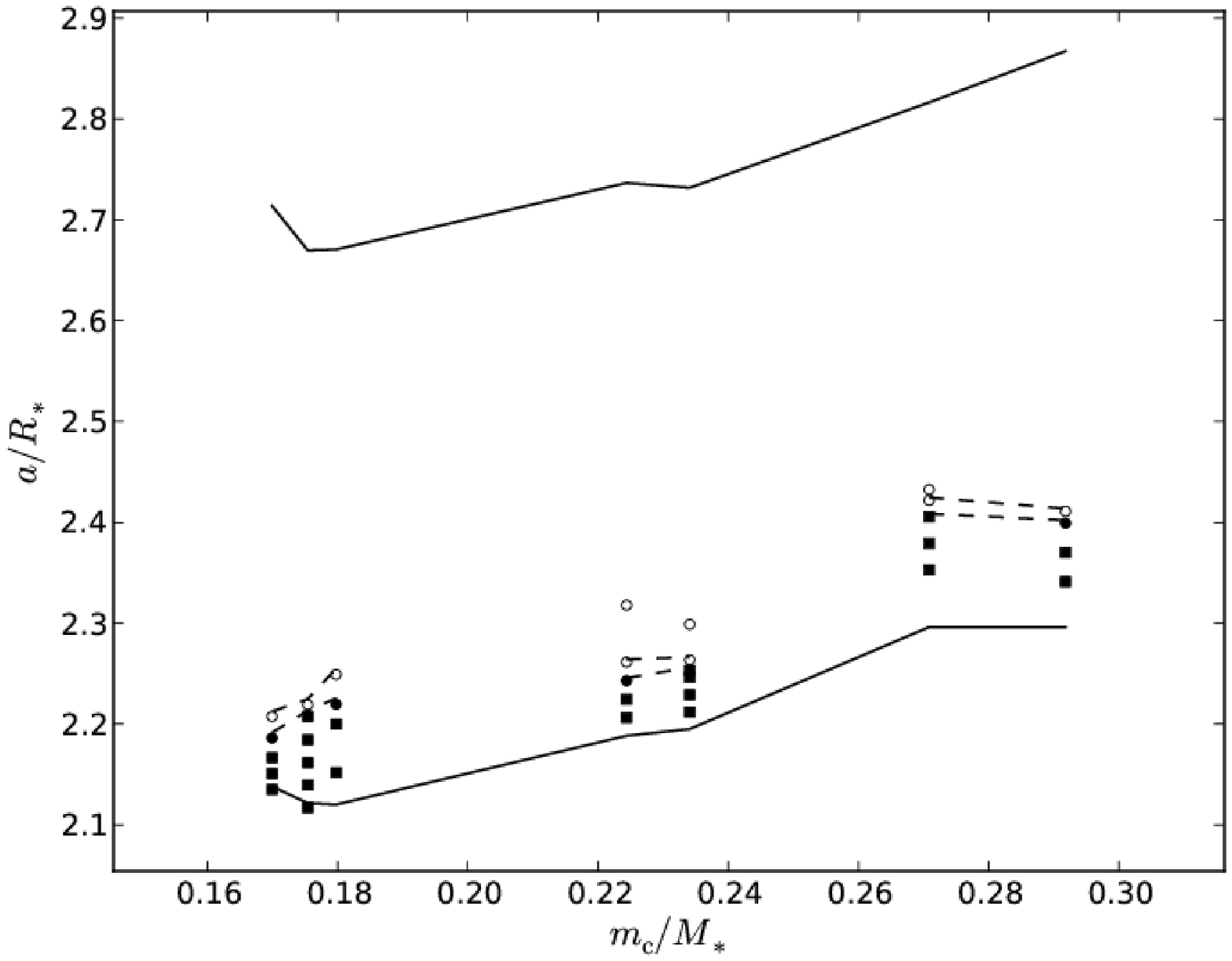}
\includegraphics[width=8.0cm]{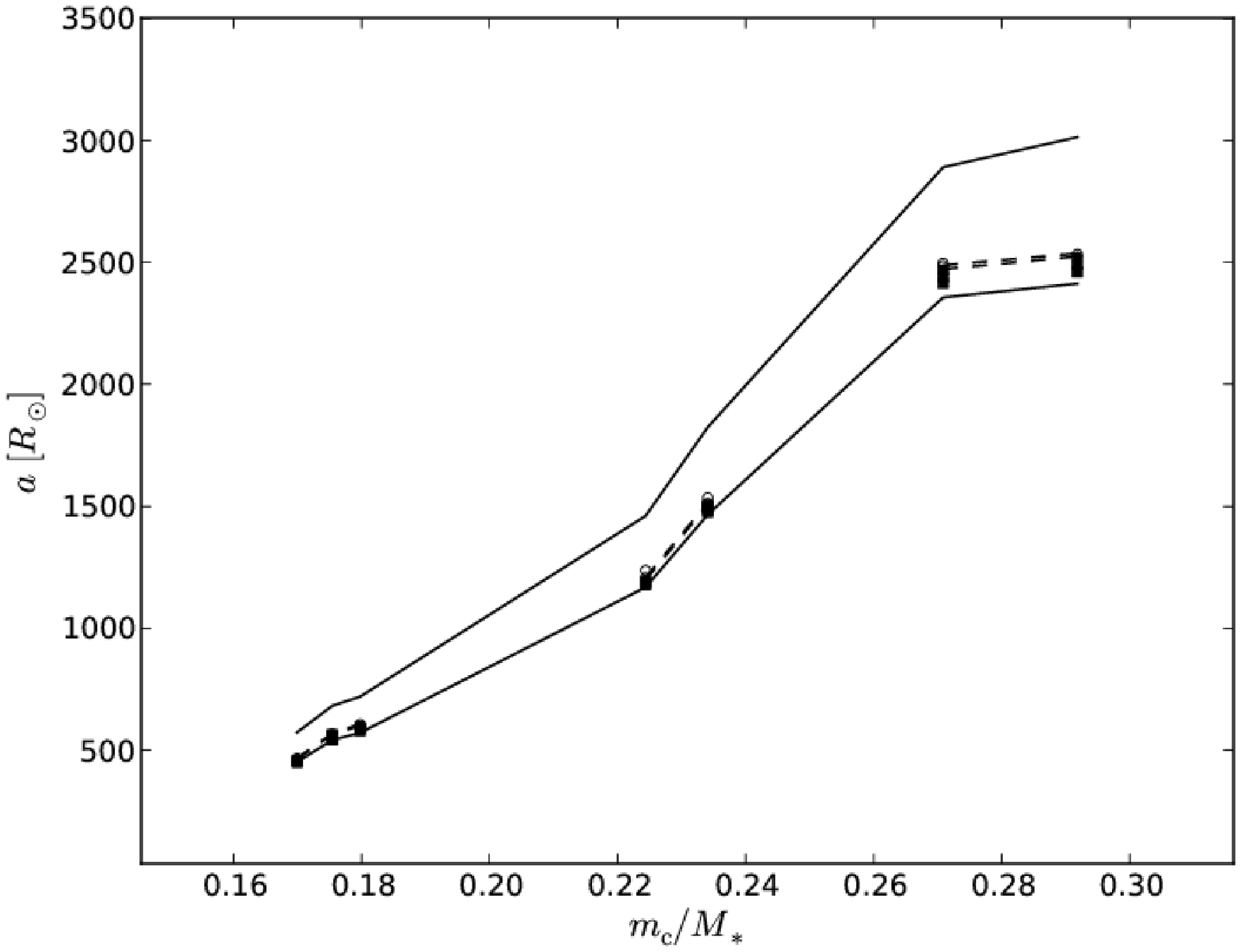}
\end{tabular}
\end{center}
\caption{Summary of the $q=1$ dynamical calculations as a function of initial separation, in units of stellar radii (left) and $R_\odot$ (right), and core mass, defined as the region where $X<0.01$, in units of stellar mass.
The three columns with the lowest core masses are the results of the $8.0\ M_\odot$ component binary calculations, the middle two columns are the results of the $14.0\ M_\odot$ component binary calculations, and the two columns with the highest core masses are the results of the $20.0\ M_\odot$ component binary calculations.
The open circles correspond to systems that did not inspiral and exhibit stable small-amplitude sinusoidal oscillations, the filled circles show systems that did not inspiral but do not have stable small-amplitude sinusoidal oscillations, and the filled squares correspond to systems that underwent an inspiral.
The solid line shows the orbital separation at first contact ($\eta=0$) and at the Roche limit ($\eta=1$). The upper and lower dashed lines mark the smallest stable and largest unstable initial separations respectively, the constraints on the dynamical stability limit as a function of mass and initial separation.}
\label{Fig:mc_r}
\end{figure*}

\begin{figure*}[htp]
\begin{center}
\begin{tabular}{cc}
\includegraphics[width=16.0cm]{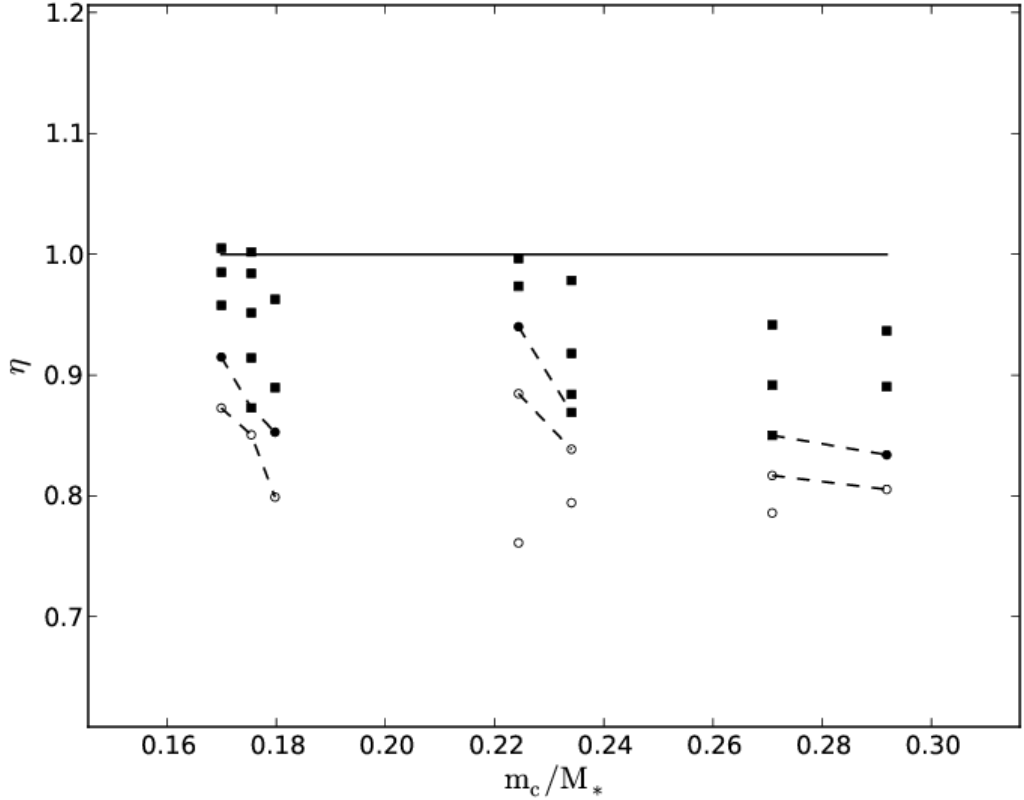}
\end{tabular}
\end{center}
\caption{Summary of the $q=1$ dynamical calculations as a function of degree of contact and core mass in units of the stellar mass.
The three columns with the lowest core masses are the results of the $8.0\ M_\odot$ component binary calculations, the middle two columns are the results of the $14.0\ M_\odot$ component binary calculations, and the two columns with the highest core masses are the results of the $20.0\ M_\odot$ component binary calculations.
The open circles correspond to systems that did not inspiral and exhibit stable small-amplitude sinusoidal oscillations, the filled circles show systems that did not inspiral but do not have stable small-amplitude sinusoidal oscillations, and the filled squares correspond to systems that underwent an inspiral.
The solid line shows the orbital separation at the Roche limit ($\eta=1$).
The upper and lower dashed lines mark the smallest stable and largest unstable initial separations, respectively, which are the constraints on the dynamical stability limit.}
\label{Fig:mc_eta}
\end{figure*}

We ran 35 dynamical integrations of $q=1$ binaries, using initial conditions generated from the binary equilibrium scans: three sets with the $m_\mathrm{i}=8.0\ M_\odot$ component binary, two sets with the $m_\mathrm{i}=14.0\ M_\odot$ component binary, and two sets with the $m_\mathrm{i}=20.0\ M_\odot$ component binary.
Each set uses a different age for the components, resulting in different core masses (see Table~\ref{TBL:Core_Masses}) and stellar envelopes, and we sample various initial orbital separations within each set (summarized in Table~\ref{TBL:Parameters}).
We note that for a given initial component mass, $m_\mathrm{i}$, the nominal core mass increases monotonically with age for the systems used in our study.
We label a run as $stable$ if we observe no inspiral and stable small-amplitude sinusoidal oscillations, $unstable$ if we observe no clear inspiral but do see a decaying core separation, and $inspiral$ if we observe a clear inspiral.
Figure~\ref{Fig:Rcore_stable} shows the evolution of the core separation for both a stable and unstable configuration, where in the stable configuration the local minimums increase through the integration while in the unstable configuration the core separation quickly decays.
The results of these dynamical integrations are tabulated in Table~\ref{TBL:Results_Dyn} and summarized in Figures~\ref{Fig:mc_r} and \ref{Fig:mc_eta}.

For each $m_\mathrm{i}$ and age we found an upper limit to the dynamical stability limit, $a_\mathrm{crit}$, discussed in more detail in \S\ref{SSec:StabilityLimit}.
The results suggest that, for a given $m_\mathrm{i}$ and $a_\mathrm{i}$, there either exists a critical age where the binary will undergo an inspiral or the binary will remain stable throughout the entire RGB. 
For each $m_\mathrm{i}$ we chose an age very close to $\tau_\mathrm{max}$, corresponding to the latest possible age and maximum initial orbital separation leading to an inspiral. The dynamical stability limit at this age provides an upper limit for the maximum initial separation that will result in hydrodynamic instability for each given component mass.
Figure~\ref{Fig:mc_r} shows the stability of the system as a function of core mass fraction, $m_\mathrm{c}/M_*$, and initial separation, and we see that while $a_\mathrm{crit}$ increases with both $m_\mathrm{i}$ and age, $a_\mathrm{crit}/R_\mathrm{*}$ increases with $m_\mathrm{i}$ but has varying dependence on the age.
Specifically, for the $m_\mathrm{i}=8.0\ M_\odot$ component binary, $a_\mathrm{crit}/R_\mathrm{*}$ increases with age, for the $m_\mathrm{i}=14.0\ M_\odot$ component binary, $a_\mathrm{crit}/R_\mathrm{*}$ stays roughly constant between the two ages, and for the $m_\mathrm{i}=20.0\ M_\odot$ component binary, $a_\mathrm{crit}/R_\mathrm{*}$ decreases slightly in the older system.
Similarly, we find that the orbital separation at first contact and at the Roche limit increases with both $m_\mathrm{i}$ and age, while the scaled orbital separation at first contact and at the Roche limit increases with $m_\mathrm{i}$ but has varying dependence on the age.
Figure~\ref{Fig:mc_eta} shows the stability of the system as a function of the core mass and degree of separation and we see that, for a given $m_\mathrm{i}$, the degree of contact at the dynamical stability limit decreases as the core mass fraction, or age, increases.
In what follows we discuss in more detail each set of runs.

\begin{deluxetable}{ccccccccccc}
\tablewidth{13.5cm}
\tabletypesize{\scriptsize}
\tablecolumns{11}
\tablecaption{Results of $q=1$ binary dynamical calculations\label{TBL:Results_Dyn}}
\tablehead{
    \colhead{$Run$} & \colhead{$t/P_\mathrm{i}$} & \colhead{$m_\mathrm{1}/M_\mathrm{tot}$} & \colhead{$m_\mathrm{2}/M_\mathrm{tot}$} & \colhead{$m_\mathrm{CE}/M_\mathrm{tot}$} & \colhead{$m_\mathrm{ej}/M_\mathrm{tot}$} & \colhead{$a_\mathrm{c}\ [R_\odot]$} & \colhead{$\Delta E_\mathrm{tot}$} & \colhead{$Result$}
   }
\startdata
$\tau = 35.22\mathrm{\ Myr}$ & \\
$1$  & $62.01$ & $0.500$ & $0.500$ & $0.000$ & $0.000$ & $466.55$ & $1.55\times10^{-6}$ & $Stable$ \\
$2$  & $22.76$ & $0.499$ & $0.501$ & $0.000$ & $0.000$ & $460.58$ & $1.54\times10^{-6}$ & $Unstable$ \\
$3$  & $30.83$ & $0.139$ & $0.134$ & $0.720$ & $0.007$ & $9.80$   & $7.72\times10^{-7}$ & $Inspiral$ \\
$4$  & $20.28$ & $0.141$ & $0.134$ & $0.713$ & $0.012$ & $9.13$   & $1.53\times10^{-4}$ & $Inspiral$ \\
$5$  & $14.75$ & $0.139$ & $0.135$ & $0.715$ & $0.012$ & $8.31$   & $2.51\times10^{-6}$ & $Inspiral$ \\
$\tau = 35.24\mathrm{\ Myr}$ & \\
$6$  & $28.78$ & $0.500$ & $0.500$ & $0.000$ & $0.000$ & $565.53$ & $9.34\times10^{-6}$ & $Stable$ \\
$7$  & $96.44$ & $0.128$ & $0.126$ & $0.740$ & $0.007$ & $10.02$  & $1.94\times10^{-5}$ & $Inspiral$ \\
$8$  & $24.69$ & $0.123$ & $0.132$ & $0.726$ & $0.018$ & $9.87$   & $5.96\times10^{-6}$ & $Inspiral$ \\
$9$  & $13.62$ & $0.129$ & $0.126$ & $0.742$ & $0.004$ & $13.43$  & $1.55\times10^{-6}$ & $Inspiral$ \\
$10$ & $10.60$ & $0.128$ & $0.126$ & $0.740$ & $0.006$ & $10.38$  & $1.21\times10^{-4}$ & $Inspiral$ \\
$11$ & $8.923$ & $0.125$ & $0.126$ & $0.742$ & $0.006$ & $10.65$  & $4.17\times10^{-4}$ & $Inspiral$ \\
$\tau = 35.26\mathrm{\ Myr}$ & \\
$12$ & $43.03$ & $0.500$ & $0.500$ & $0.000$ & $0.000$ & $605.31$ & $2.78\times10^{-5}$ & $Stable$ \\
$13$ & $87.83$ & $0.500$ & $0.500$ & $0.000$ & $0.000$ & $594.51$ & $8.52\times10^{-7}$ & $Unstable$ \\
$14$ & $35.07$ & $0.129$ & $0.122$ & $0.713$ & $0.036$ & $9.68$   & $1.37\times10^{-4}$ & $Inspiral$ \\
$15$ & $13.79$ & $0.128$ & $0.122$ & $0.701$ & $0.048$ & $8.94$   & $1.28\times10^{-4}$ & $Inspiral$ \\
$\tau = 13.30\mathrm{\ Myr}$ & \\
$16$ & $14.78$ & $0.500$ & $0.500$ & $0.000$ & $0.000$ & $1236.32$& $8.46\times10^{-4}$ & $Stable$ \\
$17$ & $16.89$ & $0.500$ & $0.500$ & $0.000$ & $0.000$ & $1205.66$& $7.67\times10^{-4}$ & $Stable$ \\
$18$ & $16.54$ & $0.500$ & $0.500$ & $0.000$ & $0.000$ & $1194.36$& $7.87\times10^{-4}$ & $Unstable$ \\
$19$ & $47.79$ & $0.192$ & $0.191$ & $0.483$ & $0.135$ & $24.02$  & $2.39\times10^{-2}$ & $Inspiral$ \\
$20$ & $26.02$ & $0.191$ & $0.192$ & $0.557$ & $0.061$ & $28.04$  & $1.24\times10^{-3}$ & $Inspiral$ \\
$\tau = 13.32\mathrm{\ Myr}$ & \\
$21$ & $54.03$ & $0.500$ & $0.500$ & $0.000$ & $0.000$ & $1532.01$& $4.45\times10^{-5}$ & $Stable$ \\
$22$ & $54.99$ & $0.500$ & $0.500$ & $0.000$ & $0.000$ & $1507.55$& $1.74\times10^{-4}$ & $Stable$ \\
$23$ & $107.04$& $0.178$ & $0.177$ & $0.638$ & $0.006$ & $62.09$  & $1.69\times10^{-4}$ & $Inspiral$ \\
$24$ & $61.19$ & $0.178$ & $0.177$ & $0.637$ & $0.008$ & $57.79$  & $1.36\times10^{-4}$ & $Inspiral$ \\
$25$ & $28.30$ & $0.177$ & $0.177$ & $0.569$ & $0.077$ & $37.81$  & $1.74\times10^{-4}$ & $Inspiral$ \\
$26$ & $17.39$ & $0.178$ & $0.177$ & $0.505$ & $0.140$ & $33.58$  & $1.04\times10^{-4}$ & $Inspiral$ \\
$\tau = 8.487\mathrm{\ Myr}$ & \\
$27$ & $46.45$ & $0.500$ & $0.500$ & $0.000$ & $0.000$ & $2488.99$& $3.58\times10^{-5}$ & $Stable$ \\
$28$ & $39.60$ & $0.500$ & $0.500$ & $0.000$ & $0.000$ & $2476.29$& $4.03\times10^{-5}$ & $Stable$ \\
$29$ & $84.79$ & $0.218$ & $0.218$ & $0.215$ & $0.349$ & $57.66$  & $1.03\times10^{-4}$ & $Inspiral$ \\
$30$ & $24.14$ & $0.218$ & $0.218$ & $0.423$ & $0.142$ & $75.05$  & $1.85\times10^{-4}$ & $Inspiral$ \\
$31$ & $14.24$ & $0.218$ & $0.218$ & $0.332$ & $0.232$ & $65.88$  & $1.05\times10^{-2}$ & $Inspiral$ \\
$\tau = 8.514\mathrm{\ Myr}$ & \\
$32$ & $13.96$ & $0.500$ & $0.500$ & $0.000$ & $0.000$ & $2525.62$& $4.32\times10^{-4}$  & $Stable$\\
$33$ & $14.43$ & $0.500$ & $0.500$ & $0.000$ & $0.000$ & $2508.09$& $1.36\times10^{-3}$  & $Unstable$\\
$34$ & $23.13$ & $0.221$ & $0.221$ & $0.431$ & $0.126$ & $83.33$  & $6.45\times10^{-4}$  & $Inspiral$\\
$35$ & $13.27$ & $0.221$ & $0.221$ & $0.266$ & $0.292$ & $71.41$  & $2.53\times10^{-2}$  & $Inspiral$\\
\enddata
\tablecomments{$t$ is the length of the run in units of $P_\mathrm{i}$, the initial orbital period, $m_\mathrm{i}$ is the initial mass of each component at birth, $\tau$ is the age of the components, $M_\mathrm{tot}$ is the total mass of the initial system, $m_\mathrm{1}$ is the mass in the first component, $m_\mathrm{2}$ is the mass in the second component, $m_\mathrm{CE}$ is the mass in the common envelope, $m_\mathrm{ej}$ is the mass of the ejecta, $a_\mathrm{c}$ is the core separation, $\Delta E$ is the fractional change in total energy, and $Result$ is the final state of the system.}
\end{deluxetable}

\subsection{Dynamical Calculations of $\mathbf{20\ M_\odot}$ Component Binaries}
\label{SSec:20.08514}
\begin{figure*}[htp]
\begin{center}
\begin{tabular}{cc}
\includegraphics[width=8.0cm]{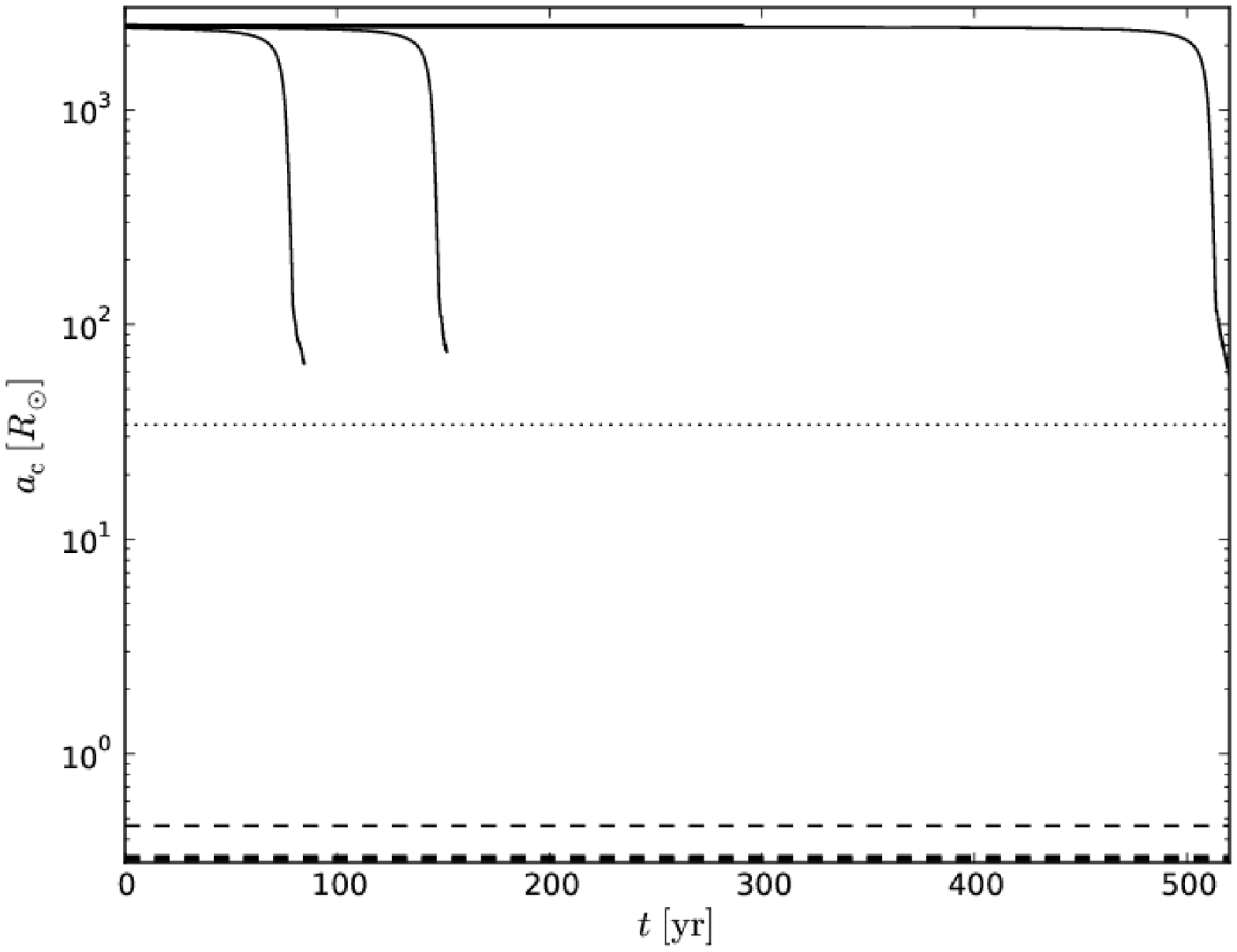}
\includegraphics[width=8.0cm]{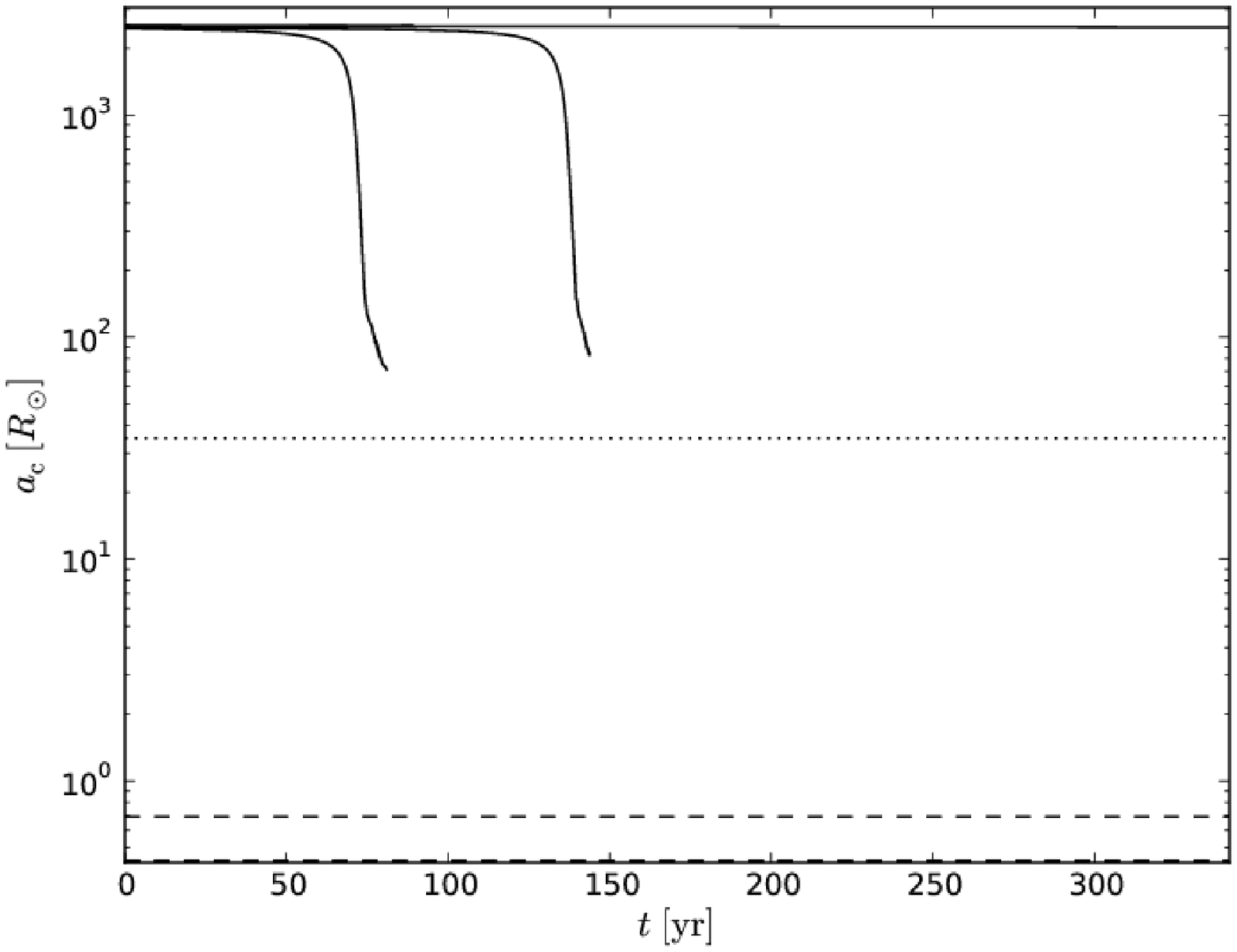}
\end{tabular}
\end{center}
\caption{Core separation as a function of time for multiple dynamical calculations of a binary with two $m_\mathrm{i}=20.0\ M_\odot$ stars at $8.487$ Myr (left) and $8.514$ Myr (right). The dashed horizontal lines show the various core radii generated by different prescriptions described in \S\ref{SSSec:Core_Masses}, the horizontal dotted line shows the resolution limit imposed by the softening length of the core.}
\label{Fig:20.0_Rcore}
\end{figure*}

We study $m_\mathrm{i}=20\ M_\odot$ component binaries at two ages, $\tau=8.487$ Myr and $\tau=8.514$ Myr.
For the younger binary we find that any configuration with $\eta\ge0.850$ ($a_\mathrm{i}\le2469\ R_\odot$) will undergo an inspiral, and for the older binary any configuration with $\eta\ge0.834$ ($a_\mathrm{i}\le2522\ R_\odot$) will undergo an inspiral.
Figure~\ref{Fig:20.0_Rcore} shows the core separation of the binary through common envelope evolution for different values of $a_\mathrm{i}$.
In this run $8.80\ M_\odot$ remains bound to each core, $10.58\ M_\odot$ is in the common envelope, and $11.61\ M_\odot$ is ejected from the system.
The final core separation for the runs ending in inspiral are between $65.88\ R_\odot$ and $83.33\ R_\odot$.

\begin{figure*}[htp]
\begin{center}
\begin{tabular}{cc}
\includegraphics[width=16.0cm]{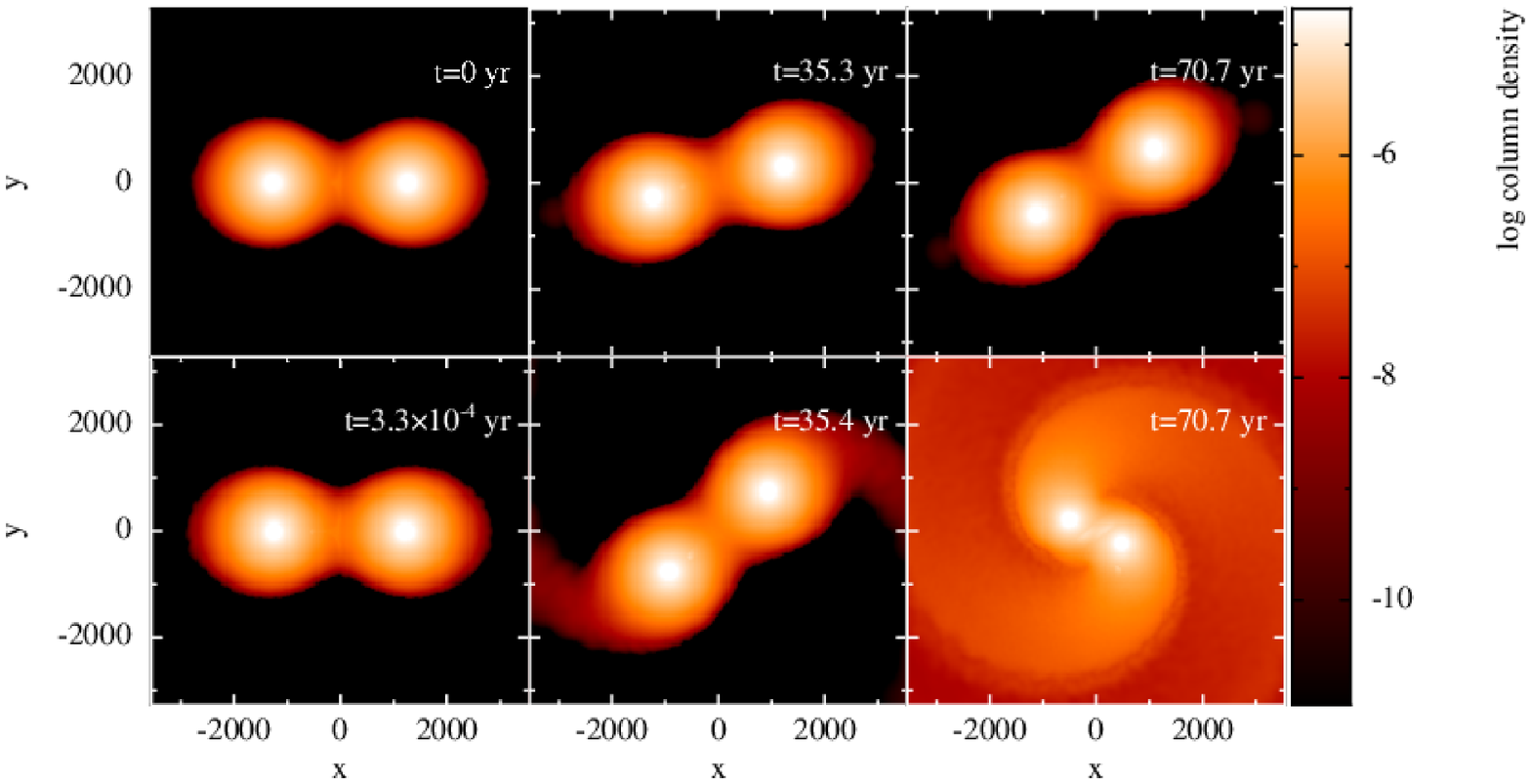}
\end{tabular}
\end{center}
\caption{Logarithm of column density in the orbital plane of a binary with $m_\mathrm{i}=20.0\ M_\odot$ components at $8.514\ \mathrm{Myr}$ for $a_\mathrm{i} = 2522\ R_\odot$, $\eta = 0.834$ (top), which remains stable, and $a_\mathrm{i} = 2461\ R_\odot$, $\eta = 0.937$ (bottom), which undergoes an inspiral. The column density is measured in units of $M_\odot R_\odot^{-2}$ and the position in units of $R_\odot$.}
\label{Fig:20.0_8514_SPH}
\end{figure*}

Figure~\ref{Fig:20.0_8514_SPH} compares the evolution of two systems at $\tau=8.514$ Myr with different initial orbital separations where one undergoes an inspiral and the other remains stable.
The top row shows the evolution of a binary with $a_\mathrm{i}=2534\ R_\odot$ ($\eta=0.805$), above the dynamical stability limit.
Nearly all of the mass remains bound to the system, and the final core separation stabilizes at $a_\mathrm{f}>0.995a_\mathrm{i}$.
The bottom row shows the evolution of a binary with $a_\mathrm{i}=2461\ R_\odot$ ($\eta=0.937$), well below the dynamical stability limit.
The system undergoes an inspiral with a final core separation of $71.41\ R_\odot$.
Compared to the lower mass binaries the stellar envelopes are more centrally dense, and we see the stellar surface quickly expanding, retaining less of their initial structure.
In both the unstable and stable binaries, the gas near the surface of the stars quickly fills the Roche lobe of each component, and mass flow occurs through the $L_\mathrm{1}$ Lagrangian point.

\pagebreak

\subsection{Dynamical Calculations of $\mathbf{14\ M_\odot}$ Component Binaries}
\label{SSec:14.0_1330}
\begin{figure*}[htp]
\begin{center}
\begin{tabular}{cc}
\includegraphics[width=8.0cm]{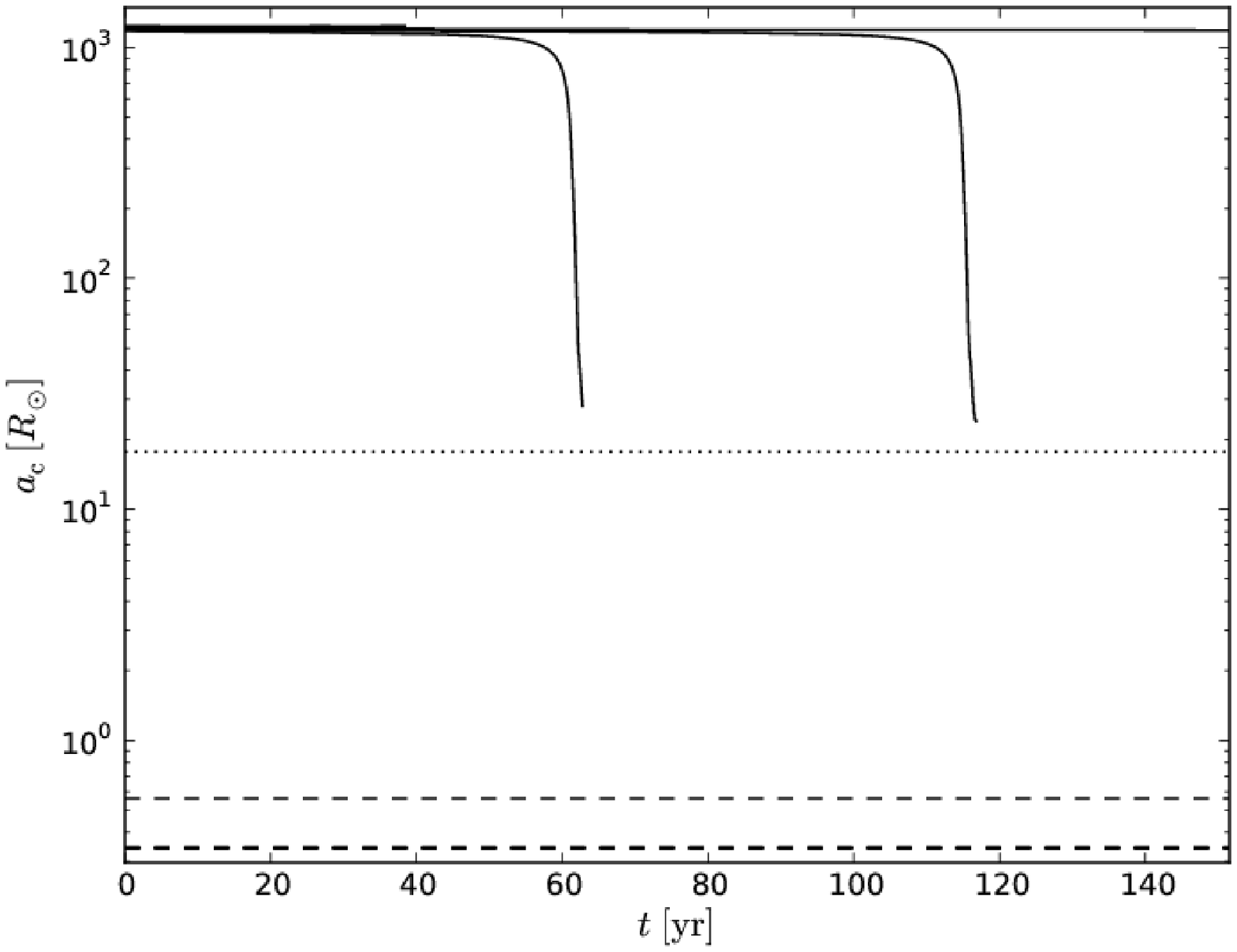}
\includegraphics[width=8.0cm]{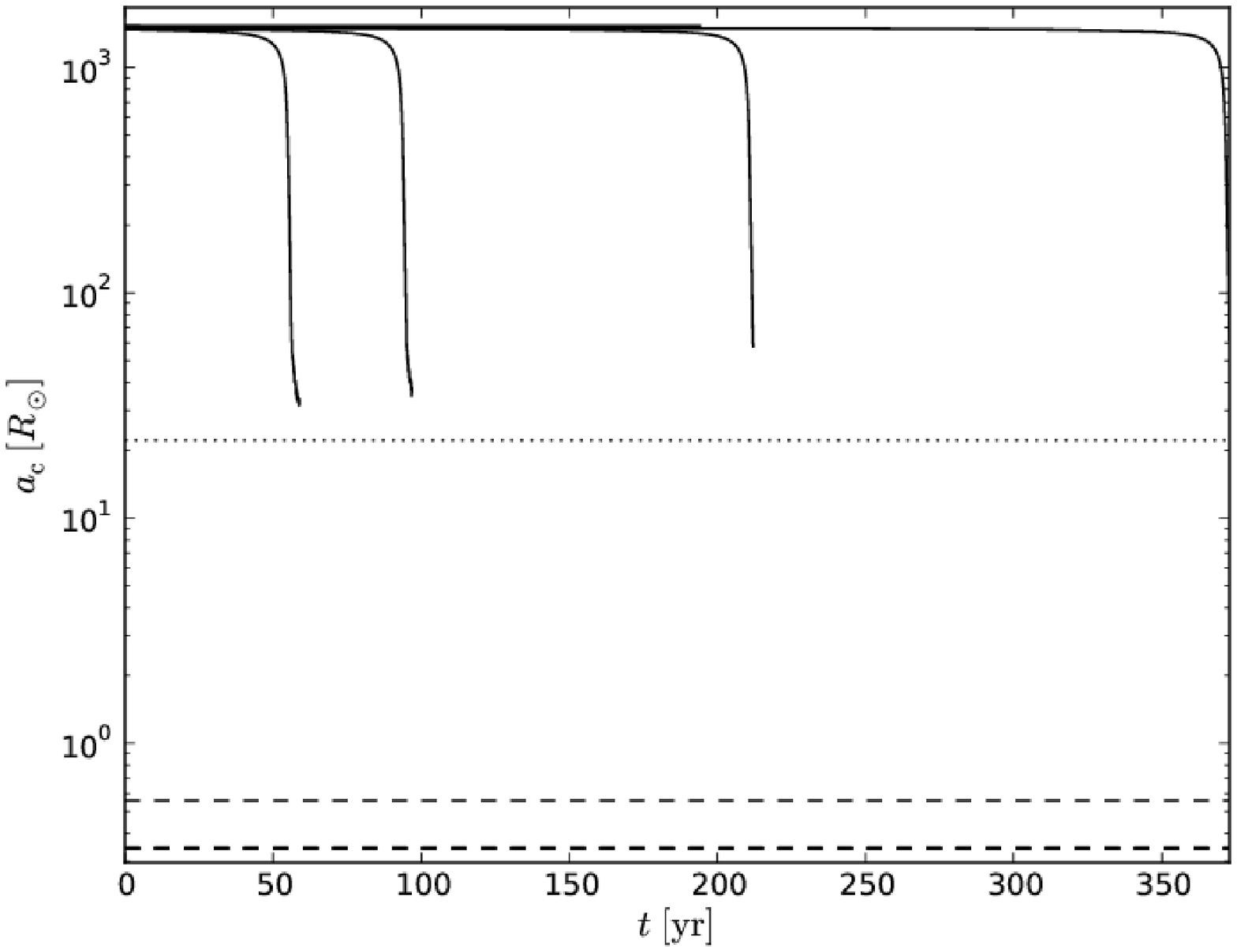}
\end{tabular}
\end{center}
\caption{Core separation as a function of time for multiple dynamical calculations of a binary with two $m_\mathrm{i}=14.0\ M_\odot$ stars at $13.30$ Myr (left) and $13.32$ Myr (right). The dashed horizontal lines show the various core radii generated by different prescriptions described in \S\ref{SSSec:Core_Masses}, the horizontal dotted line shows the resolution limit imposed by the softening length of the core.}
\label{Fig:14.0_Rcore}
\end{figure*}

We study $m_\mathrm{i}=14.0\ M_\odot$ component binaries at two ages, $\tau=13.30\ \mathrm{Myr}$ and $\tau=13.32\ \mathrm{Myr}$.
For the younger binary we find that any configuration with $\eta\ge0.885$ ($a_\mathrm{i}\le1207\ R_\odot$) will undergo an inspiral, and for the older binary any configuration with $\eta\ge0.869$ ($a_\mathrm{i}\le1505\ R_\odot$) will undergo an inspiral.
Figure~\ref{Fig:14.0_Rcore} shows the core separation of the binary through common envelope evolution for different values of $a_\mathrm{i}$.
The final core separation for the runs ending in inspiral are between $28.04\ R_\odot$ and $62.09\ R_\odot$.

\begin{figure*}[htp]
\begin{center}
\begin{tabular}{cc}
\includegraphics[width=16.0cm]{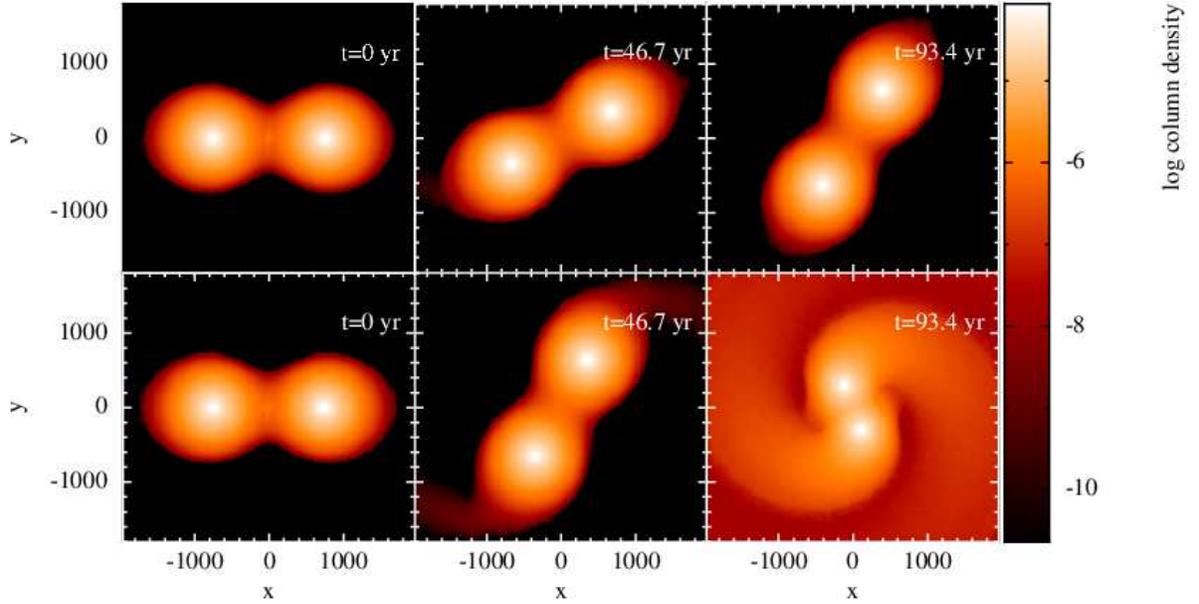}
\end{tabular}
\end{center}
\caption{Logarithm of column density in the orbital plane of a binary with $m_\mathrm{i}=14.0\ M_\odot$ components at $13.32\ \mathrm{Myr}$ for $a_\mathrm{i} = 1512\ R_\odot$, $\eta = 0.839$ (top), which remains stable, and $a_\mathrm{i} = 1489\ R_\odot$, $\eta = 0.918$ (bottom), which undergoes an inspiral. The column density is measured in units of $M_\odot R_\odot^{-2}$ and the position in units of $R_\odot$.}
\label{Fig:14.0_1332_SPH}
\end{figure*}

Figure~\ref{Fig:14.0_1332_SPH} compares the evolution of two systems at $13.32$ Myr with different initial orbital separations where one undergoes an inspiral and the other remains stable.
The top row shows the evolution of a binary with $a_\mathrm{i}=1512\ R_\odot$ ($\eta=0.839$), just above the dynamical stability limit.
Nearly all of the mass remains bound to the system, and the final core separation stabilizes at $a_\mathrm{f}>0.995a_\mathrm{i}$.
The bottom row shows the evolution of a binary with $a_\mathrm{i}=1489\ R_\odot$ ($\eta=0.918$), below the dynamical stability limit.
The system undergoes an inspiral with a final core separation of $37.81\ R_\odot$. In this run $4.94\ M_\odot$ remains bound to each core, $15.88\ M_\odot$ is in the common envelope, and $2.14\ M_\odot$ is ejected from the system.
Compared to the $20.0\ M_\odot$ component binaries, the $14.0\ M_\odot$ binaries remain stable deeper into contact.

\pagebreak

\subsection{Dynamical Calculations of $\mathbf{8\ M_\odot}$ Component Binaries}
\label{SSec:8.03522}
\begin{figure*}[htp]
\begin{center}
\begin{tabular}{cc}
\includegraphics[width=8.0cm]{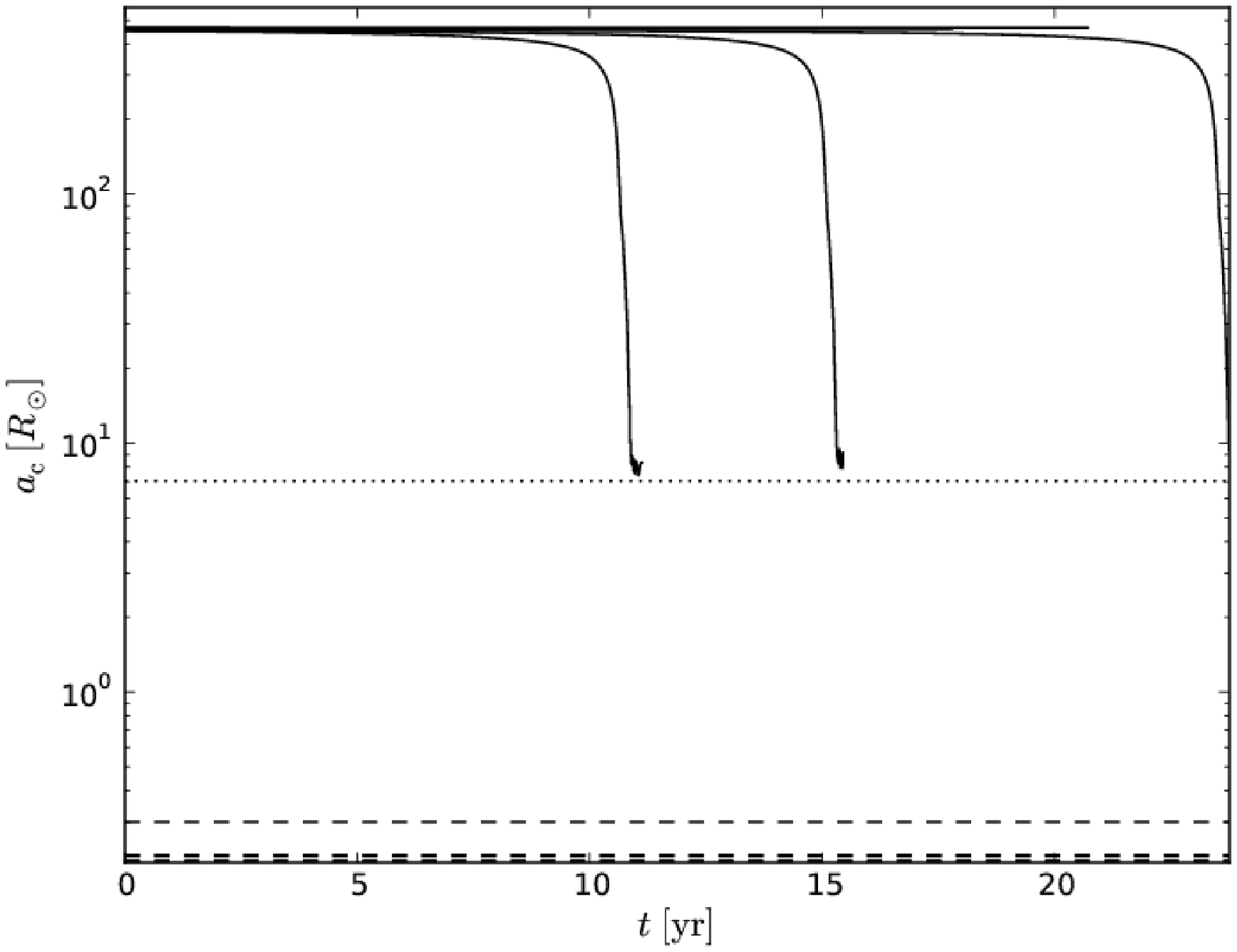}
\includegraphics[width=8.0cm]{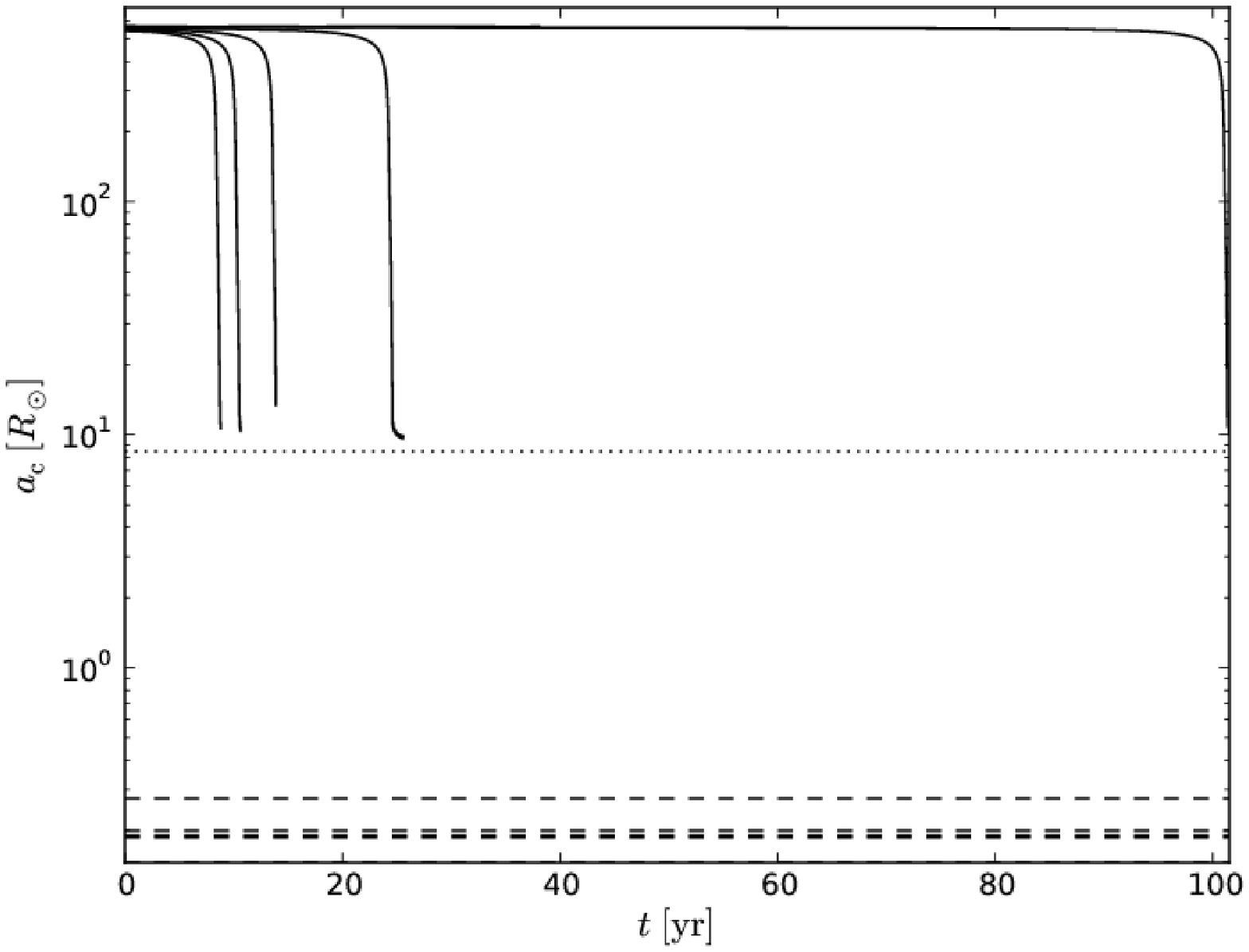} \\
\includegraphics[width=8.0cm]{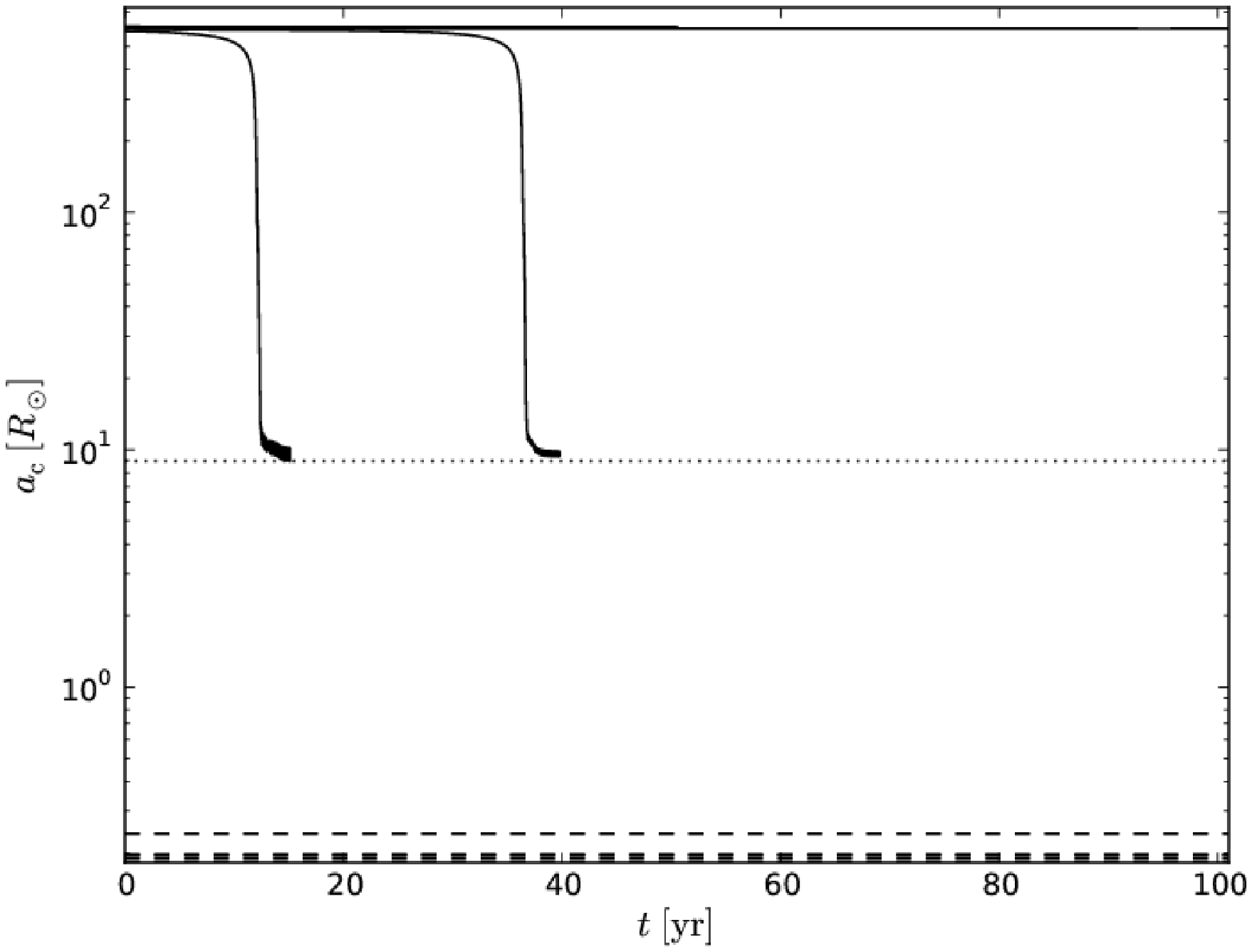}
\color{white}
\rule{8.0cm}{5.0cm}
\color{black}
\end{tabular}
\end{center}
\caption{Core separation as a function of time for multiple dynamical calculations of a binary with two $m_\mathrm{i}=8.0\ M_\odot$ stars at $35.22$ Myr (top left), $35.24$ Myr (top right), and $35.26$ Myr (bottom). The dashed horizontal lines show the various core radii generated by different prescriptions described in \S\ref{SSSec:Core_Masses}, the horizontal dotted line shows the resolution limit imposed by the softening length of the core.}
\label{Fig:8.0_Rcore}
\end{figure*}

\begin{figure*}[htp]
\begin{center}
\begin{tabular}{cc}
\includegraphics[width=16.0cm]{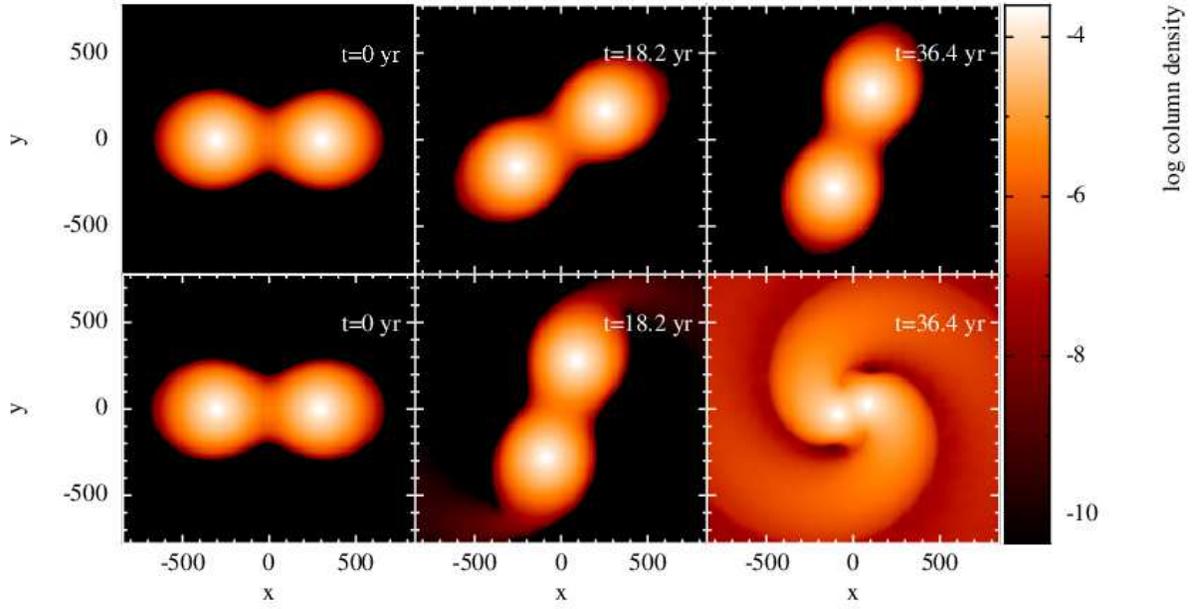}
\end{tabular}
\end{center}
\caption{Logarithm of column density in the orbital plane of a binary with $m_\mathrm{i}=8.0\ M_\odot$ components at $35.26$ Myr for $a_\mathrm{i} = 606\ R_\odot$, $\eta = 0.799$ (top), which remains stable, and $a_\mathrm{i} = 593 R_\odot$, $\eta = 0.886$ (bottom), which undergoes an inspiral. The column density is measured in units of $M_\odot R_\odot^{-2}$ and the position in units of $R_\odot$.}
\label{Fig:8.0_3522_SPH}
\end{figure*}

We study $m_\mathrm{i}=8.0\ M_\odot$ component binaries at three ages, $\tau=35.22$ Myr, $\tau=35.24$ Myr, and $\tau=35.26$ Myr and find an upper limit for the dynamical stability limit at $\eta\ge0.915$ ($a_\mathrm{i}\le462\ R_\odot$), $\eta\ge0.873$ ($a_\mathrm{i}\le564\ R_\odot$), and $\eta\ge0.853$ ($a_\mathrm{i}\le598\ R_\odot$) respectively.
Figure~\ref{Fig:8.0_Rcore} shows the core separation of the binary through common envelope evolution for different values of $a_\mathrm{i}$.
Figure~\ref{Fig:8.0_3522_SPH} compares the evolution of two systems at 35.26 Myr with different initial orbital separations where one undergoes an inspiral and the other remains stable.
The final core separation for the runs ending in inspiral are between $8.31\ R_\odot$ and $13.43\ R_\odot$.

The top row shows the evolution of a binary with $a_\mathrm{i}=606\ R_\odot$ ($\eta=0.799$), just above the dynamical stability limit.
Nearly all of the mass remains bound to the system, and the final core separation stabilizes at $a_\mathrm{f}>0.997a_\mathrm{i}$.
The bottom row shows the evolution of a binary with $a_\mathrm{i}=593\ R_\odot$ ($\eta=0.886$), just below the dynamical stability limit.
In this run $4.00\ M_\odot$ remains bound to the cores, $11.4\ M_\odot$ is in the common envelope, and $0.57\ M_\odot$ is ejected from the system.
Compared to the higher mass binaries the binary retains a well defined surface until the initial inspiral, after which most of the gas forms a common envelope around the tightly bound cores.

\pagebreak

\subsection{Dynamical Calculations of a $\mathbf{q=0.997}$ Binary}
\label{SSec:qne1}
\begin{figure*}[htp]
\begin{center}
\begin{tabular}{cc}
\includegraphics[width=16.0cm]{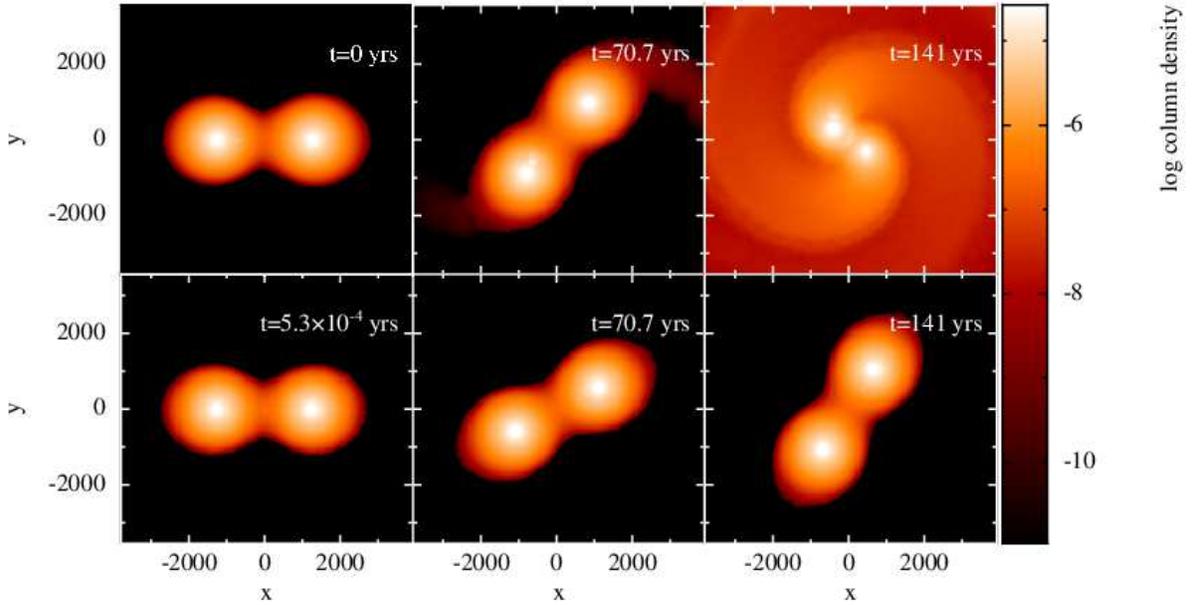}
\end{tabular}
\end{center}
\caption{Logarithm of column density in the orbital plane of two binaries with $m_\mathrm{i} = 20.0\ M_\odot$ components. The $q=0.997$ binary (top) with component ages of $8.514$ Myr and $8.487$ Myr at $a_\mathrm{i} = 2553\ R_\odot$ and $\eta = 0.794$ undergoes an inspiral. The $q=1$ binary (bottom) with components at $8.487$ Myr with $a_\mathrm{i} = 2497\ R_\odot$ and $\eta = 0.786$ is stable. The column density is measured in units of $M_\odot R_\odot^{-2}$ and the position in units of $R_\odot$.}
\label{Fig:qne1}
\end{figure*}

We perform a series of dynamical calculations of a binary system with $m_\mathrm{i} = 20\ M_\odot$ components, the secondary at $8.514$ Myr and the primary at $8.487$ Myr ($q=0.997$), a realization of a binary with components forming from the same cloud at slightly different times.
We summarize the initial conditions and results of these dynamical calculations in Table~\ref{TBL:qne1}.
We find that any configuration with $\eta\ge0.794$ ($a_i\le2556\ R_\odot$) will undergo an inspiral, and, even with the mass ratio being very close to unity, the secondary transfers a significant amount of mass to the primary; the stable configuration has a final mass ratio $q=0.965$.
Figure \ref{Fig:qne1} compares the dynamical calculations of this system and a $q=1$ binary system with $m_\mathrm{i} = 20\ M_\odot$ components, both at $8.487$ Myr, at roughly the same initial degree of contact, $\eta=0.794$ for the $q=0.997$ binary and $\eta=0.786$ for the $q=1$ binary.
We see that even though the $q=1$ binary has a smaller initial separation, $a_\mathrm{i}=2497$, than the $q=0.997$ binary, $a_\mathrm{i}=2556$, the $q=1$ binary is stable while the $q=0.997$ binary undergoes an inspiral.

\begin{deluxetable}{cccccccccccc}
\tablewidth{16cm}
\tabletypesize{\small}
\tablecolumns{12}
\tablecaption{Results of $q=0.997$ binary dynamical calculations\label{TBL:qne1}}
\tablehead{
    \colhead{$a_\mathrm{i}\ [R_\odot]$} & \colhead{$\eta$} & \colhead{$t/P_\mathrm{i}$} & \colhead{$m_\mathrm{1}/M_\mathrm{tot}$} & \colhead{$m_\mathrm{2}/M_\mathrm{tot}$} & \colhead{$m_\mathrm{CE}/M_\mathrm{tot}$} & \colhead{$m_\mathrm{ej}/M_\mathrm{tot}$} & \colhead{$a_\mathrm{c}\ [R_\odot]$} & \colhead{$\Delta E_\mathrm{tot}$} & \colhead{$Result$}
   }
\startdata
$2584$ & $0.716$ & $37.57$ & $0.491$ & $0.509$ & $0.000$ & $0.000$ & $2572.98$ & $3.65\times10^{-5}$ & $Stable$ \\
$2556$ & $0.794$ & $21.80$ & $0.248$ & $0.261$ & $0.477$ & $0.004$ & $269.60$  & $8.51\times10^{-5}$ & $Inspiral$ \\
$2522$ & $0.846$ & $22.74$ & $0.248$ & $0.250$ & $0.497$ & $0.004$ & $234.51$  & $8.99\times10^{-5}$ & $Inspiral$ \\
$2491$ & $0.895$ & $15.52$ & $0.221$ & $0.218$ & $0.475$ & $0.085$ & $89.38$   & $1.27\times10^{-4}$ & $Inspiral$ \\
$2461$ & $0.944$ & $12.43$ & $0.221$ & $0.218$ & $0.364$ & $0.197$ & $78.45$   & $9.33\times10^{-5}$ & $Inspiral$ \\
\enddata
\tablecomments{$a_\mathrm{i}$ is the initial orbital separation, $\eta$ is the degree of contact, $t$ is the length of the run in units of $P_\mathrm{i}$, the initial orbital period, $M_\mathrm{tot}$ is the total mass of the initial system, $m_\mathrm{1}$ is the mass in the first component, $m_\mathrm{2}$ is the mass in the second component, $m_\mathrm{CE}$ is the mass in the common envelope, $m_\mathrm{ej}$ is the mass of the ejecta, $a_\mathrm{c}$ is the core separation, $\Delta E$ is the fractional change in total energy, and $Result$ is the final state of the system.}
\end{deluxetable}

In the unstable configurations we observe a higher rate of mass loss through $L_\mathrm{2}$ than through $L_\mathrm{3}$, as the secondary component has a larger radius, compared to the $q=1$ binaries, where the mass flow is equal through $L_\mathrm{2}$ and $L_\mathrm{3}$.
We find that the $q=0.997$ binary system becomes unstable at a smaller degree of contact compared to $q=1$ binary systems composed of either component, and we suspect that the difference is even more exaggerated as the mass ratio moves further from unity, as symmetry enhances the stability of the binary.
\pagebreak

\subsection{Resolution Test}
\label{SSec:Resolution_Test}
\begin{figure*}[htp]
\begin{center}
\begin{tabular}{cc}
\includegraphics[width=10.0cm]{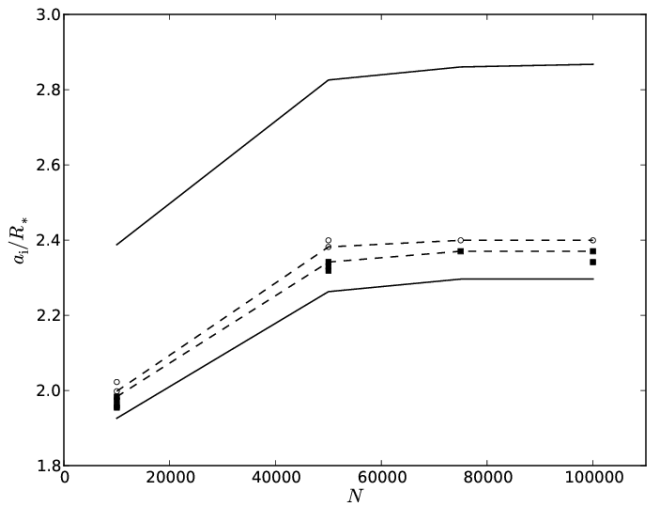}
\end{tabular}
\end{center}
\caption{Summary of dynamical calculations as a function of the number of particles in the single-star model. The open circles correspond to systems that did not inspiral and exhibit stable small-amplitude oscillations and the filled squares correspond to systems that underwent an inspiral. The solid lines show the orbital separation at first contact ($\eta=0$) and at the Roche limit ($\eta=1$). The upper and lower dashed lines mark the smallest stable and largest unstable initial separations respectively, the constraints to the dynamical stability limit.}
\label{Fig:Res_eta}
\end{figure*}

\begin{figure*}[htp]
\begin{center}
\begin{tabular}{cc}
\includegraphics[width=10.0cm]{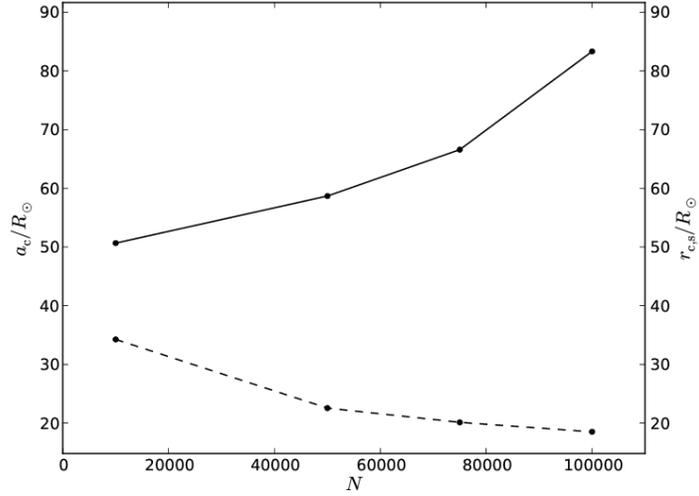}
\end{tabular}
\end{center}
\caption{Final core separation of the inspiraling binary closest to the dynamical stability limit (solid line) and softening length, $r_\mathrm{c,s}$, of the core particle (dotted line) as a function of the number of particles in the single-star model. We see that the final core separation decreases as the core particle softening length increases, suggesting that the final core separation seen in the simulations is not caused by running into the resolution limit.}
\label{Fig:Res_rfin}
\end{figure*}

To test the effects of varying the number of particles used to generate the single-star models, we ran additional sets of dynamical calculations using $N=10^4$, $N=5\times10^4$, and $N=7.5\times10^4$ particles to model the $m_\mathrm{i}=20\ M_\odot$ component at $8.514$ Myr in a $q=1$ binary.
Table~\ref{TBL:Resolution_Test} shows the softening length and mass of the core particle, dynamical stability limit, and efficiency of ejection as a function of the number of particles used in the single-star model.
Figure~\ref{Fig:Res_eta} shows the results of the dynamical integrations as a function of the number of particles in the single-star model.
We see that as the resolution increases, the dynamical stability limit quickly converges to the value found in the calculations with $N=10^5$ particles.
Figure~\ref{Fig:Res_rfin} shows the final core separation of the inspiraling binary closest to the dynamical stability limit and softening length of the core particle as a function of the number of particles used in the single-star model.
We find that the core separation after the {\it plunge-in} decreases as the resolution decreases, suggesting that the final separation is not limited by the resolution limit of the simulation and may be the physical transition from the {\it plunge-in} to the {\it slow spiral-in}.
The smaller core separations at the onset of the {\it slow spiral-in} seen in the lower resolution simulations may be due to the larger discrete masses of the particles being unbound from the system, leading to more mass being unbound at lower efficiencies.

\begin{deluxetable}{cccccccc}
\tablewidth{10.0cm}
\tabletypesize{\small}
\tablecolumns{7}
\tablecaption{Results of Resolution Test\label{TBL:Resolution_Test}}
\tablehead{
    \colhead{$N$} & \colhead{$r_{\mathrm{c,s}}\ [R_\odot]$} & \colhead{$m_{\mathrm{c,s}}/m_\mathrm{f}$} & \colhead{$a_{\mathrm{c}}\ [R_\odot]$} & \colhead{$\alpha_\mathrm{CE}$} & \colhead{$a_{\mathrm{crit}}\ [R_\odot]$} & \colhead{$\eta_\mathrm{crit}$}}
\startdata
$10000$  & $34.26$ & $0.482$ & $50.67$ & $0.795$ & $2102$ & $0.764$ & \\
$50000$  & $22.56$ & $0.436$ & $58.69$ & $0.706$ & $2525$ & $0.763$ & \\
$75000$  & $20.13$ & $0.427$ & $66.59$ & $0.778$ & $2525$ & $0.834$ & \\
$100000$ & $18.52$ & $0.422$ & $83.33$ & $0.760$ & $2461$ & $0.805$ & \\
\enddata
\tablecomments{$N$ is the number of particles in the single-star model, $r_\mathrm{c,s}$ is the softening length of the core particle, $m_\mathrm{c,s}/m_\mathrm{f}$ is the ratio of the core particle mass to the total mass, $a_\mathrm{c}$ is the final separation of the core particles in the unstable binary closest to the dynamical stability limit, $\alpha_\mathrm{CE}$ is the calculated efficiency of ejection of the unstable binary closest to the dynamical stability limit, $a_\mathrm{crit}$ is the upper bound on the dynamical stability limit (see \S\ref{SSec:Dynamical_Run}), and $\eta_\mathrm{crit}$ is the lower limit of the lower bound on the critical degree of contact (see \S\ref{Sec:Results}).}
\end{deluxetable}
\pagebreak

\section{Analysis}
\label{Sec:Analysis}

\subsection{Dynamical Stability Limit}
\label{SSec:StabilityLimit}
For each set of dynamical calculations of a binary with given components, there exists a dynamical stability limit, $a_\mathrm{crit}$, such that the behavior of binaries with $a_\mathrm{i}\ge a_\mathrm{crit}$ deviates greatly from that of binaries with $a_\mathrm{i}<a_\mathrm{crit}$.
Binaries with $a_\mathrm{i}\ge a_\mathrm{crit}$ lose at most a very small amount of mass and exhibit stable small-amplitude sinusoidal oscillations in core separation.
Binaries with $a_\mathrm{i}<a_\mathrm{crit}$ experience mass loss at the $L_\mathrm{2}$ and $L_\mathrm{3}$ Lagrangian points and eventually undergo inspiral.
These are expected to eject most of the gas and leave two core particles orbiting within the remaining common envelope with a separation much smaller than the initial orbital separation.
From our dynamical calculations, we identify an upper limit for $a_\mathrm{crit}$ as the stable configuration with the largest degree of contact and investigate further the behavior of these systems.

\begin{figure*}[htp]
\begin{center}
\begin{tabular}{cc}
\includegraphics[width=8.0cm]{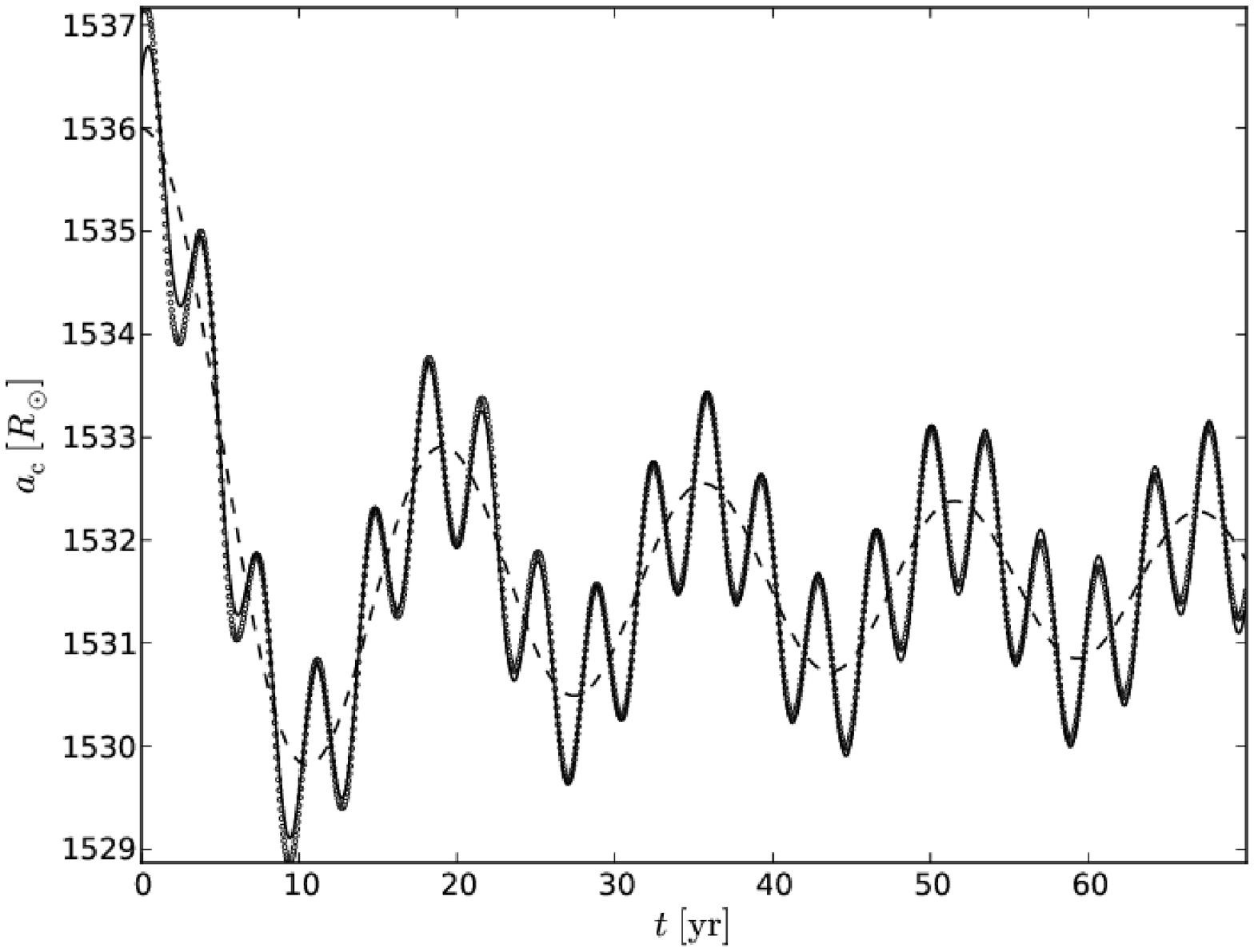} \\
\includegraphics[width=8.0cm]{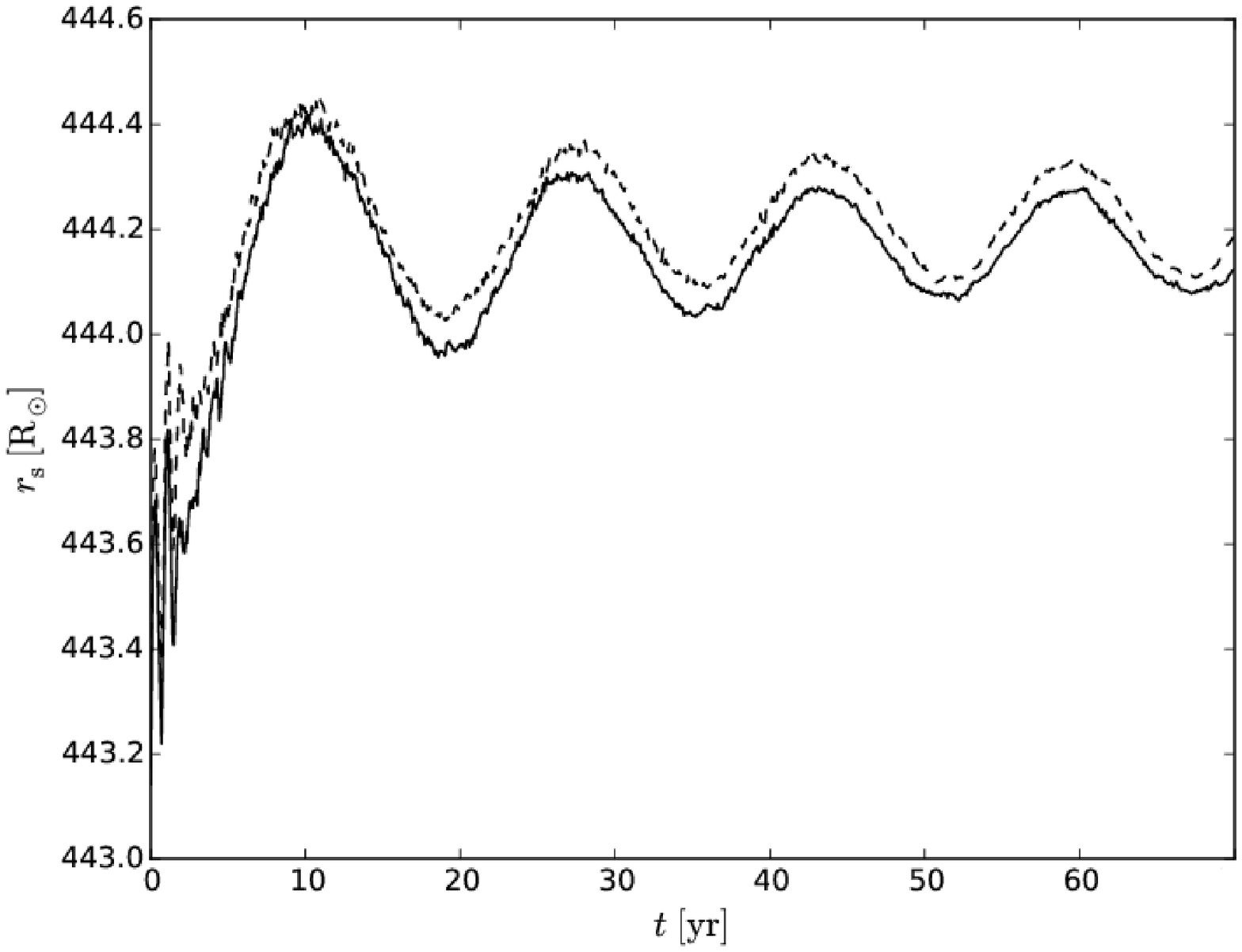}
\end{tabular}
\end{center}
\caption{Core separation and fits as a function of time for a stable $q=1$ binary with $m_\mathrm{i}=14.0\ M_\odot$ components at $13.32$ Myr and $a_\mathrm{i}=1535$ ($\eta=0.794$) (top). The circles show the core separation, the dashed line is the fit using only the mass-profile oscillation term and the constant core separation term, $a_\mathrm{c}=A_\mathrm{m}J_0(\omega_\mathrm{m}t)+a_{c,0}$, from \eqref{EQ:StableFit}, and the solid line is the fit using \eqref{EQ:StableFit}, which adds an additional sinusoidal term, $A_\mathrm{e}\sin(\omega_\mathrm{e}t+\phi)$, for the epicyclic oscillations. Radius encompassing $90\%$ of the mass for each component (bottom). We find that the radial pulsations match the mass-profile oscillations seen in the core separation.}
\label{Fig:StabilityLimit}
\end{figure*}

In the stable binary systems we observe an initial perturbation in the core separation, caused by oscillations in the stellar components, as the models are not in perfect equilibrium.
The perturbation is slowly damped by the artificial viscosity and becomes a stable small-amplitude oscillation with a period of many tens of dynamical timescales.
In addition, we see smaller-amplitude oscillations with a period roughly equal to the orbital period of the binary, which are essentially epicyclic oscilations occuring because the orbit is not perfectly circular \citep{1994ApJ...432..242R}.
In fitting the core separation versus time, we characterize the larger-period oscillations as a $\mathrm{0^{th}}$ order Bessel function and the smaller-period oscillations as a sine wave, \begin{equation}a_\mathrm{c}=A_\mathrm{m}J_0(\omega_\mathrm{m}t)+A_\mathrm{e}\sin(\omega_\mathrm{e}t+\phi)+a_{c,0},\label{EQ:StableFit}\end{equation} where $A_\mathrm{m}$ and $\omega_\mathrm{m}$ are the amplitude and frequency of the mass-profile oscillations (corresponding to radial pulsations of the components), $A_\mathrm{e}$ and $\omega_\mathrm{e}$ are the amplitude and frequency of the epicyclic oscillations, $\phi$ is the phase of the epicyclic oscillations, and $a_\mathrm{c,0}$ is the stable mean core separation.
We find that in the first few mass-oscillations the frequency increases very slightly, which we account for in the fit using $\omega_\mathrm{m}=\omega_\mathrm{m,0}+\omega'_\mathrm{m}t,$ where $\omega_\mathrm{m,0}$ is the initial angular frequency and $\omega'_\mathrm{m}$ is the rate at which the angular frequency increases $(\omega_\mathrm{m,0}\gg\omega'_\mathrm{m})$.
Figure~\ref{Fig:StabilityLimit} shows the core separation with the fitted equation and $90\%$ mass fraction radii as a function of time for a stable binary system.
Table~\ref{TBL:Dynamical_Stability_Limit} lists the parameter fits for the stable binary configurations closest to the dynamical stability limit, where $P_\mathrm{m}=2\pi/\omega_\mathrm{m,0}$ and $P_\mathrm{e}=2\pi/\omega_\mathrm{e}$.

\begin{deluxetable}{ccccccccccccc}
\tablewidth{17.5cm}
\tabletypesize{\small}
\tablecolumns{11}
\tablecaption{Stable Simulations Close to the Dynamical Stability Limit\label{TBL:Dynamical_Stability_Limit}}
\tablehead{
    \colhead{$m_\mathrm{i}\ [M_\odot]$} & \colhead{$\tau\ [\mathrm{Myr}]$} & \colhead{$m_\mathrm{c}/m_\mathrm{f}$} & \colhead{$a_\mathrm{crit}/R_\mathrm{*}$} & \colhead{$a_\mathrm{crit}\ [R_\odot]$} & \colhead{$\eta_\mathrm{crit}$} & \colhead{$A_\mathrm{m}/R_*$} & \colhead{$P_\mathrm{m}/P_\mathrm{orb}$} & \colhead{$A_\mathrm{e}/R_*$} & \colhead{$P_\mathrm{e}/P_\mathrm{orb}$} & \colhead{$a_\mathrm{c,0}\ [R_\odot]$}
}
\startdata
$8.0$  & $35.22$  & $0.170$ & $2.207$ & $466.5$ & $0.873$ & $1.07\times10^{-3}$ & $13.95$ & $2.78\times10^{-5}$ & $1.01$ & $466.6$\\
$8.0$  & $35.24$  & $0.175$ & $2.219$ & $566.8$ & $0.837$ & $3.87\times10^{-3}$ & $11.23$ & $1.73\times10^{-4}$ & $1.01$ & $565.6$ \\
$8.0$  & $35.26$  & $0.180$ & $2.255$ & $606.1$ & $0.799$ & $3.60\times10^{-3}$ & $11.09$ & $2.66\times10^{-4}$ & $1.01$ & $605.3$\\
$14.0$ & $13.30$  & $0.224$ & $2.261$ & $1207$  & $0.885$ & $2.35\times10^{-3}$ &  $4.64$ & $2.89\times10^{-4}$ & $0.98$ & $1205.6$\\
$14.0$ & $13.32$  & $0.234$ & $2.264$ & $1512$  & $0.839$ & $2.50\times10^{-3}$ &  $6.62$ & $5.01\times10^{-4}$ & $0.98$ & $1507.9$\\
$20.0$ & $8.487$  & $0.271$ & $2.422$ & $2486$  & $0.817$ & $3.69\times10^{-3}$ &  $5.11$ & $2.69\times10^{-4}$ & $0.98$ & $2476.9$\\
$20.0$ & $8.514$  & $0.292$ & $2.411$ & $2534$  & $0.805$ & $3.60\times10^{-3}$ &  $5.29$ & $2.10\times10^{-4}$ & $0.99$ & $2524.4$\\
\enddata
\tablecomments{$m_\mathrm{i}$ is the mass of the star at birth, $\tau$ is the age of the star, $m_\mathrm{c}/m_\mathrm{f}$ is the ratio of the nominal core mass to the total mass at $\tau$, $a_\mathrm{crit}$ is the upper bound on the dynamical stability limit (see \S\ref{SSec:Dynamical_Run}), $R_\mathrm{*}$ is the radius of the star at $\tau$, $\eta_\mathrm{crit}$ is the lower bound on the critical degree of contact (see \S\ref{Sec:Results}), $A_\mathrm{m}$ is the amplitude of the mass-profile oscillations, $P_\mathrm{m}/P_\mathrm{orb}$ is the ratio of the mass-profile oscillations period to the orbital period, $A_\mathrm{e}$ is the amplitude of the epicyclic oscillations, $P_\mathrm{e}/P_\mathrm{orb}$ is the ratio of the epicyclic period to the orbital period, and $a_\mathrm{c,0}$ is the mean stable core separation.}
\end{deluxetable}
\pagebreak

\subsection{Mass Flow in Unstable Binaries}
\label{SSec:Mass_Flow}

In our dynamical calculations of unstable binaries, we see three distinct phases of mass flow (see \S\ref{SSec:Dynamical_Run}), the {\it loss of corotation}, the {\it plunge-in}, and the beginning of the {\it slow spiral-in} \citep{2001ASPC..229..239P}.
We characterize the mass flow by determining the mass fraction of each stellar component, the common envelope, and the ejecta, along with the efficiency of ejection in order to extrapolate the final state of our calculation to a potential remnant binary.

We determine the mass fraction of each component using a prescription from \citet{2006ApJ...640..441L}.
We consider an {\it SPH} particle $i$ bound to the system if the particle's energy with respect to the center of mass is negative. The particle's energy is defined as \begin{equation}\label{EQ:Ecom}E_\mathrm{com}=m_i\left(\frac{1}{2}v_{i}^2+u_i+\Phi_{\mathrm{grav,}i}\right),\end{equation} where $v_{i}$ is the velocity and $d_{i}$ is the distance of the particle relative to the center of mass, $u_i$ is the specific internal energy of the particle, and $\Phi_{\mathrm{grav,}i}$ is the gravitational potential of the particle.
Of the bound particles, we consider a particle $i$ bound to the core $1$ if the following conditions are met:
\\
\\
\indent1) $E_{\mathrm{1,}i} < 0$

2) $d_{\mathrm{1,}i}<a_\mathrm{c}$

3) $E_{\mathrm{1,}i} < E_{\mathrm{2,}i}$
\\
\\
where $v_{\mathrm{1,}i}$ is the velocity and $d_{\mathrm{1,}i}$ is the distance of the particle relative to core 1, $a_\mathrm{c}$ is the separation between the two cores, and $E_{\mathrm{1,}i}$ is the energy with respect to core 1, defined as \begin{equation}\label{EQ:Ecore}E_{\mathrm{1,}i} = m_i\left(\frac{1}{2}v_{\mathrm{1,}i}^2+u_i-\frac{GM_1}{d_{\mathrm{1,}i}}\right).\end{equation}
Analogous conditions are used to test if a particle is bound to core 2.
We consider particles bound to the system, but not bound to a specific core, to be part of the common envelope and particles not bound to the system to be ejected.
Table~\ref{TBL:Results_Dyn} and Table~\ref{TBL:qne1} list the final mass of each component, the common envelope, and the ejecta for the $q=1$ binary and the $q=0.997$ binary dynamical calculations, respectively.

We follow the prescription described in \citet{2011ApJ...730...76I} to calculate the efficiency of the common envelope ejection from loss of orbital separation of the cores, $\alpha_\mathrm{CE}$, defined by \begin{equation}\alpha_\mathrm{CE} = \frac{\Delta E_\mathrm{CE}}{\Delta E_\mathrm{c}}\label{EQ:AlphaCE},\end{equation} where $\Delta E_\mathrm{CE}$ is the change in the binding energy of the common envelope and $\Delta E_\mathrm{c}$ is the change in orbital energy of the cores.
The efficiency $\alpha_\mathrm{CE}$ is always less than unity since the ejected mass have velocities greater than the escape speed.
Table~\ref{TBL:Remnant_Properties} summarizes the derived values of $\alpha_\mathrm{CE}$ from our dynamical calculations.

\begin{figure*}[htp]
\begin{center}
\begin{tabular}{cc}
\includegraphics[width=12.0cm]{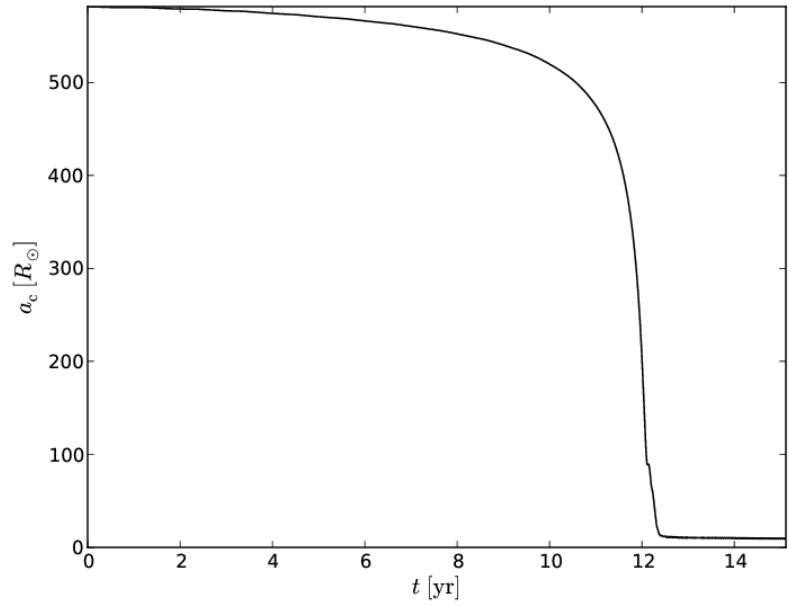}
\end{tabular}
\end{center}
\caption{Core separation of a $q=1$ binary with $8.0\ M_\odot$ components at $35.26$ Myr with $a_\mathrm{i}=580\ R_\odot$ ($\eta=0.963$). We see the core separation decrease steadily until the {\it plunge-in} where the core separation decreases to $a_\mathrm{c}\sim9\ R_\odot$ before stabilizing.}
\label{Fig:RCore_unstable}
\end{figure*}

\begin{figure*}[htp]
\begin{center}
\begin{tabular}{cc}
\includegraphics[width=12.0cm]{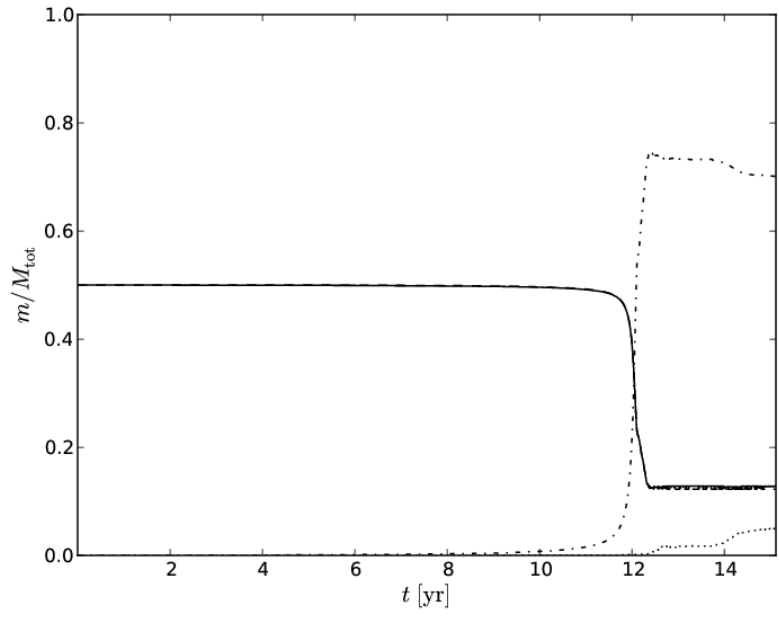}
\end{tabular}
\end{center}
\caption{Mass flow of a $q=1$ binary with $8.0\ M_\odot$ components at $35.26$ Myr with $a_\mathrm{i}=580\ R_\odot$ ($\eta=0.963$). The solid and dashed lines show the mass fraction bound to each core (since $q=1$, these values are nearly identical), the dash-dotted line shows the mass fraction in the common envelope and the dotted line shows the mass fraction in the ejecta.}
\label{Fig:Mass_Flow}
\end{figure*}

\begin{figure*}[htp]
\begin{center}
\begin{tabular}{cc}
\includegraphics[width=12.0cm]{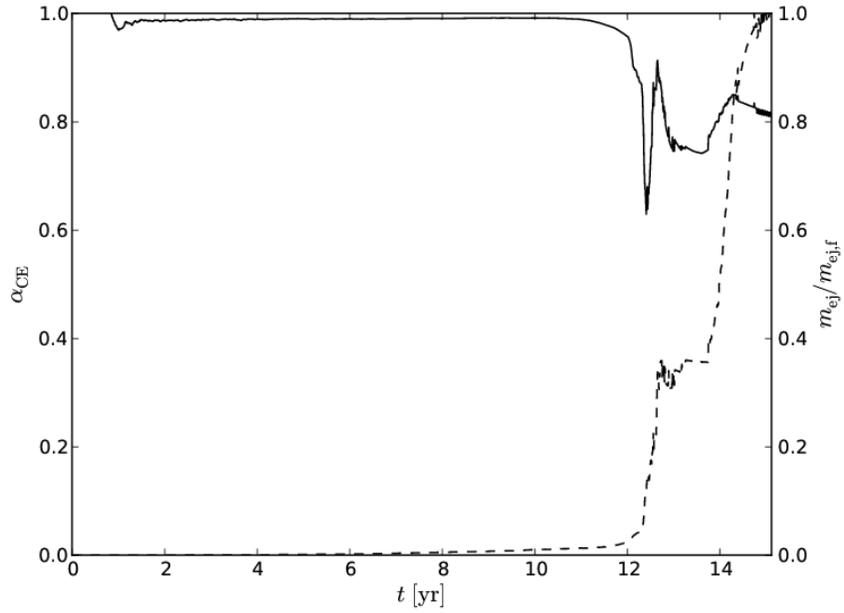}
\end{tabular}
\end{center}
\caption{Efficiency coefficient of the ejecta as a function of time for a $q=1$ binary with $8.0\ M_\odot$ components at $35.26$ Myr with $a_\mathrm{i}=580\ R_\odot$ ($\eta=0.963$). The efficiency coefficient, $\alpha_\mathrm{CE}$, is defined as $\Delta E_\mathrm{CE}/\Delta E_\mathrm{c}$, where $\Delta E_\mathrm{CE}$ is the change in energy in the common envelope and $\Delta E_\mathrm{c}$ is the change in orbital energy in the core particles. The solid line shows $\alpha_\mathrm{CE}$ and the dashed line shows $m_\mathrm{ej}/m_\mathrm{ej,f}$, the mass ejected by the system scaled by the total mass ejected by the system at the end of the dynamical calculation. $\alpha_\mathrm{CE}$ starts very close to unity until the {\it plunge-in}, where gas is unbound at a low efficiency ($\alpha_\mathrm{CE}\sim0.6$), followed by a period of higher efficiency ($\alpha_\mathrm{CE}>0.9$) mass loss during the {\it slow spiral-in}.}
\label{Fig:Alpha}
\end{figure*}

\begin{figure*}[htp]
\begin{center}
\begin{tabular}{cc}
\includegraphics[width=12.0cm]{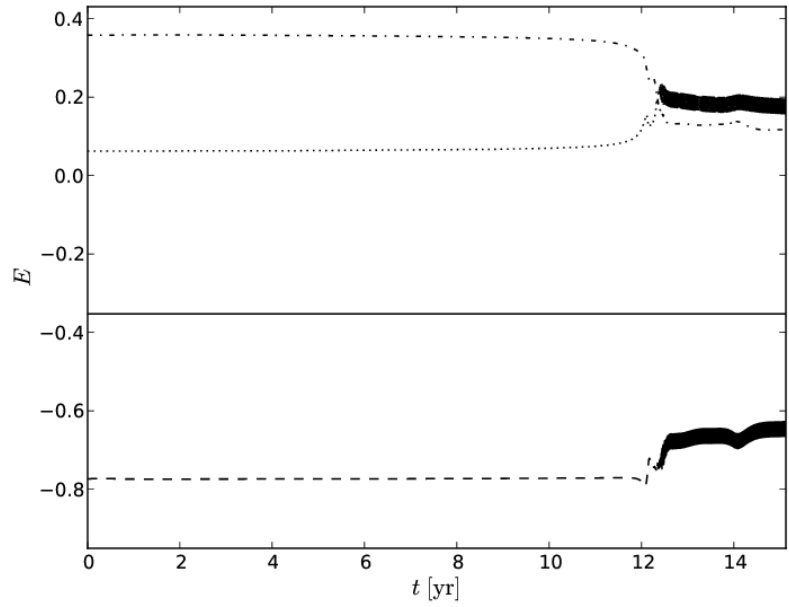}
\end{tabular}
\end{center}
\caption{Energy of a $q=1$ binary with $8.0\ M_\odot$ components at $35.26$ Myr with $\eta=0.886$ ($a_\mathrm{i}=592.9\ R_\odot$). The solid line is the total energy, the dashed line is the potential energy, the dotted line is the kinetic energy, and the dash-dotted line is the internal energy.}
\label{Fig:Energy}
\end{figure*}

Figures~\ref{Fig:RCore_unstable}--\ref{Fig:Energy} show the core separation, mass flow, efficiency of ejection, and energy as a function of time for an unstable $q=1$ binary with $m_\mathrm{i}=8.0\ M_\odot$ components at $35.26$ Myr with $\eta=0.886$ ($a_\mathrm{i}=592.9\ R_\odot$).
During the {\it loss of corotation}, we see mass flow from the components to the common envelope with a very small amount of unbound mass at high efficiency ($\alpha_\mathrm{CE}>0.9$).
The core separation steadily decreases as internal energy is transferred to kinetic and potential energy.
This is followed by the {\it plunge-in}, where the core separation suddenly decreases to $0.16a_\mathrm{i}$, and nearly all of the gas bound to an individual component transfers to the common envelope, with a small fraction becoming unbound with a lower efficiency ($\alpha_\mathrm{CE}\sim0.6$).
During this phase we observe a sharp transfer from the internal energy to the kinetic and potential energy due to shocks causing the gas to expand.
Finally, during the {\it slow spiral-in}, we observe the mass in the common envelope slowly being ejected from the system at a high efficiency ($\alpha_\mathrm{CE}>0.9$), suggesting that there exists a phase of mass loss after the {\it plunge-in} that occurs on the dynamical timescale.
The core separation oscillates about a relatively stable final separation, accompanied by rapid exchanges between kinetic and potential energy alongside a steady decrease in the thermal energy as the shocks from the cores slowly unbind the gas in the common envelope.

\pagebreak

\subsection{Properties of the Remnant Binary}
\label{SSec:Remnant_Properties}
Since the full ejection of the remaining envelope after an inspiral likely occurs on the thermal timescale \citep{2013A&ARv..21...59I}, we extrapolate the final separation of the remnant binary using a prescription described in \citet{2011ApJ...730...76I}, modified to fit a $q=1$ binary system, assuming that the energy required to unbind the remaining envelope is extracted from the orbital energy of the cores.
We use the efficiency coefficient, $\alpha_\mathrm{CE}$, the core separation, $a_\mathrm{c}$, and mass of each core particle, $m_{\mathrm{c,}1}$ and $m_{\mathrm{c,}2}$, from the final snapshot to approximate the final separation of the core particles, $a_\mathrm{f,sim}$, after completely ejecting the remaining bound particles using \begin{equation}\label{EQ:Afin}a_\mathrm{f,sim}=\frac{\alpha_\mathrm{CE}\mu a_\mathrm{c}}{\alpha_\mathrm{CE}\mu-E_\mathrm{CE}a_\mathrm{c}},\end{equation} where $\mu=Gm_{\mathrm{c,}1}m_{\mathrm{c,}2}$, and $E_\mathrm{CE}$ is the energy of the remaining bound particles.

Next, we investigate the sensitivity of the core separation of the remnant binary on the core mass.
Due to the resolution limit of the simulation, the masses of the core particles are larger than the core masses calculated in \S\ref{SSSec:Core_Masses}.
As the energy of the core particle does not change during the dynamical calculation, the energy required to unbind the excess mass will be the same as the binding energy of the excess mass using the initial mass profile, even if the mass profile within the core particle should change through the calculation.
We extrapolate the final core separation after fully ejecting the remaining envelope, leaving only the two helium cores, as \begin{equation}a_\mathrm{f}=\frac{\alpha_\mathrm{CE}\mu a_\mathrm{f,sim}}{\alpha_\mathrm{CE}\mu-E_\mathrm{bind}a_\mathrm{f,sim}},\end{equation} where the energy required to unbind the excess mass is calculated as \begin{equation}E_\mathrm{bind}=2\int^{m_\mathrm{c,sim}}_{m_\mathrm{c}}\left(\frac{Gm_\mathrm{c}}{r}+u\right)dm=8\pi\int^{r_\mathrm{c,sim}}_{r_\mathrm{c}}\left(\frac{Gm_\mathrm{c}}{r}+u\right)r^2\rho dr,\end{equation} where $u$ is the internal energy per unit mass at radius $r$, and $\rho$ is the density at $r$.
We assume a crude criterion that the final orbital separation must be larger than twice the core radius to survive as a tightly bound binary, and otherwise consider the system to have merged.

\begin{figure*}[htp]
\begin{center}
\begin{tabular}{cc}
\includegraphics[width=8.0cm]{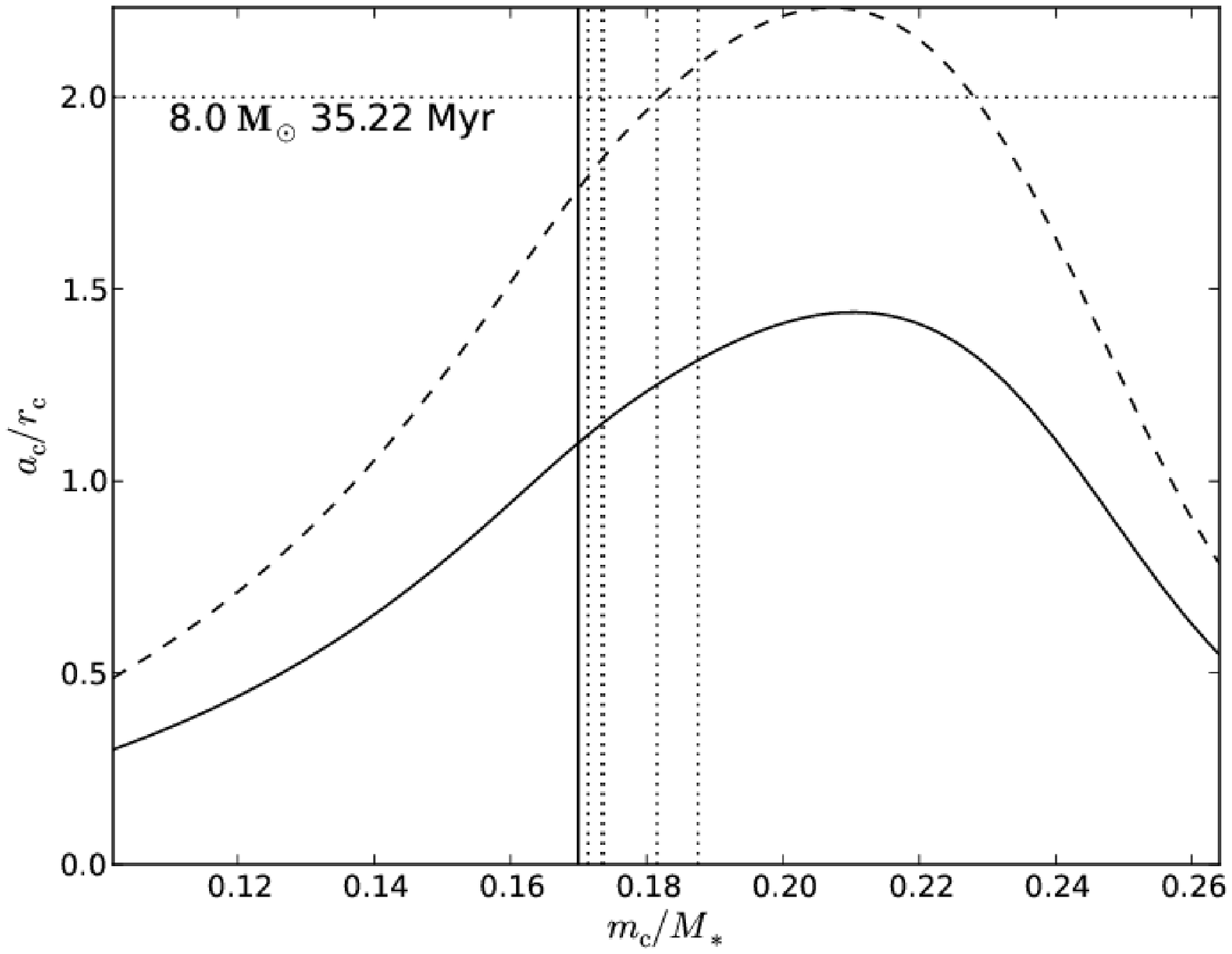}
\includegraphics[width=8.0cm]{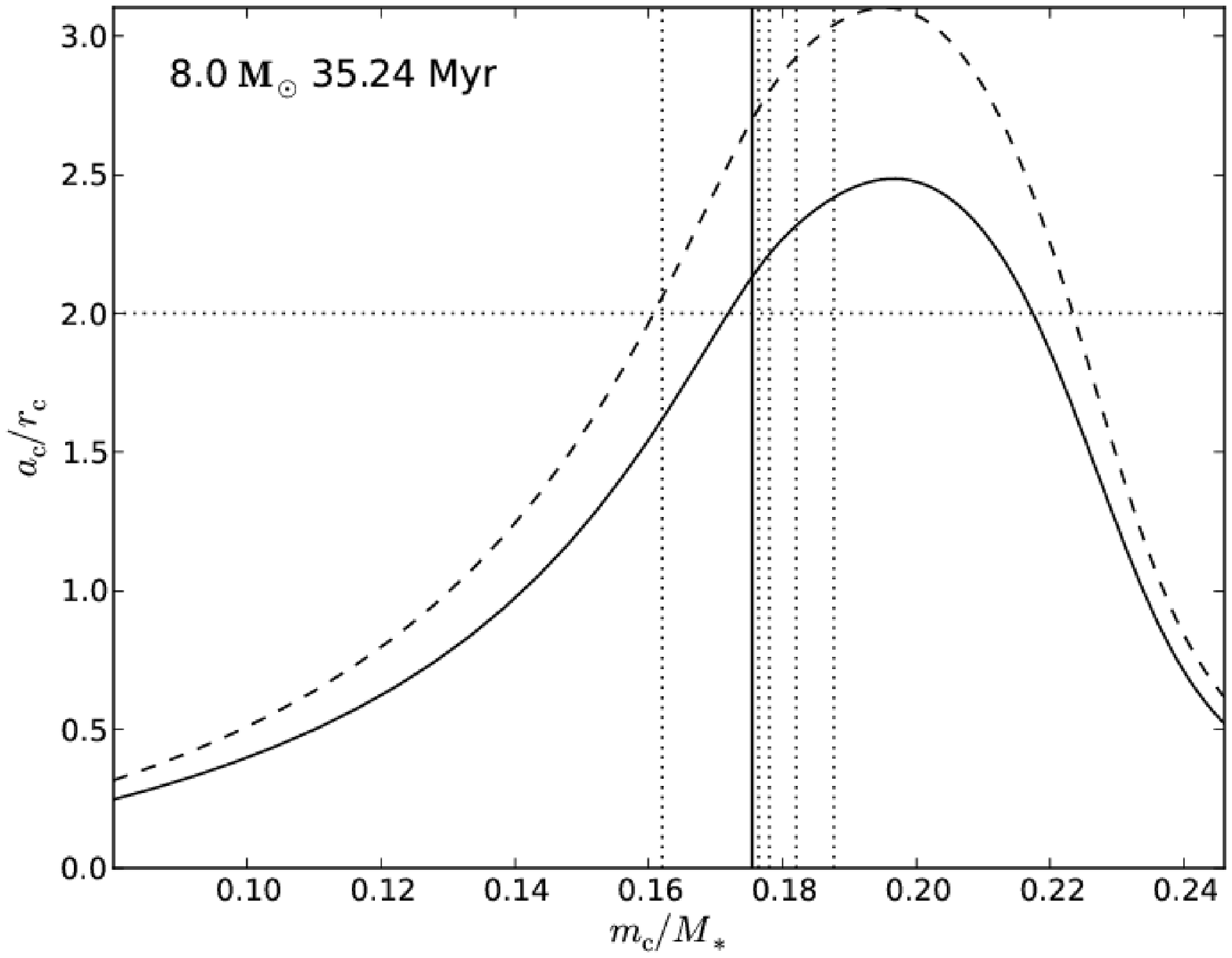} \\
\includegraphics[width=8.0cm]{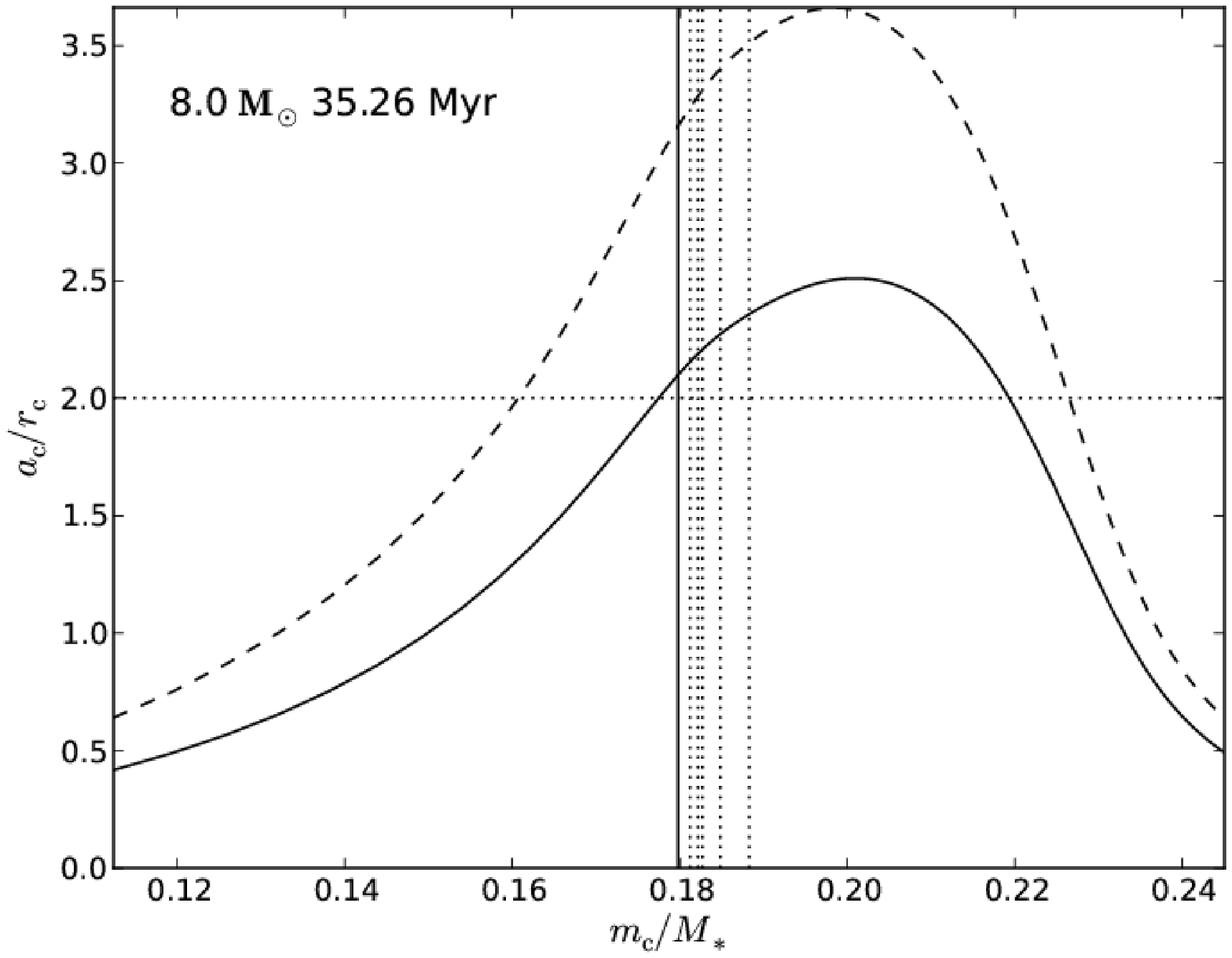}
\color{white}
\rule{8.0cm}{5.0cm}
\color{black}
\end{tabular}
\end{center}
\caption{Ratio of final core separation to the core radius as a function of core mass for the inspiraling binary closest to the dynamical stability limit for $m_\mathrm{i}=8.0\ M_\odot$ components at $35.22$ Myr (top left), $35.24$ Myr (top right), and $35.26$ Myr (bottom). The solid and dashed lines are the final separation as a function of core mass using $\alpha_\mathrm{CE}$ calculating using \eqref{EQ:AlphaCE} and $\alpha_\mathrm{CE}=1$ respectively. The horizontal dotted line shows the orbital separation where the two cores come into contact and are considered to be merged. The vertical solid line shows the nominal core mass and the vertical dotted lines show the core masses calculated using other prescriptions in \S\ref{SSSec:Core_Masses}.}
\label{Fig:afin}
\end{figure*}

\begin{figure*}[htp]
\begin{center}
\begin{tabular}{cc}
\includegraphics[width=8.0cm]{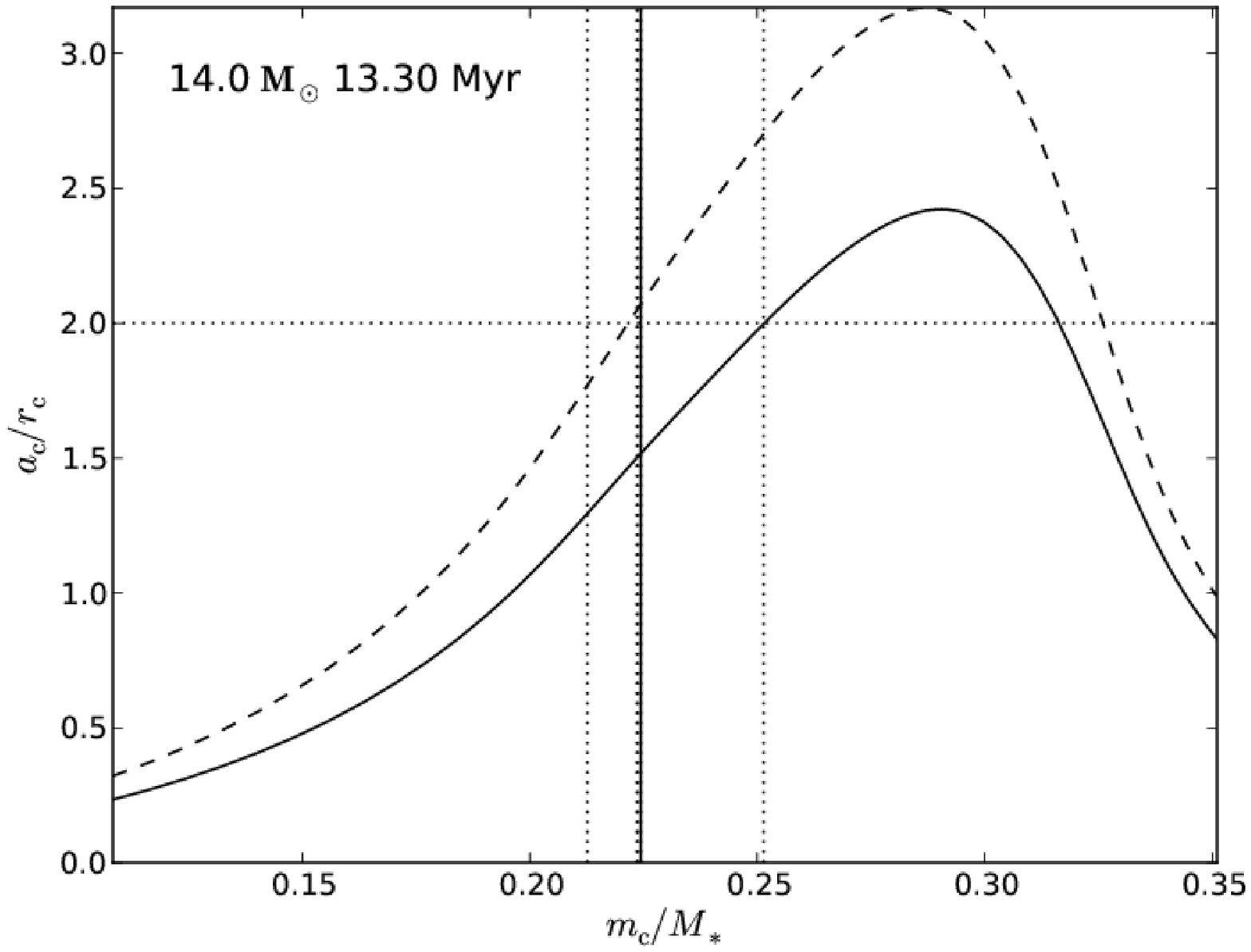}
\includegraphics[width=8.0cm]{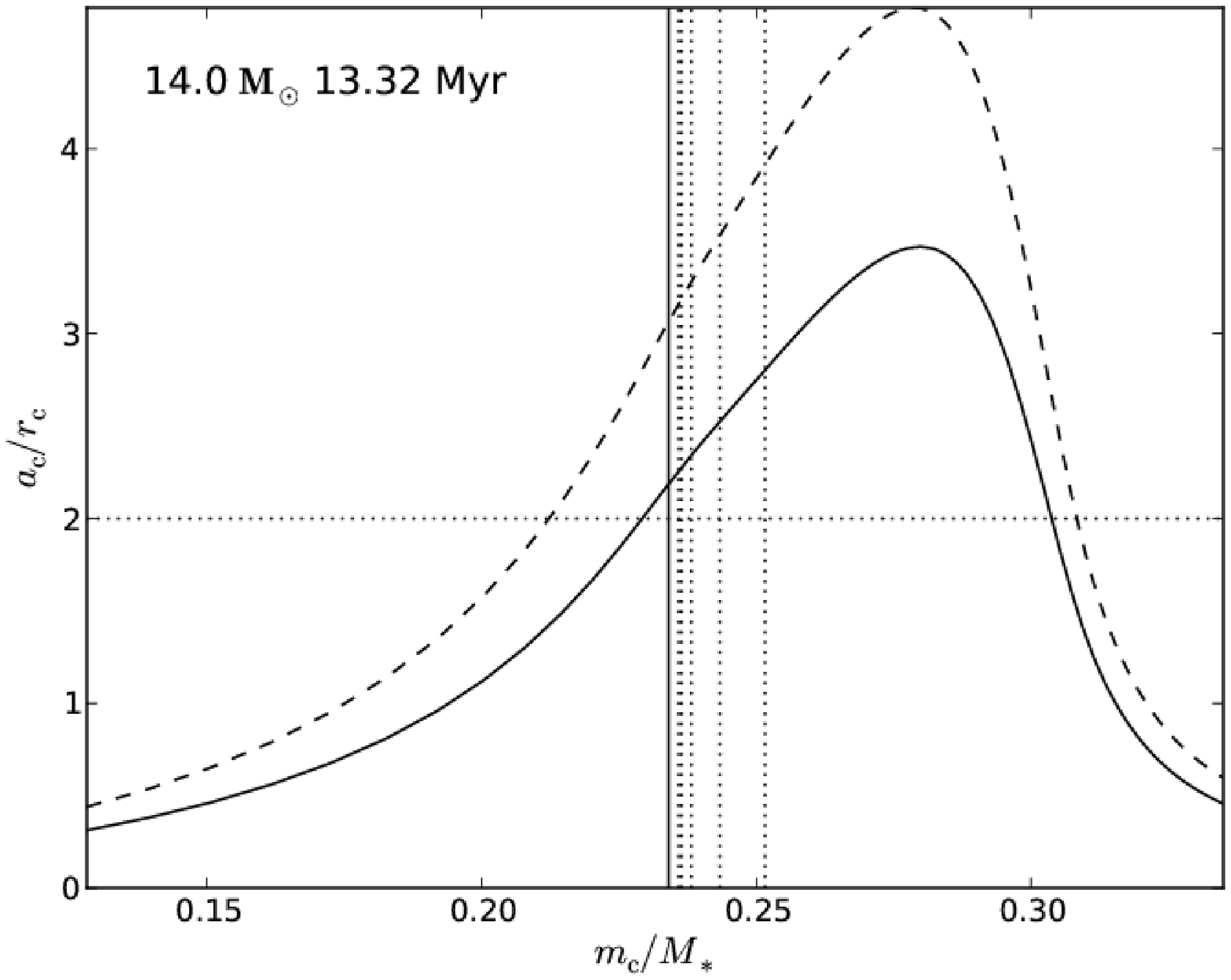} \\
\includegraphics[width=8.0cm]{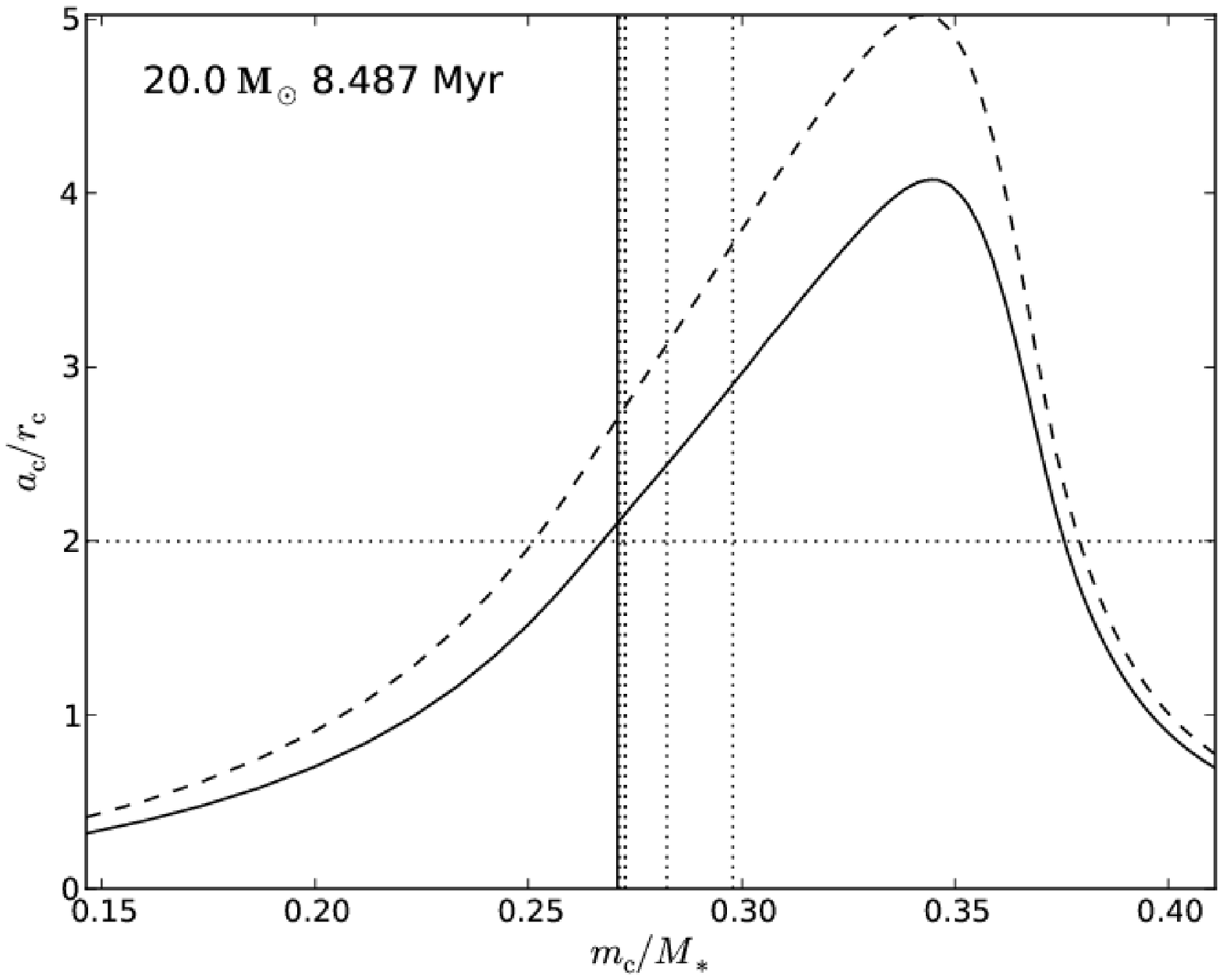}
\includegraphics[width=8.0cm]{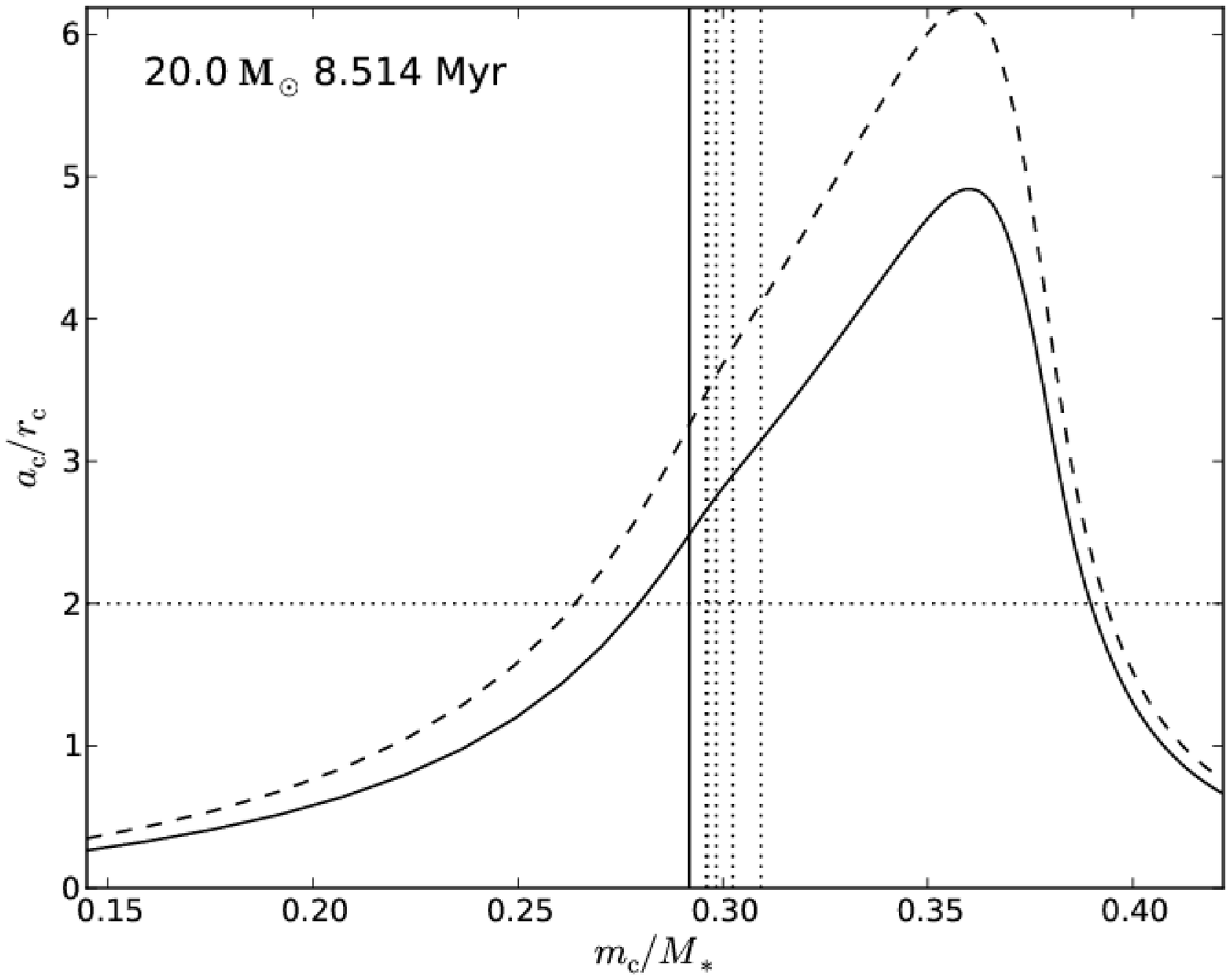}
\end{tabular}
\end{center}
\caption{Ratio of final core separation to the core radius as a function of core mass for the inspiraling binary closest to the dynamical stability limit for $m_\mathrm{i}=14.0\ M_\odot$ components at $13.30$ Myr (top left) and $13.32$ Myr (top right),  and $m_\mathrm{i}=20.0\ M_\odot$ components at $8.487$ Myr (bottom left) and $8.514$ Myr (bottom right). The solid and dashed lines are the final separation as a function of core mass using $\alpha_\mathrm{CE}$ calculating using \eqref{EQ:AlphaCE} and $\alpha_\mathrm{CE}=1$ respectively. The horizontal dotted line shows the orbital separation where the two cores come into contact and are considered to be merged. The vertical solid line shows the nominal core mass and the vertical dotted lines show the core masses calculated using other prescriptions in \S\ref{SSSec:Core_Masses}.}
\label{Fig:afin2}
\end{figure*}

Table~\ref{TBL:Remnant_Properties} summarizes the final core separations of the inspiraling binary closest to the dynamical stability limit for each set of dynamical calculations, using both $\alpha_\mathrm{CE}$ calculated using \eqref{EQ:AlphaCE} and the canonical value, $\alpha_\mathrm{CE}=1$, and both nominal core mass, $m_\mathrm{c,1}$, and the maximum calculated core mass (see Table~\ref{TBL:Core_Masses}).
Since we found that the mass loss after the {\it plunge-in} has a high ejection efficiency, the final separation assuming $\alpha_\mathrm{CE}=1$ may provide a more accurate prediction, unless the system experiences a second {\it plunge-in}, in which case, the efficiency depends on the mass fraction lost in the low-efficiency {\it plunge-in}.
The final core separation calculated using the maximum calculated core mass represents the outcome if the cores are able to retain some amount of gas after the common envelope has been ejected, as one of the methods used to calculate the core mass generally overestimates the size of the core for massive stars.
Figures~\ref{Fig:afin} and \ref{Fig:afin2} show the final separation as a function of core radius for these runs, and we see that the final separation is very sensitive to the size of the core, with both upper and lower constraints on producing a surviving helium core binary.
We find that for the nominal core size calculated in \S\ref{SSSec:Core_Masses}, in all but one binary the helium cores may survive common envelope evolution with an orbital period between $0.266$ hr and $1.01$ hr.

We find that older component binaries are less tightly bound, as the mass profile becomes more centrally dense as the stars age, resulting in less excess mass to be unbound.
Specifically, in the $m_\mathrm{i}=8.0\ M_\odot$ component mass binaries, the $\tau=35.22$ Myr system would undergo a merger while the $\tau=35.24$ Myr and $\tau=35.26$ Myr systems survive as a tightly bound binary.
In the $m_\mathrm{i}=14.0\ M_\odot$ component mass binaries, the $\tau=13.30$ Myr system requires that the excess mass be unbound at a minimum efficiency or it will undergo a merger, while the $\tau=13.32$ Myr system survives as a tightly bound binary at both values of $\alpha_\mathrm{CE}$.
In the $m_\mathrm{i}=20.0\ M_\odot$ component mass binaries, we predict both systems survive as a tightly bound binary.
For binaries with smaller component masses, there may be a minimum orbital separation such that the binary becomes unstable only after reaching a minimum age such that the cores are dense enough to avoid a merger.

\begin{deluxetable}{ccccccccccccc}
\tablewidth{17cm}
\tabletypesize{\small}
\tablecolumns{11}
\tablecaption{Remnant binary properties\label{TBL:Remnant_Properties}}
\tablehead{
    \colhead{$m_\mathrm{i}$} & \colhead{$\tau\ [\mathrm{Myr}]$} & \colhead{$a_\mathrm{i}\ [R_\odot]$} & \colhead{$r_\mathrm{c,1}\ [R_\odot]$} & \colhead{$\alpha_\mathrm{CE}$} & \colhead{$a_{\mathrm{f,}\alpha=1}\ [R_\odot]$} & \colhead{$P_{\mathrm{f,}\alpha=1}\ [\mathrm{hr}]$} & \colhead{$a_\mathrm{f}\ [R_\odot]$} & \colhead{$P_\mathrm{f}\ [\mathrm{hr}]$} & \colhead{$a_\mathrm{max}\ [R_\odot]$} & \colhead{$P_\mathrm{max}\ [\mathrm{hr}]$}
   }
\startdata
$8.0$  & $35.22$ & $458$  & $0.206$ & $0.615$ & $0.362$ & $0.215$ & $0.226$ & $0.106$ & $0.626$ & $0.244$ & \\
$8.0$  & $35.24$ & $564$  & $0.187$ & $0.781$ & $0.503$ & $0.352$ & $0.397$ & $0.247$ & $0.831$ & $0.529$ & \\
$8.0$  & $35.26$ & $593$  & $0.182$ & $0.651$ & $0.573$ & $0.427$ & $0.381$ & $0.231$ & $0.846$ & $0.421$ & \\
$14.0$ & $13.30$ & $1188$ & $0.344$ & $0.727$ & $0.711$ & $0.446$ & $0.521$ & $0.280$ & $1.506$ & $0.875$ & \\
$14.0$ & $13.32$ & $1505$ & $0.311$ & $0.710$ & $0.953$ & $0.693$ & $0.680$ & $0.311$ & $1.806$ & $1.094$ & \\
$20.0$ & $8.487$ & $2469$ & $0.429$ & $0.774$ & $1.158$ & $0.778$ & $0.901$ & $0.533$ & $2.564$ & $1.769$ & \\
$20.0$ & $8.514$ & $2522$ & $0.424$ & $0.760$ & $1.376$ & $1.010$ & $1.050$ & $0.673$ & $2.367$ & $1.521$ & \\
\enddata
\tablecomments{$m_\mathrm{i}$ is the initial mass of each component at birth, $\tau$ is the age of the components, $a_\mathrm{i}$ is the largest initial orbital separation of an unstable binary, $r_\mathrm{c,1}$ is the nominal core radius (see \S\ref{SSSec:Core_Masses}, $\alpha_\mathrm{CE}$ is the calculated efficiency of ejection, used in finding the final orbital separation, $a_\mathrm{f}$, and period, $P_\mathrm{f}$, assuming the common envelope is fully ejected. $a_{\mathrm{f,}\alpha=1}$ is the final orbital separation and $P_{\mathrm{f,}\alpha=1}$ is the period, assuming $\alpha_\mathrm{CE}=1$. $a_\mathrm{max}$ is the final orbital separation and $P_\mathrm{max}$ is the period, assuming $\alpha_\mathrm{CE}=1$ and the maximum calculated core mass. $Result$ specifies the final outcome of the systems, if the cores merge or remain as a tightly bound helium star binary using the canonical value of $\alpha_\mathrm{CE}=1$ and the nominal core mass.}
\end{deluxetable}
\pagebreak

\section{Conclusions}
\label{Sec:Conclusions}

\subsection{Summary of Results}
We performed 35 dynamical integrations of $q=1$ binary systems and found the dynamical stability limit as a function of age and the initial component mass.
The main conclusions of our study are as follows:
\begin{itemize}\setlength{\itemsep}{-4pt}
  \item[1)]For $q=1$ massive binaries there are two divergent outcomes: if $a_\mathrm{i}\ge a_\mathrm{crit}$ we expect that the binary remains stable through the contact phase and if $a_\mathrm{i}<a_\mathrm{crit}$ we expect that the binary will undergo an inspiral and become a short period helium star binary, where $a_\mathrm{crit}$ is the dynamical stability limit found for the binary at $\tau_\mathrm{max}$.
  Specifically, we find $a_\mathrm{crit}\approx2530\ R_\odot$ for the $m_\mathrm{i}=20.0\ M_\odot$ component binary, $a_\mathrm{crit}\approx1510\ R_\odot$ for the $m_\mathrm{i}=14.0\ M_\odot$ component binary, and $a_\mathrm{crit}\approx600\ R_\odot$ for the $m_\mathrm{i}=8.0\ M_\odot$ component binary.
  \item[2)]We ran a set of dynamical calculations of a binary with mass ratio very close to unity ($q=0.997$) and found that the results deviate from the $q=1$ binaries, becoming unstable at a smaller degree of contact, and suspect that this will be exaggerated as the mass ratio moves further from unity.
  \item[3)]We performed a resolution test, varying the particle per star from $N=10^4$ up to our standard value of $N=10^5$.
  We are particularly interested in the behavior of the system at the end of the {\it plunge-in}, where the orbital separation becomes relatively stable.
  The core separation at this transition actually decreases as the softening length increases (see Figure \ref{Fig:Res_rfin}), indicating that this transition is in fact a physical change from the {\it plunge-in} to the {\it self-regulated spiral-in}.
  \item[4)]Although resolution and timescale issues prevent us from accurately integrating the entirety of the spiral-in phase, we present a simple prescription to extrapolate the final state of remnant binaries (see \S\ref{SSec:Remnant_Properties}).
  We find that the final core separation is sensitive to the core size of the components.
  With the nominal core masses adopted, the $m_\mathrm{i}=20.0\ M_\odot$ binaries studied will not merge and will eventually become a very short period helium core binary ($0.53\ \mathrm{hr}<P_\mathrm{f}<1.01\ \mathrm{hr}$).
  For the $m_\mathrm{i}=14.0\ M_\odot$ component binary, at $\tau=13.30$ Myr, there is a minimum ejection efficiency required to prevent the cores from merging, whereas at $\tau=13.32$ Myr, the cores do not merge for the calculated $\alpha_\mathrm{CE}$ ($0.28\ \mathrm{hr}<P_\mathrm{f}<0.69\ \mathrm{hr}$).
  Finally, in the $m_\mathrm{i}=8.0\ M_\odot$ component binary, at $\tau=35.22$ Myr, the helium cores may merge, while at $\tau=35.24$ Myr and $\tau=35.26$ Myr the helium cores remain in a tightly bound binary ($0.23\ \mathrm{hr}<P_\mathrm{f}<0.43\ \mathrm{hr}$).
  We predict that there may be a maximum initial orbital separation as a function of $m_\mathrm{i}$ required to produce a surviving helium core binary in order for the stellar components to evolve a centrally-dense enough mass profiles to avoid merger of the cores.
\end{itemize}

\subsection{Comparisons to Paper 1}
We repeat the investigation of dynamical stability in $q=1$ binaries performed in {\it Paper 1} with an updated equation of state, taking into account both ideal gas and radiation pressure, and more realistic stellar models, generated from the stellar evolution code, {\it EZ}.
Approximating the stars as condensed polytropes with $\Gamma=5/3$, {\it Paper 1} reports an anti-correlation between the core mass and scaled orbital separation at both first contact, $r_{\eta=0}\equiv a_{\eta=0}/R_*$, and at the Roche limit, $r_{\eta=1}\equiv a_{\eta=1}/R_*$.
Conversely, we find that $r_{\eta=0}$ and $r_{\eta=1}$ increase with core mass, as higher-mass components are both more centrally dense and have larger regions dominated by radiation pressure, leading to gas crossing both the inner and outer Lagrangian points at larger orbital separations than in lower-mass component binaries.
As expected, the binary scan of the youngest low-mass component binary, $m_\mathrm{i}=8\ M_\odot$ at $\tau=35.22\ \mathrm{Myr}$, agrees very well with the results seen in {\it Paper 1}, with $r_{\eta=0}\simeq2.7$ and $r_{\eta=1}\simeq2.1$, while the higher-mass components have larger values of $r_{\eta=0}$ and $r_{\eta=1}$; specifically, for the $m_\mathrm{i}=20\ M_\odot$ binary at $\tau=8.514\ \mathrm{Myr}$ we find $r_{\eta=0}\simeq2.9$ and $r_{\eta=1}\simeq2.3$.

{\it Paper 1} varied the core mass from $0\le m_\mathrm{c}/M_\mathrm{*}\le1$ and finds that massive star binaries with $m_\mathrm{c}/M_\mathrm{*}\gtrsim0.15$ are both securlarly and dynamically stable from first contact through the Roche limit ($0\le\eta\le1$) while massive star binaries with $m_\mathrm{c}/M_\mathrm{*}\lesssim0.15$ reach the secular instability limit before the Roche limit.
In our dynamical calculations we also find that dynamical stability is possible even in very deep contact (typically, $\eta_\mathrm{crit}\sim0.8$ to $0.9$) but not all the way down to the Roche limit ($\eta=1$), where the dynamical stability limit depends on both the mass and age of the components.
Specifically, we find that systems are stable deeper into contact at younger ages and that they are able to survive at smaller scaled orbital separations, $a_\mathrm{i}/R_*$, at lower masses.
We remain in agreement with the criterion for secular stability reported in {\it Paper 1}, $m_\mathrm{c}/M_\mathrm{*}\gtrsim0.15$, as we do not observe secular instability in any of our binary systems, which all have $m_\mathrm{c}/M_\mathrm{*}\ge0.264$.

Due to our increased resolution and longer computation time, we are able to resolve the beginning of the {\it slow spiral-in} while the core separation is larger than the gravitational softening length of the cores.
We observe a phase of high-efficiency ($\alpha_{CE}\gtrsim0.9$) mass loss that occurs on the dynamical timescale from the tightly bound cores orbiting within the common envelope, with the subsequent shocks dumping thermal energy into the gas.
In one particularly long integration (see {\it Run 29} in Table~\ref{TBL:Results_Dyn}) we observe $\sim60\%$ by mass of the gas becoming unbound.

\subsection{Future Work}
Further dynamical calculations sampling the parameter space may uncover a prescription for the age at which a binary will undergo hydrodynamic instability as a function of initial component mass and orbital separation, and exploration into $q\ne1$ binaries may reveal a limit, below which the common envelope evolution deviates significantly from the $q=1$ case presented in this paper.
A more accurate treatment of the helium-core binary evolution after reaching the {\it slow spiral-in}, replacing the crude estimates presented in this paper, will provide a better set of initial conditions to predict the orbital parameters and masses of the eventual twin-neutron star binary.

\subsection{ACKNOWLEDGEMENTS}
We thank Evghenii Gaburov for work done to develop {\it StarSmasher}.
We also thank the referee for very thorough and constructive comments on improving the paper.
JH acknowledges support from an NSF GK-12 Fellowship funded through NSF Award DGE-0948017 to Northwestern University.
This work was supported by NSF Grant PHY-0855592 at Northwestern University and the Extreme Science and Engineering Discovery Environment (XSEDE), which is supported by National Science Foundation grant number OCI-1053575.
This material is based upon work supported by the NSF under grant No. AST-1313091.
The computations in this paper were performed on Northwestern University's HPC cluster Quest and a private cluster at Allegheny University.
This work used the SPLASH visualization software \citep{2007PASA...24..159P}.

\appendix

\section{Appendix}
\subsection{Artificial Viscosity in StarSmasher}
\label{A:Artificial Viscosity}
Our {\it SPH} evolution equations are described by \citep{2010MNRAS.402..105G}, but with a different artificial viscosity (AV) implementation.
In particular, the AV contribution to the acceleration of particle $i$ is calculated as
\begin{equation}
\dot{\bf v}_{{\rm AV},i} = -\sum_j \frac{1}{2} m_j\left[
\Pi_{ij} {\bf \nabla}_i W_{ij}(h_i)+
\Pi_{ji} {\bf \nabla}_i W_{ij}(h_j)
\right]\,. \label{vdotAV}
\end{equation}
We use the AV form
\begin{equation}
\Pi_{ij}=2 
{P_i\over\rho_i^2}
\left(-\alpha\mu_{ij}+
      \beta\mu_{ij}^2\right)\,,
\label{piDB}
\end{equation}
where
\[\mu_{ij}=\begin{cases}
  \medskip\displaystyle
  \frac{(\mathbf{v}_i-\mathbf{v}_j)\cdot(\mathbf{r}_i-\mathbf{r}_j)}{c_i|\mathbf{r}_i-\mathbf{r}_j|}f_i  \,,&\mathrm{if}\ \ (\mathbf{v}_i-\mathbf{v}_j)\cdot(\mathbf{r}_i-\mathbf{r}_j)<0\,\mathrm{;}\cr
  0                                                                                                      \,,&\mathrm{if}\ \ (\mathbf{v}_i-\mathbf{v}_j)\cdot(\mathbf{r}_i-\mathbf{r}_j)\ge0\,.\label{muDB}
\end{cases}\]
Here $c_i$ is the sound speed at the location ${\bf r}_i$ of particle $i$.  The Balsara switch $f_i$ for particle $i$ is defined by
\begin{equation}
f_i={|{\bf\nabla}\cdot{\bf v}|_i\over
     |{\bf\nabla}\cdot{\bf v}|_i+
     |{\bf\nabla}\times{\bf v}|_i+
     \eta'c_i/h_i}\,,
\label{fi}
\end{equation}
with $\eta'=10^{-5}$ preventing numerical divergences \citep{1995JCoPh.121..357B}.
The function $f_i$ approaches unity in regions of strong compression ($|{\bf\nabla}\cdot{\bf v}|_i>>|{\bf\nabla}\times{\bf v}|_i$) and vanishes in
regions of large vorticity ($|{\bf \nabla}\times {\bf v}|_i >>|{\bf\nabla}\cdot {\bf v}|_i$). Consequently, our evolution equations have the advantage that the AV is suppressed in shear layers.
We note that the AV term is not symmetric under interchange of the indices $i$ and $j$ (that is, $\Pi_{ij}\ne \Pi_{ji}$).
Such an approach reduces the number of arrays shared among parallel processes.
As the term in square brackets in equation (\ref{vdotAV}) is antisymmetric under the interchange of particles $i$ and $j$, momentum conservation is maintained.

The rate of change of the specific internal energy due to AV is
\begin{equation}
  \left({du_i\over dt}\right)_{\rm AV} = 
  \sum_j \frac{1}{2} \Pi_{ij}  m_j ({\bf v_i}-{\bf v_j})\cdot  
\nabla_iW_{ij}(h_i),
  \label{eq:dudtAV}
\end{equation}
which guarantees conservation of entropy in the absence of shocks.
It is straightforward to show total energy is conserved by our AV treatment: $\sum_i m_i({\bf v}_i\cdot{\bf \dot v}_{{\rm AV},i}+du_i/dt_{\rm AV})=0$.

\end{document}